%% file: mnras.tex
\documentclass[fleqn,usenatbib]{mnras}

\usepackage{newtxtext,newtxmath}

\usepackage{newtxtext,newtxmath}

\usepackage{graphicx}	
\usepackage{amsmath}	

\usepackage{amssymb}	
\usepackage{enumerate}
\usepackage{tablefootnote}
\usepackage[dvipsnames]{xcolor} 
\usepackage[normalem]{ulem} 

\title[PSR+BH binaries]{Modelling Neutron Star-Black Hole Binaries: Future Pulsar Surveys
and Gravitational Wave Detectors}

\author[Debatri Chattopadhyay et al.]{
\newauthor  Debatri Chattopadhyay,$^{1,2}$\thanks{E-mail: dchattopadhyay@swin.edu.au}
Simon Stevenson,$^{1,2}$
Jarrod R. Hurley,$^{1,2}$
Matthew Bailes,$^{1,2}$
\newauthor and Floor Broekgaarden$^{3}$
\\
$^{1}$ Centre for Astrophysics and Supercomputing, Swinburne University of Technology, John St., Hawthorn, Victoria- 3122, Australia \\
$^{2}$ The ARC Centre of Excellence for Gravitational Wave Discovery,  OzGrav\\
$^{3}$ Harvard-Smithsonian Center for Astrophysics, 60 Garden St., Cambridge, MA-02138, USA}

\date{Accepted XXX. Received YYY; in original form ZZZ}

\pubyear{2020}

\begin{document}
\label{firstpage}
\pagerange{\pageref{firstpage}--\pageref{lastpage}}
\maketitle
\begin{abstract}
Binaries comprised of a neutron star (NS) and a black hole (BH) have so far eluded observations as pulsars and with gravitational waves (GWs). We model the formation and evolution of these NS+BH binaries—including pulsar evolution—using the binary population synthesis code COMPAS. We  predict the presence of a total of 50-2000 binaries containing a pulsar and a BH (PSR+BHs) in the Galactic field.
We find the  population observable by the next-generation of radio telescopes, represented by the SKA and MeerKAT, current (LIGO/Virgo) and future (LISA) GW detectors. We conclude that the SKA will observe 1-80 PSR+BHs, with 0–4 binaries containing millisecond pulsars. MeerKAT is expected to observe 0-40 PSR+BH systems. Future radio detections of NS+BHs will constrain uncertain binary evolution processes such as BH natal kicks. We show that systems in which the NS formed first (NSBH) can be distinguished from those where the BH formed first (BHNS) by their pulsar and binary properties. We find 40\% of the LIGO/Virgo observed NS+BHs from a Milky-Way like field population will have a chirp mass $\geq 3.0$\,M$_\odot$. We estimate the spin distributions of NS+BHs with two models for the spins of BHs. The remnants of BHNS mergers will have a spin of $\sim$0.4, whilst NSBH merger remnants can have a spin of $\sim$0.6 or $\sim$0.9 depending on the model for BH spins. We estimate that approximately 25-1400 PSR+BHs will be radio alive whilst emitting GWs in the LISA frequency band, raising the possibility of joint observation by the SKA and LISA.
\end{abstract}

\begin{keywords}
pulsar -- black hole -- neutron star -- radio -- gravitational waves -- compact binary coalescence
\end{keywords}



\section{Introduction}
\label{sec:intro}

Pulsar-black hole (PSR+BH) binaries are theorized to form unique systems of extreme gravity 
where the precise rotation and detectable radio signal of the pulsar may help to probe and test theories related to the companion black hole. 
The intense curvature of space-time around such systems provides the perfect conditions to conduct tests of general relativity, alternate theories of gravity  \citep[e.g.][]{Kramer:2004hd, Simonetti:2010mk, Liu:2014uka,Shao:2014wja,Seymour:2018bce} and to probe quantum gravity \citep{Estes_2017}.

There are currently no Galactic PSR+BH binaries within the population of $\gtrsim 2000$ pulsars \citep{2005AJ....129.1993M} observed by radio telescopes around the world.
Surveys conducted with the next generation of radio telescopes with greatly enhanced sensitivity are expected to discover new pulsar systems. 
These include ongoing surveys with MeerKAT \citep[][]{Booth:2009} in South Africa 
and the Five-Hundred Metre Aperture Spherical Radio Telescope \citep[FAST;][]{2011IJMPD..20..989N} in China, as well as surveys planned in the near future with the Square Kilometre Array \citep[SKA;][]{Kramer:2004hd}, for example.  
The first observation of a PSR+BH binary is a key science target for these telescopes.

PSR+BH binaries form the radio-observable subset of the larger population of neutron star-black hole (NS+BH) binaries.
Mergers within this NS+BH population are promising gravitational-wave sources for the Advanced Laser Interferometer Gravitational-wave Observatory  \citep[aLIGO;][]{TheLIGOScientificDetector:2014jea}, Virgo \citep{TheVirgoDetector:2014hva} and the Kamioka Gravitational Wave Detector \citep[KAGRA;][]{Akutsu:2018axf}. 
No NS+BH mergers were observed in the first two observing runs of Advanced LIGO and Virgo, allowing the NS+BH merger rate to be constrained to $\mathcal{R}_\mathrm{NS+BH} < 610$\,Gpc$^{-3}$\,yr$^{-1}$ \citep{LIGOScientific:2018mvr}. 
The third LIGO/Virgo observing run (O3) has detected several gravitational-wave sources consistent with NS+BH mergers, including GW190814 \citep{Abbott:2020khf}, GW190425 \citep{Abbott:2020uma,Han:2020qmn,2020ApJ...890L...4K} and the low confidence NS+BH merger candidate GW190426 \citep{Abbott:2020niy}. However none of these candidates represent a confident detection of a canonical NS+BH.
In the future, NS+BH binaries are expected to be observed by gravitational-wave observatories such as the space-based Laser Interferometer Space Antenna \citep[LISA;][]{2017arXiv170200786A} and the Einstein Telescope \citep{Punturo_2010, Maggiore:2019uih}.

These potential gravitational-wave observations of NS+BH mergers raise the need to re-evaluate our previous understanding of their possible formation channels and subsequent evolutionary pathways \citep[see][for a review]{Siggurdson:2003}. 
Dynamical formation of NS+BH binaries in stellar triples \citep{2019ApJ...881...41L,Fragione:2019zhm,Fragione:2019mpq,2019ApJ...883...23H}, dense stellar environments such as star clusters \citep{2006NatPh...2..116G,2013MNRAS.428.3618C,2014MNRAS.442..207C} or the Galactic centre \citep{2011MNRAS.415.3951F,Fragione:2018yrb,Stephan:2019fhf,McKernan:2020lgr} has been shown to be possible. 
However, formation in star clusters is predicted to play an insignificant role in the net NS+BH merger rate \citep{Ye:2019xvf}. 

Most studies conclude that the formation of NS+BH binaries in clusters is highly inefficient owing to the population of black hole binaries suppressing the formation of neutron star binaries \citep[e.g.][]{2013MNRAS.428.3618C,Ziosi:2014sra,Ye:2019xvf,ArcaSedda:2020NatCommPhys,Fragione:2020wac,Hoang:2020ApJ},  
although some recent work has suggested that these environments could contribute a substantial fraction of NS+BH mergers observed with gravitational waves \citep[e.g.][]{2020arXiv200302277R, Santoliquido:2020bry}. 
For this paper, we do not consider the dynamical formation channel of NS+BH mergers.

We instead focus on the formation of NS+BHs (and PSR+BHs) from isolated massive binary stars in the Galactic field.
This evolutionary channel is analogous to a more massive variant of the canonical double neutron star (DNS) formation channels \citep[e.g.][]{Vigna-Gomez:2018dza,Vigna-Gomez:2020bgo,Chattopadhyay:DNS2019}. Many population synthesis studies have previously explored NS+BH formation from isolated massive binaries \citep[e.g.][]{1993MNRAS.260..675T,1994ApJ...423L.121L,1998A&A...332..173P,2002ApJ...572..407B,2003MNRAS.342.1169V,2005MNRAS.359.1517L,Dominik:2012,2014A&A...564A.134M,Mennekens:2016jcs}. 
There has recently been a renewed interest in population synthesis studies of NS+BH formation, largely motivated by gravitational-wave observations \citep[][]{Mapelli:2018wys,Giacobbo:2018etu,Giacobbo:2019fmo,Mapelli:2019bnp,Drozda:2020}. 

The current formation rate of NS+BH binaries remains uncertain. \citet{1991ApJ...379L..17N} estimated that the birth rate and number of NS+BH binaries in the Galaxy is comparable to the number of DNS binaries, while \citet{2005ApJ...628..343P} use population synthesis simulations to argue that the NS+BH birth rate is likely $< 10^{-7}$\,yr$^{-1}$, with $<10$ NS+BHs existing in the Galaxy at present, corresponding to one NS+BH per 100--1000 DNSs. \citet{2018MNRAS.481.1908K} find that NS+BH binaries are the most common double compact objects that form in their population synthesis simulations. 
Importantly though for our present study, they find that the NS+BH binaries where the NS forms first (NSBH) are 3--4 orders of magnitude more rare than the ones where the BH forms first (BHNS). 
This is because they assume highly inefficient mass transfer. \citet{Sipior:2004ck} discussed a formation channel where mass transfer reverses the mass ratio in the binary, leading to the formation of the NS before the black hole, allowing for the possibility of pulsar recycling.
Additionally, NS+BH binaries may form from Pop III binaries \citep[e.g.][]{Kinugawa:2016skw}, or from very massive, close binaries in low metallicity environments through chemically homogeneous evolution \citep{2017A&A...604A..55M} or from binary driven hypernovae in ultra-compact binaries \citep{2015PhRvL.115w1102F}. 

A number of potential electromagnetic counterparts to NS+BH mergers have been proposed \citep[for a review, see][]{2017LRR....20....3M} including short gamma-ray bursts \citep[e.g.][]{1984SvAL...10..177B,1991AcA....41..257P,1993Natur.361..236M,1999ApJ...527L..39J,2007PhR...442..166N,2008ApJ...675..566O} optical transients and kilonovae \citep{1998ApJ...507L..59L,Barbieri:2019sjc,Barbieri:2019kli}, and maybe even fast radio bursts \citep{2015ApJ...814L..20M, Bhattacharyya:2017orq, 2018PhRvD..98l3002L}. Such possible electromagnetic signatures have been studied using population synthesis \citep{Postnov:2020hqh} and disk ejecta outflow of such mergers have also been modelled \citep{Fernandez:2020}. 
No electromagnetic counterparts to the NS+BH merger candidates from O3 were observed  \citep[e.g.][]{Coughlin:2019xfb,Dobie:2019ctw,Andreoni:2019qgh,Ackley:2020qkz,Vieira:2020lze,Kawaguchi:2020osi,Anand:2020NatAs}. 

NS+BH mergers observed both electromagnetically and with gravitational waves may be used to probe cosmology, astrophysical processes or thermodynamic equations of matter. 
For example, the Hubble constant measurement that embeds the expansion rate of the universe can be computed from  multi-messenger observations of NS+BH mergers  \citep[e.g.][]{Nissanke:2009kt,Cai:2016sby,Vitale:2018wlg,Feeney:2020kxk}. 
Furthermore, 
gravitational wave observations of NS+BH mergers can be used to probe the putative mass gap between neutron stars and black holes \citep{Bailyn:1997xt,2010ApJ...725.1918O,2011ApJ...741..103F,2012ApJ...757...36K}, though recent observations suggest that low mass black holes do exist \citep{Thompson:2019Science,Abbott:2020khf}.  The GW190814 signal; generated from the merger of $\approx$23 M$_\odot$ BH and a $\approx$2.6 M$_\odot$ compact object has ignited the possibility of the heaviest NS or lightest BH ever recorded, 2.6 M$_\odot$ being in the region of the presumed mass gap \citep{Abbott:2020khf}.
Significant uncertainties in estimating component masses makes distinguishing DNS and NS+BH  binaries difficult in practice \citep[][]{Hannam:2013uu,Littenberg:2015tpa,Abbott:2020uma,Chen:2020aiw,Tang:2020lfn,Fasano:2020arxiv}.

The gravitational-wave signal observed from NS+BH mergers encodes the NS equation of state, hence such detections can be used to determine the mass-radius relation of NSs \citep[][]{Vallisneri:2000PRL,2011PhRvD..84j4017P,2014ApJ...791L...7P,2012PhRvD..85d4061L,2014PhRvD..89d3009L,Kumar:2016zlj,Ascenzi:2018mwp}.
NS+BH mergers are also potentially important sources of $r$-process enrichment \citep{1974ApJ...192L.145L,1976ApJ...210..549L,2004NewA....9....1D,Korobkin:2012uy,2018ApJ...869...50D}. 
The abundance of $r$-process elements in the Galaxy have been used to place an upper limit on the NS+BH merger rate \citep{2014ApJ...795L...9B}. 

Finally, predictions about the population of NS+BH mergers can inform the input parameters for detailed numerical relativity simulations \citep[e.g.][]{Paschalidis:2014qra,Kawaguchi:2015bwa,Kiuchi:2015qua,Ruiz:2018wah,Foucart:2019bxj}. 
These simulations have enabled detailed gravitational waveform modelling for NS+BH systems \citep{Thompson:2020nei,Matas:2020wab} 
with recent studies showing that statistical uncertainties should dominate over systematic uncertainties for estimating the source properties of typical NS+BH mergers with Advanced LIGO and Virgo \citep[][]{Huang:2020pba}. 

In this paper we model NS+BH binaries to make predictions about population numbers and merger statistics. 
We do this with the added feature of detailed pulsar evolution so that we can also model the subset of PSR+BH binaries. 
Previous binary population synthesis studies in this vein include  
\citet{2009MNRAS.395.2326K} and \citet{Kiel:2010MNRAS} who modelled NS+BH formation using a binary evolution code coupled with a model for pulsar evolution \citep{Kiel:2008xw} to probe the properties of Galactic PSR+BH binaries, while \citet{2018MNRAS.477L.128S} estimated an upper limit of $<80$ Galactic PSR+BH binaries and showed that FAST is expected to detect $\sim10$\% of the population.

In our approach we combine a detailed model for pulsar evolution with 
the rapid binary population synthesis code COMPAS \citep{Stevenson:2017tfq} to allow us to probe how different assumptions in the pulsar model affect population outcomes. 
We also use the code NIGO \citep{2015A&C....12...11R, RossiHurley:2015} to distribute the binaries in a Milky-Way-like gravitational potential and evolve their orbits, accounting for the effects of natal kicks. 
As a final step in the process 
we then segregate the radio-alive PSR+BHs (i.e. one of the members is a pulsar, a subset of the net population of NS+BHs) and account for the radio selection effects. 
In our analysis we use mock-surveys of Parkes, MeerKAT and the SKA to determine the radio observable properties of PSR+BHs 
while 
the net NS+BH population is analysed from the perspective of gravitational wave detectors LIGO and LISA.
We present a comparative exploration of the NS+BH binaries in light of radio and gravitational waves, from the background population to the observable sub-population for current and future detectors. 

This paper is structured as follows: in Section~\ref{sec:methods} we describe our methods and in Section~\ref{subsec:formation_channels} we describe the main formation channels for PSR+BH binaries. 
We present results for PSR+BHs observable by future pulsar surveys in Section~\ref{sec:results_radio}, and gravitational-wave observables in Section~\ref{sec:gravitational_waves}.
We discuss our results in Section~\ref{sec:discussion}. 
Finally we conclude in Section~\ref{sec:conclusions}.



\section{Binary population synthesis including pulsar evolution}
\label{sec:methods}

We utilise the COMPAS suite \citep{Stevenson:2017tfq,Vigna-Gomez:2018dza,Stevenson:2019rcw,Neijssel:2019,Vigna-Gomez:2020bgo,Chattopadhyay:DNS2019} to simulate the evolution of a population of massive binaries. 
COMPAS includes a rapid binary population synthesis code which uses parameterised models of stellar and binary evolution \citep[][]{Hurley:2000pk,2002MNRAS.329..897H}.

We make standard assumptions regarding the initial conditions of massive binaries; the primary mass $m_1$ is drawn from an initial mass function \citep{Kroupa:2000iv}, the mass ratio of the binary $q=m_2/m_1$ is drawn from a uniform distribution \citep[e.g.][]{Sana:2012} and the initial binary separations are drawn from a flat-in-the-log distribution \citep{1924PTarO..25f...1O,1983ARA&A..21..343A,Sana:2012} within the range of 0.1\,AU and 1000\,AU. 
We assume all binaries initially have circular orbits \citep{2002MNRAS.329..897H}.
COMPAS uses Monte Carlo methods to sample binaries with initial conditions drawn from the distributions described above.

As in \citet{Chattopadhyay:DNS2019} we assume a constant star formation history for the Milky Way for the past 10\,Gyr at solar metallicity, $Z=0.0142$ \citep{Asplund:2009}, appropriate for PSR+BHs formed in the Milky Way. 

We evolve the motion of PSR+BH binaries in the Galactic potential using the Numerical Integrator of Orbits \citep[NIGO;][]{2015A&C....12...11R}. 
Details of our model assumptions for the Galactic potential can be found in \citet{Chattopadhyay:DNS2019}. 

In the following subsections we describe some of the most important assumptions we make in modelling supernovae (Section~\ref{subsec:supernovae}), pulsar evolution (Section~\ref{subsec:pulsarPhysics}) and radio selection effects (Section~\ref{subsec:radio_selection_effects}). We describe our suite of models in Section~\ref{subsec:models}.

\subsection{Supernovae}
\label{subsec:supernovae}

Neutron stars and black holes are formed from massive stars in different types of supernovae (SNe) depending on the formation channel. 
The type of supernova impacts the resultant binary evolution by determining its orbital period and also the eccentricity of the double compact object through natal kicks. 

\subsubsection{Compact object masses and radii}
\label{subsubsec:compact_object_masses_radii}

We determine the mass of neutron stars and black holes at birth based on the pre-SN mass and final core mass of the collapsing star, using the `delayed' prescription from \citet{Fryer:2012}. 
In our model, we delineate between whether a NS or BH is formed as the remnant of a massive star based solely upon the compact object mass. 
Causality demands that the maximum neutron star mass be below $\sim 3$\,M$_\odot$ \citep{Rhoades:1974, Kalogera:1996}, whilst the most massive pulsars currently known are $\sim 2$\,M$_\odot$ \citep[][]{Demorest:2010bx,Antoniadis:2013pzd,Cromartie:2019kug}.
Gravitational wave observations of GW170817 have constrained the maximum stable and non-rotating neutron star mass to $\lesssim 2.3$\,M$_\odot$, assuming prompt collapse of the merger product \citep[][]{Margalit:2017dij, Ruiz:2017due, LIGOScientific:2018mvr, Shibata:2019ctb, LIGOScientific:2019eut,Chatziioannou:2020pqz}. The maximum mass of  uniformly rotating stable NSs calculated analytically is approximately 20\% more massive than the non-rotating case \citep{Cook:1992, Friedman:1987}. For a deferentially rotating NS, equilibrium solutions can be $\gtrapprox$3M$_\odot$ \citep{Baumgarte:1999cq} to as high as ``{\"u}bermassive'' NSs of $\approx4.5-7$M$_\odot$ \citep{Espino:2019ebx}. Our model assumes a maximum neutron star mass of $2.5$\,M$_\odot$.

We have explored the effect of changing the remnant mass prescription to `rapid' \citep{Fryer:2012} for one of our models. 
In contrast to `delayed', the `rapid' prescription allows SNe explosions to occur within 250ms after the bounce shock from the collapsing star, creating energetic explosions ($>10^{41}$ Joules). 
By construction, the `rapid' prescription creates a mass-gap for NSs and BHs between $\approx$ 2 M$_\odot$ and $\approx$ 5 M$_\odot$ as apparently perceived from the observations of X-ray binaries \citep{Bailyn:1997xt, Ozel:2010}.
However, recent observations in electromagnetic \citep{Wyrzykowski:2016, Giesers:2019, Thompson:2019Science, Liu:2019lfc, Jayasinghe:2021uqb}
and gravitational waves  \citep{Abbott:2020khf, Abbott:2020niy} suggest a need to revisit the perceived mass gap.
See \citet{Broekgaarden:2020NSBH} for further discussions.

We assume all NSs have radii of 12\,km, consistent with recent measurements from gravitational waves \citep{Abbott:2018exr,Capano:2019eae}, gravitational wave and pulsar surveys \citep{Landry:2020vaw} and the Neutron star Interior Composition Explorer  \citep[NICER;][]{Riley:2019yda,Miller:2019cac,Raaijmakers:2019dks}. 
The moment of inertia is computed using an equation of state independent relation from \citet{Lattimer:2004nj}.


\subsubsection{Natal kicks of neutron stars and black holes}
\label{subsubsec:neutronStarBlackHoleKicks}

Each NS acquires a kick at its birth time, 
referred to as the NS natal kick \citep{GunnOstriker:1970, Helfand:1977, Lyne:1994}, with the magnitude of the kick velocity drawn from a distribution according to the type of SN it originates from. 
These kick distributions can be constrained empirically with observations of the space velocities of Galactic pulsars \citep[e.g.][]{Hobbs:2005, Verbunt:2017zqi}, by comparing population synthesis predictions to these observations \citep[Willcox et al. in prep]{Pfahl:2002ApJ,2009MNRAS.395.2326K} 
and through detailed modelling \citep[e.g.][]{Muller:2018utr}. 
On average, core collapse (CC) SNe \citep{Fryer:2012} are expected to generate a stronger birth kick than electron capture (EC)  \citep{Nomoto:1984, Nomoto:1987, Gessner:2018ekd} and ultra-stripped (US) SNe \citep{Tauris:2013, Tauris:2015}. 
In turn, because both EC and USSNe are either related to or enhanced by mass-transfer processes, 
it is expected that NSs in interacting binaries are more likely to obtain smaller kicks than their isolated counterparts. 
ECSNe explosions occur for stars at the low-mass end of the SN progenitor mass spectrum. 
These explosions are thought to be more symmetric less energetic and faster than standard CCSNe, leading to typically lower natal kick velocities than the latter \citep{Podsiadlowski:2004, Gessner:2018ekd}. 
For massive stars in binaries the potential loss of the envelope after hydrogen burning owing to the presence of a close companion leads to a widening of the mass range permissible for ECSNe relative to single stars \citep{Poelarends:2017dua}.
In the USSN case, the presence of a compact object companion in a close binary can cause the SN progenitor to lose most of its envelope,  leading to the `ultra-stripping' \citep{Tauris:2013, Tauris:2015}. 
This lower-mass envelope has a lower binding energy and creates less gravitational pull on the core during the SN, resulting in a smaller magnitude kick velocity than otherwise expected \citep{Tauris:2015, Suwa:2015saa, Muller:2018utr}. The natal kick sustained solely due to mass loss \citep[termed as 'Blaauw kick,][]{Blaauw:1961}, can be high enough to increase the orbital eccentricity of the binary.
For more details on the modelling of SNe kicks in COMPAS, see \citet{Vigna-Gomez:2018dza}.

For all three types of SNe we assume that natal kicks are drawn from a Maxwellian distribution \citep{Hansen:1997zw} with a one-dimensional root-mean-square $\sigma$. 
For CCSNe we assume $\sigma_\mathrm{CCSNe} = 265$\,km s$^{-1}$ \citep{Hobbs:2005}, while for both USSNe and ECSNe we assume $\sigma_\mathrm{ECSNe} = \sigma_\mathrm{USSNe} = 30$\,km s$^{-1}$ \citep{Pfahl:2002, Podsiadlowski:2004,Gessner:2018ekd, Suwa:2015saa, Muller:2018utr}. 
For more details on NS natal kicks in binary population synthesis simulations, see \citet{2010MNRAS.407.1245B} and \citet{Vigna-Gomez:2018dza}.

Similarly to NSs, BHs are also expected to obtain a birth kick velocity. 
There have been efforts to determine BH natal kicks from the observations of low mass X-ray binaries \citep{Repetto:2012, Repetto:2015kra,Mandel:2015eta,Repetto:2017gry}. 
However, debate still remains on the magnitude of the BH kick and the factors (explosion, asymmetric mass ejection, initial-final mass relation) that play key roles in determining it \citep[e.g.][]{Repetto:2012,Janka:2013hfa,Sukhbold:2015wba}. 

Our default assumption is that black hole kicks are drawn from a similar kick distribution as for NSs but are reduced by the fraction of ejected mass which falls back on to the proto-compact object \citep[][]{Fryer:2012} often termed as the `fallback mass'. We also present results from two variations, where we either give black holes no kicks at birth, or give black holes the same kicks as neutron stars at birth without scaling by the fallback mass (see Table~\ref{tab:NSBH}). In the frame of reference of the NS or BH progenitor star undergoing a SN, its natal kick is assumed to be isotropic.

\subsection{Mass transfer}
\label{subsec:mass_transfer}

Mass transfer is important for the formation of NSBH binaries as it can lead to a mass ratio reversal of the binary \citep{Sipior:2004ck}.
The efficiency of mass transfer is given by the ratio $\beta_\mathrm{MT}$ of the mass accreted $\Delta M_\mathrm{acc}$ by a star to the mass donated $\Delta M_\mathrm{don}$
\begin{equation}
    \beta_\mathrm{MT} \equiv \frac{\Delta M_\mathrm{acc}}{\Delta M_\mathrm{don}} \,  . 
    \label{eq:beta_def}
\end{equation}
There is observational evidence for a range of mass transfer efficiencies in massive binaries depending on the masses and orbital period of the binary \citep{deMink:2007af}. 
\citet{2005A&A...435.1013P} study three Galactic post mass-transfer Wolf-Rayet-O-star binaries. 
They show that highly inefficient ($\beta_\mathrm{MT} < 0.1$) mass transfer is required to explain the current properties of these binaries.  
This finding is corroborated by \citet{2016ApJ...833..108S}. 
See also early work by \citet{Vanbeveren:1982A&A}.
Other systems are consistent with having undergone almost fully conservative mass transfer \citep{2018A&A...615A..30S}.

In our standard model, $\beta_\mathrm{MT}$ is calculated based upon the ratio of the thermal timescales for the donor and accretor \citep{2002MNRAS.329..897H,Shao:2014aca,Schneider:2015ApJ...805...20S,Stevenson:2017tfq,2020arXiv200300195V}
\begin{equation}
    \beta_\mathrm{MT} = \mathrm{min}(1, 10 \frac{\tau_\mathrm{acc}}{\tau_\mathrm{don}}) \, ,
    \label{eq:beta_thermal}
\end{equation}
where, $\tau_\mathrm{acc}$ and $\tau_\mathrm{don}$ are the thermal (Kelvin-Helmholtz) timescales for the accretor and donor respectively. 
The progenitors of most of our NS+BH binaries initiate the first episode of MT either late on the main sequence (case AB), or early on the Hertzsprung gap (case B). In both of these cases, our method for determining the efficiency of mass transfer (see section~\ref{subsec:mass_transfer}) results in close to completely conservative mass transfer ($\beta_\mathrm{MT} \approx 1$). Wider interacting binaries experience less conservative mass transfer ($\beta_\mathrm{MT} < 0.1$) \citep[see][]{Schneider:2015ApJ...805...20S}.

While our model generally predicts more conservative mass transfer in shorter orbital period binaries \citep[][]{Schneider:2015ApJ...805...20S,Shao:2014aca}, recent work has suggested that mass transfer in Be X-ray binaries may be more efficient than assumed in our default model \citep{2020arXiv200300195V}.
In all models accretion onto a compact object is limited to the Eddington rate \citep{Chattopadhyay:DNS2019}, resulting in $\beta_\mathrm{MT} \approx 0$.

\subsection{Pulsar Physics}
\label{subsec:pulsarPhysics}

We model and evolve the properties of pulsars with time using the pulsar code implemented in COMPAS  \citep[as discussed in detail by][]{Chattopadhyay:DNS2019}, based on the earlier works of \citet{FaucherGiguere:2005ny}, \citet{Kiel:2008xw} and \citet{2011MNRAS.413..461O}.

We assume all neutron stars are born as pulsars. Every NS is assigned a natal pulsar spin and surface magnetic field at its formation. 
In our Fiducial model, they have spin periods  drawn from an uniform distribution within the range of 10--100\,ms and magnetic field strengths drawn from an uniform distribution between the range of 10$^{10}$--10$^{13}$\,G, motivated by observations of young pulsars \citep[see][for details]{Chattopadhyay:DNS2019}. 
The radio luminosity of pulsars at 1400\,MHz is drawn from a log normal luminosity distribution as described by \citet{Szary:2014dia}. 

Pulsars are assumed to be rotation powered. The magnetization and the angular frequency decay over time as
\begin{equation}
    B = (B_0 - B_\mathrm{min})\times\exp(-t/\tau_d)+B_\mathrm{min} ,
    \label{MagneticFieldIsolated} 
\end{equation}

and
\begin{equation}
     \dot{\Omega} = -\frac{8\pi B^2R^6\Omega^3\sin^2\alpha}{3 \mu_0 c^3 I } ,
     \label{SpinDownIsolated}
 \end{equation}

where $B$ is the surface magnetic field of the pulsar (in Tesla), $\tau_d$ is the magnetic field decay timescale (a free parameter in our model) in the same time units as $t$, $B_0$ is the initial surface magnetic field (Tesla) and $B_\mathrm{min}$ (Tesla) is the minimum surface magnetic field strength at which we assume the magnetic field decay ceases. 
The angular frequency $\Omega$ (s$^{-1}$) changes at a rate $\Dot{\Omega}$ (s/s), $R$ is the radius of the pulsar (m), $\alpha$ is the angle between the axis of rotation and the magnetic axis, $c$ is the speed of light (m/s), $\mu_0$ is the permeability of free space (Tesla-m/Ampere) and $I$ is the moment of inertia of the pulsar (kg m$^2$). 
We assume $B_\mathrm{min}= 10^8$\,G, i.e. $10^4$\,Tesla, for all models in this paper \citep{ZhangKojima:2006, 2011MNRAS.413..461O, Chattopadhyay:DNS2019}. 
The angular frequency $\Omega$ is related to the spin period of the pulsar $P$ (s) through $P = \frac{2\pi}{\Omega}$ while $\Dot{\Omega}$ is related to spin down rate 
$\Dot{P}$ (s/s) through $\dot{P}= -\frac{\dot{\Omega}P}{\Omega}$.

The pulsar may accrete matter from its still-evolving companion and spin up due to the exchange of angular momentum \citep{1994MNRAS.269..455J, Kiel:2008xw, Chattopadhyay:DNS2019}.
We assume that the magnetic field strength of the pulsar is reduced (or buried) by the accumulation of accreted matter \citep{ZhangKojima:2006}. 
The increase in the rotational velocity of the pulsar and the decrease in its surface magnetic field due to accretion is described by
\begin{equation}
   \dot{\Omega}_\mathrm{acc} = \frac{\epsilon V_\mathrm{diff} R_A^2 \dot{M}_\mathrm{NS}}{I} , 
   \label{AngularMomentumAccretion}
\end{equation}
and
\begin{equation}
   B = (B_0 - B_\mathrm{min})\times\exp(-\Delta M_\mathrm{NS}/\Delta M_d)+B_\mathrm{min} , 
   \label{MagneticFieldAccretion}
\end{equation}
where $\dot{\Omega}_\mathrm{acc}$ is the angular acceleration of the pulsar due to accretion, $\epsilon$ is the accretion efficiency factor, $V_\mathrm{diff}$ (m/s) is the difference between Keplerian angular velocity at the magnetic radius $\Omega_K|_{R_m}$ and the co-rotation angular velocity $\Omega_\mathrm{co}$, $R_A$ is the magnetic radius (m), $\dot{M}_\mathrm{NS}$ is the rate of change of mass of the neutron star due to accretion (kg/s), $\Delta M_\mathrm{NS}$ is the amount of mass accreted by the neutron star and $\Delta M_d$ is called the magnetic field decay mass-scale of the pulsar (in the same units as $\Delta M_\mathrm{NS}$). 
The magnetic radius is assumed to be half the Alfven radius $R_A = R_\mathrm{Alfven}/2$. 

The magnetic field decay mass scale $\Delta M_d$ is also a free parameter in our simulations. 
In this paper we allow for pulsar mass accretion via the formation of an accretion disk as well as during common envelope. 
The common envelope mass accretion is modelled using linear fitting from Fig.(4) of \citet{MacLeod:2014yda} as described by
Equation (18) of \citet{Chattopadhyay:DNS2019}. The angle between the rotational and magnetic axis of the pulsar $\alpha$ is assumed to be 45 degrees for all our models. 
This makes sin$^2\alpha = 0.5$ in equation.~\ref{SpinDownIsolated}.

The evolution of the pulsar is highly sensitive to the unconstrained magnetic field decay time-scale and mass-scale parameters. 
Our Fiducial model uses values of these parameters chosen to match the Galactic DNS population. 
We also vary these parameters and demonstrate the impact of uncertainties on our predictions. 
Our models are phenomenological, with the {\it Fiducial} model parameters constrained to produce a good match with the Galactic DNS population \citep{Chattopadhyay:DNS2019}.

The loss of magnetization of a slowly-rotating pulsar over time becomes insufficient to produce electron-positron pairs \citep{Chen:1993, RudakRitter1994, Medin:2010}, rendering the pulsar radio-dead. 
This may be described by utilising empirical death lines \citep{RudakRitter1994} or by imposing a radio efficiency threshold limit as described by \citet{Szary:2014dia}. 
We use a hybrid approach of the two as described in detail in Section 2.3 of \citet{Chattopadhyay:DNS2019}. 

\subsubsection{Common Envelope evolution}
\label{subsec:common_envelope_evolution}

Dynamically unstable mass transfer leads to common envelope (CE) evolution \citep{1976IAUS...73...75P,Ivanova2013}. We treat CE evolution using the standard energy prescription \citep{Webbink:1984,deKool:1990,Ivanova2013}, where a fraction $\alpha_\mathrm{CE} = 1$ of the orbital energy is available to unbind the CE. 
We use a fit to detailed single stellar models \citep{XuLi:2010} to compute stellar envelope binding energies \citep[for more details, see][]{Howitt:2019vik}. 
Since CE evolution occurs on a timescale shorter than the time resolution in COMPAS, we assume that it is an instantaneous process.

It is theoretically unclear whether Hertzsprung gap (HG) stars are able to survive CE evolution \citep[e.g.][]{2007ApJ...662..504B,Dominik:2012}. 
We choose to allow such binaries to survive CE evolution, following \citet{Vigna-Gomez:2018dza}, terming the scenario as `optimistic' CE evolution. 
We explore the situation where such binaries do not survive the CE and merge, the `pessimistic' case, through one of our models (see section~\ref{subsec:models}) and show the impact of this assumption on our results in Section~\ref{sec:results_radio}. 

Mass accretion onto NSs during CE evolution is modelled using a fit to results from \citet{MacLeod:2014yda} as detailed in \citet{Chattopadhyay:DNS2019}. 
In our model, this results in spin up of the pulsar and burial of its magnetic field (c.f. Section~\ref{subsec:pulsarPhysics}).

It remains uncertain whether NSs can also accrete angular momentum (and thus be spun up) during CE evolution or not \citep[see e.g. discussion and references in][]{Barkov:2011MNRAS.415..944B}. 
In our Fiducial model, we allow for the recycling of pulsars during CE evolution. The amount of angular momentum gained by the pulsar depends on the accreted mass, unlike during RLOF where the rate of accretion is the deciding variable (for details see \citet{MacLeod:2014yda}). We calculate the typical case of CE angular momentum gain in a NS from a BH-progenitor to be $\sim\mathcal{O}(10^{41})$ kg-m$^2$/sec. 
Due to the uncertainty, we also present results for a model where no accretion of mass or angular momentum is allowed during CE evolution (see Section~\ref{subsec:models}).
NSs accreting during CE evolution may also eject processed material \citep{Keegans:2019gje}.

\citet{Barkov:2011MNRAS.415..944B} study a scenario in which pulsars are recycled during CE evolution leading to a dramatic \emph{increase} in their magnetic field strength to magnetar levels ($B \gtrsim 10^{15}$\,G) powering a supernova-like explosion. 
We do not allow for an increase of the magnetic field during CE evolution in our models. \citet{1998ApJ...506..780B} study the formation of NS+BH binaries in a scenario where a NS collapses to a black hole due to hypercritical accretion during CE accretion \citep[see also][]{2002ApJ...571..394B}. 
However recent work \citep{Ricker:2007cq,MacLeod:2014yda,De:2019rfn} has suggested that accretion onto a compact object during CE evolution is limited to $< 0.1$\,M$_\odot$. 
We adopt these recent results and thus find this channel contributes negligibly to the formation of NS+BH binaries. 

\subsection{Radio selection effects}
\label{subsec:radio_selection_effects}

\input{telescope_specs}

We compute the populations of PSR+BH binaries observable by both current and future radio telescopes including Parkes, MeerKAT, and the SKA.
We list our assumed survey parameters for each telescope in Table~\ref{table:telescope_specs}. As in \citet{Chattopadhyay:DNS2019}, `Parkes' stands as a representative for all the past pulsar surveys and hence does not have any cut-off limit of the sky-coverage area. MeerKAT$_\mathrm{F}$ has different bandwidth, gain, integration time ($t_\mathrm{int}$), and sky-coverage than MeerKAT, representing a full-scale (F) survey with similar parameters as the SKA. To show the effect of the survey parameters on detection rates, we have also varied our assumption of the same survey specifications by adopting lower integration time (T) and changing the sky coverage to the Galactic-plane (G) for MeerKAT$_\mathrm{T}$, MeerKAT$_\mathrm{G}$ and MeerKAT$_\mathrm{GT}$.
The survey defined as SKA in Tab.~\ref{table:telescope_specs} is similar to SKA 1-mid \citep[phase-1, mid frequency i.e. 350MHz-15GHz operational range; ][]{Grainge:2017,schediwy_gozzard:2019}.
The estimates from the simulated surveys given in Tab.~\ref{table:telescope_specs} should be sufficient to construct a picture of other survey combinations, for example a lower integration time Galactic-plane SKA survey. 

The code PSREvolve \citep{2011MNRAS.413..461O, Chattopadhyay:DNS2019} is used to calculated the signal-to-noise ratio of the pulsars using the radiometer equation \citep{Dewey:1985, Lorimer:2004}
\begin{equation}
    S_\mathrm{min} = \beta\frac{(S/N_\mathrm{min})(T_\mathrm{rec} + T_\mathrm{sky})}{G\sqrt{n_\mathrm{p}t_\mathrm{int}\Delta f}}\sqrt{\frac{W_\mathrm{e}}{P - W_\mathrm{e}}} ,
    \label{eq:Radiometer}
\end{equation}
to determine the minimum flux $S_\mathrm{min}$ a radio source must have in order to be observed with a signal-to-noise ratio $(S/N)_\mathrm{min}$. 
The parameter $\beta$ accounts for errors that increase the noise in the signal (digitisation errors, radio interference, band-pass distortion), $T_\mathrm{rec}$ and $T_\mathrm{sky}$ represent the receiver noise temperature and sky temperature in the direction of the particular pulsar respectively, $G$ is the gain of the telescope, $n_\mathrm{p}$ is the number of polarizations in the detector, 
$\Delta f$ is the receiver bandwidth, $W_\mathrm{e}$ is pulse width and $P$ is the period of the pulsar. 
The sky temperature $T_\mathrm{sky}$ is determined by the location of the pulsar in the galaxy (PSREvolve inputs the information calculated by NIGO) while the pulse period $P$ is computed by COMPAS. 
We assume $\beta=1$ and $(S/N)_\mathrm{min} = 10$.

Our assumed telescope parameters imply (approximate) limiting fluxes of $0.1$\,mJy for Parkes and $0.01$\,mJy for MeerKAT and the SKA, as shown in Fig.~\ref{fig:SvP}.
\citet{Calore:2015bsx} present a similar plot for millisecond pulsars observed by the SKA and MeerKAT.

Not all pulsars will have their beams point towards the Earth. We use a fit for the pulsar beaming fraction $f_{\mathrm{beaming}}$ from \citet{Tauris:1998}, which gives 
\begin{equation}
    f_\mathrm{beaming} = 0.09\left(\log\frac{P/s}{10}\right)^2 + 0.03 \, , \quad 0 \leq  f_\mathrm{beaming} \leq 1
    \label{equ:fBeaming}
\end{equation}
also described in section 2.7.1 of \citet{Chattopadhyay:DNS2019}. We weight the pulsars by their beaming fraction to get statistically robust data-sets while also accounting for the observational bias. 

\subsubsection{Eccentric binaries}
\label{subsubsec:eccentric_binaries}

Due to the Doppler effect, the frequency modulation of pulse signals from binary pulsar systems with high eccentricity and short orbital period have lower signal-to-noise ratio. 
We use a fitting formulae derived from the results of \citet{Bagchi:2013wga} for pulsar-black hole binaries to account for this radio selection effect. 
The $\gamma_\mathrm{1m}$ factor from the paper is the first order estimate of the loss of efficiency in a standard pulsar search.

For this calculation, we assume a pulsar mass of 1.4\,$\mathrm{M_\odot}$, a black hole mass of 10\,$\mathrm{M_\odot}$, $1000\,$s duration of observation and 60$^{\circ}$ orbital inclination angle of the pulsar and generalise the results for all cases.
If $P_\mathrm{orb}$ is the orbital period of the pulsar binary system in days, $P$ is the spin period of the pulsar in seconds and $e$ is the eccentricity of the system, we then define a cut-off limit for radio detectability as
\begin{equation}
    P_\mathrm{orb}/d \geq m\times P/s + c ,
    \label{Bagchi_main}
\end{equation}
where
\begin{equation}
    m = m_m\times e + c_m  ,
    \label{Bagchi_m}
\end{equation}
and
\begin{equation}
    c = m_c\times e + c_c . 
    \label{Bagchi_c}
\end{equation}
By linear regression fitting for $e = 0.1, 0.5, 0.8$, we obtain $m_m = -26.42$, $c_m = -18.31$, $m_c = -2.53$ and $c_c = 4.51$. 

For DNSs, assuming individual masses to be 1.4\,$\mathrm{M_\odot}$ each, $1000\,$s duration of observation and 60$^{\circ}$ orbital inclination angle of the pulsar we obtain $m_m = -8.90$, $c_m = -27.68$, $m_c = -3.40$ and $c_c = 5.72$. 

It is to be noted that Eqn.~\ref{Bagchi_main} is realistically not a hard cut-off. 
Such non-detectable pulsars by standard searches can still be discovered by acceleration-jerk-search algorithms \citep{Bagchi:2013wga, Andersen:2018}.


\subsection{Models}
\label{subsec:models}

\subsubsection{Fiducial model}
\label{subsubsec:fiducial_model}

Unlike DNSs, there are no current observations of Galactic PSR+BH binaries. 
We define our Fiducial model as the one that best matches the Galactic DNS population after taking radio selection effects into account as in \citet{Chattopadhyay:DNS2019}. 
The inclusion of an eccentric binary radio selection effect (see section~\ref{subsubsec:eccentric_binaries}), which was not accounted for in \citet{Chattopadhyay:DNS2019}, causes a small shift in our `best-fit' Galactic DNS model relative to that paper. 
Utilising the same Kolmogorov–Smirnov (KS) test described in \citet{Chattopadhyay:DNS2019} to compare with the observed Galactic DNSs, we find that a magnetic field decay mass-scale $\Delta$M$_\mathrm{d} = 0.15$\,M$_\mathrm{\odot}$ provides the best match, compared to the $\Delta$M$_\mathrm{d} = 0.2$\,M$_\mathrm{\odot}$ found in \citet{Chattopadhyay:DNS2019}. 
All other parameters remain the same. 
We note that the shift is small, remaining within the same order-of-magnitude for our DNS predictions compared to previous work.  

\subsubsection{Model variations}
\label{subsubsec:model_variations}

\input{modelsTable}

To explore the impact of uncertainties in modelling the physics described in Section~\ref{sec:methods} on predictions for PSR+BH binaries, we create an ensemble of ten models including the Fiducial model (see Table~\ref{tab:NSBH} for details). 
The other nine models are designed by varying only one parameter per model from the Fiducial, allowing us to analyse the effect of each on the resultant population. 
The nomenclature of the models is as follows - i) the prefix of the name is the abbreviation of the initial parameter that has been changed from Fiducial and ii) the suffix denotes the altered magnitude or distribution. 
For example, the model in which the black hole kick (BHK) prescription is changed to zero (Z) is model BHK-Z. 
BHK-F represents the model with full (F) BH natal kicks, same as for the NSs. 
In CE-P we explore the pessimistic (P) common-envelope assumption, where the HG donor star involved in a CE phase always merges with the companion and hence the binary does not survive. 
We also explore the effect of metallicity (ZM) on the resultant NS+BH population by models ZM-001 and ZM-02. 
\citet{Chattopadhyay:DNS2019} showed that the magnetic field decay time (FDT) and mass (FDM) scales, $\tau_d$ and $\Delta M_d$ play key roles in determining the properties of the modelled simulation. 
We hence inspect models FDT-500 and FDM-20 with $\tau_d=500$\,Myr and $\Delta M_d=0.02$\,M$_\odot$. 
The birth magnetic field (BMF) period distributions has been shown to affect the final population of DNSs \citep{Chattopadhyay:DNS2019}; we probe these through models which assume a Flat-in-Log (FL) birth magnetic field distribution. 
In our Fiducial model, we use the `delayed' prescription \citep{Fryer:2012} to determine the remnant masses (c.f. Section~\ref{subsubsec:compact_object_masses_radii}). 
We have varied the remnant mass prescription to the `rapid' prescription \citep{Fryer:2012} in model RM-R.

\subsubsection{Re-scaling our simulation to the Milky Way}
\label{subsubsec:rescaling}

Each model has been simulated with 10$^7$ initial zero-age main sequence binaries with primaries in the mass range 4--100\,M$_\odot$, according to the initial mass function distribution given by \citet{Kroupa:2000iv}. 
For saving computational time as well as producing robust statistics, we reuse each NS+BH binary 100 times per simulation by assigning each NS+BH binary 100 birth-times drawn according to the uniform star formation history of the Milky Way in the range 0--13\,Gyr   \citep[][]{2015A&A...578A..87S}. 

This gives us an effective population size of $10^{9}$ binaries \citep[see][for more details]{Chattopadhyay:DNS2019}.
Our $10^{9}$ massive binaries represent around $10^{11}$ binaries including low mass stars according to our chosen initial mass function \citep{Kroupa:2000iv}.
Assuming a binary fraction $f_\mathrm{bin}$, this population represents $2 N_\mathrm{bin} / f_\mathrm{bin}$ stars. 
For a binary fraction of 20\% \citep{Lada:2006dc} suitable for low mass stars (which make up the majority of the stars), our population corresponds to around $10^{12}$ stars.
The Milky Way contains around $10^{11}$ stars, with around 20\% located in the bulge \citep{Flynn:2006tm}. 
Our evolved population of stars therefore represents around 10 Milky Ways worth of stars. 
Hence, to calculate the number of predicted observations for our models, scaled to a population representative of the Milky Way, we divide the number of detections by 10. We discuss uncertainties in this rescaling in Section~\ref{sec:discussion}.

\subsubsection{Distributing binaries in the Galaxy}
\label{subsubsec:distributing}

The NS+BH binaries are distributed in a 3-dimensional Milky-Way like potential, accounting for the supernova kicks, using the code NIGO \citep{2015A&C....12...11R, RossiHurley:2015}. 
The Galaxy is modelled with the central bulge as a Plummer sphere \citep{Plummer:1911, MiyamotoNagai:1975}, an exponential disc formed by linear superposition of three Miyamoto-Nagai potentials \citep{MiyamotoNagai:1975, Flynn:1996ej} and a Navarro–Frenk–White (NWF) dark matter halo \citep{Navarro:1996gj}. 
The numerical values of the Galactic potential equation variables are implemented from \citet{Irrgang:2013} and \citet{SmithFlynn:2015}. 
Section 2.6.2 of \citet{Chattopadhyay:DNS2019}
shows the detailed equations and the parameters used for the Galactic potential.

The properties of NS+BH binaries at the current observation time are then analysed. 
Reprocessing each binary multiple times (with unique birth times) allows us to extract more information from individual binaries. 
Though the process initiates some systematic error as the initial parameters of the reprocessed binaries are identical, the benefit of studying different phases of evolution of the binary and computational efficiency makes us lean towards this method. 
For more details see Section 2.5 and Fig. 3 of \citet{Chattopadhyay:DNS2019}.


\section{Formation channels}
\label{subsec:formation_channels}

\input{ZAMS}
\input{NetNumbers}
\input{radioLifetime}

The formation of NS+BH binaries through isolated binary evolution can be broadly categorized into two channels: i) where the NS forms first (NSBH) and ii) where the BH is formed first (BHNS). 
When we are referring to the entire population of neutron star - BH binaries, irrespective of which formed first, we will use NS+BH as before, which comprises both NSBHs and BHNSs.

The majority of NS+BH binaries are formed from initially massive, wide binaries (see \citealp{Broekgaarden:2020NSBH} and references therein for more details). 
The initially more massive star (the primary) evolves off of the Main Sequence (MS) first and fills its Roche Lobe whilst crossing the Hertzsprung Gap (HG). 
Stable case B mass transfer proceeds, stripping the hydrogen envelope of the primary and leaving behind a naked helium star. 
The helium star lives for only $\sim10^{5}$\,yr before exploding in a supernova and forming a NS or a BH. 
The Galactic high-mass X-ray binary Cyg X-1 is currently at this stage (BH-MS) and may eventually form a wide BHNS binary \citep{2011ApJ...742L...2B}.
In this dominant channel, the secondary star then ends its MS, expanding and filling its Roche Lobe as an evolved star.
The secondary is typically a HG star at this point for all the models except CE-P, where it must be a core helium burning (CHeB) star since in that model we do not allow HG stars to survive CE evolution (sec.~\ref{subsec:common_envelope_evolution}).
The combination of our prescription of Eddington limited accretion onto compact objects, and
the possibility of having binaries with highly asymmetric masses---even for a binary with a BH and a NS progenitor---can result in unstable mass transfer,  leading to CE evolution (see \citealp{Broekgaarden:2020NSBH} for further details).
A small fraction of the secondary's envelope ($<0.1$\,M$_\odot$) may be accreted onto the primary compact object during this stage \citep{MacLeod:2014yda}.
The orbital separation is dramatically reduced during the CE phase as orbital energy is used to unbind the CE.
If the CE is ejected, a compact binary consisting of a NS or BH with a stripped Wolf-Rayet star remains. 
The Galactic X-ray binary Cyg X-3 may be a BHNS progenitor in this stage of evolution \citep{Belczynski:2012jc}. 
Depending on the orbital separation and the stellar masses, there can be an event of stable mass transfer from the Wolf-Rayet (naked Helium) star onto the compact object. This is called case BB mass transfer \citep{Delgado81, Tauris:2015}, and if the NS is formed first, it can be further spun up.
The Wolf-Rayet star finally forms a NS or BH; if the binary remains bound subsequent to the supernova explosion, a  NS+BH binary is formed. 
The orbit of the binary then decays due to the emission of GWs, eventually leading to a NS+BH merger.

If the initial ZAMS mass ratio of the binary is close to unity, the stars evolve on a very similar timescale. 
When both stars move from the MS to the giant branch and also have distinct core-envelope composition, depending on the orbital separation, the primary can initiate RLOF. 
The mass transfer soon becomes unstable and a double core CE \citep{1995ApJ...440..270B,1998ApJ...506..780B,Dewi:2006bx,Vigna-Gomez:2018dza,Vigna-Gomez:2020bgo,Broekgaarden:2020NSBH} is formed. 
Although the double core CE channel can lead to BHNS mergers due to reduced orbital separation, the NS is formed after the CE, and thus in this channel the NS cannot be spun up during the CE, and hence no recycled pulsars can form through this channel.
The other infrequent formation channels of NS+BH binaries through COMPAS have been discussed in sec.3.1 of \citet{Broekgaarden:2020NSBH}.

Throughout this paper we will refer to the overall NS+BH binary population, the combined NSBH and BHNS sub-populations, as the `net' population. 
The systems within each sub-population that contain a pulsar are denoted as the `radio' population, noting that a NS is considered to be a pulsar when it is still radio-alive and has not passed through to the graveyard region according to our definition in section~\ref{subsec:pulsarPhysics}.
Thus the `radio' population contains the PSR+BH binaries, 
irrespective of whether the pulsar is detected by a pulsar-survey, 
and is a subset of the 'net' population. 
We refer to the combined radio and non-radio populations as a whole for each of these sub-populations as `total' (e.g. total NSBH = radio NSBH + non-radio NSBH).

In Table~\ref{tab:ZAMS_radioVSnet} we show the mean values of various binary properties for NS+BHs in the Fiducial model, comparing the NSBH and BHNS total and radio sub-populations. The radio and the total populations have significantly different parameter values to each other, as confirmed by a Kolmogorov-Smirnov (KS) test with all the tabulated parameter $p$-values $<10^{-6}$, hence effectively 0.

Table~\ref{tab:netNumbers} shows the aggregate of different sub-populations of NS+BH binaries model-wise, including the net background population as well as pulsar survey predictions for our ensemble of models. 
For each model, the numbers comprise the NS+BH systems expected to be produced from 
the Milky-Way (see Sec.~\ref{subsec:models} for scaling details).

For our Fiducial model (top row of Tab.~\ref{tab:netNumbers}), BHNSs dominate the `net' or overall NS+BH binary population, accounting for more than 97\% of it. 
Approximately 20\% of the net NSBHs are radio-alive, compared to $\approx$ 2\% of the BHNSs. 
The higher fraction of NSBHs rather than BHNSs being radio-alive is because the NSs of NSBHs, being a primary, experience mass accretion. More than 96\% of the radio NSBHs of our Fiducial model are recycled.
This results in recycling of the pulsar, the higher spin extending its life as a radio luminous binary. 
The quantitative dominance of BHNSs in the net population is still reflected on the radio sub-population, $\approx$ 80\% being BHNSs.

The initial mass and semi-major axis distributions play the key roles in determining the possibility of a NS+BH binary appearing as a PSR+BH binary at the time of observation. 
For BHNSs, the mean zero-age main sequence (ZAMS) primary (BH progenitor) and secondary (NS progenitor) masses for the radio population are larger by $\approx$ 1.17 and 1.82 times compared to the net population as shown in Tab.~\ref{tab:ZAMS_radioVSnet}. 
This is reflected in the final BH and NS masses as well with the radio distributions having more massive double compact object masses. 
The larger masses of the pulsars result in a larger moment of inertia. 
We use an equation-of-state insensitive relation for the moment of inertia \citep{Lattimer:2004nj,Raithel:2016vtt}
\begin{equation}
    I = 0.237 M_\mathrm{NS} R_\mathrm{NS}^2 
    \bigg[1 + 4.2 \frac{M_\mathrm{NS}}{M_\odot} \frac{\mathrm{km}}{R_\mathrm{NS}} 
    + 90\Big(\frac{M_\mathrm{NS}}{M_\odot} \frac{\mathrm{km}}{R_\mathrm{NS}} \Big)^4 
    \bigg] ,
    \label{eq:moment_of_inertia}
\end{equation}
assuming a constant NS radius of 12\,km for all of our models (c.f. Sec.~\ref{subsubsec:compact_object_masses_radii}). 
Due to their larger moment of inertia, more massive pulsars spin down more slowly (c.f. Eqn.~\ref{SpinDownIsolated}). 
Hence, radio BHNSs are biased towards heavier masses. 
In addition, radio BHNSs have a mean ZAMS separation that is $\approx$ 0.33 times smaller than that of the net BHNS population (see Tab.~\ref{tab:ZAMS_radioVSnet}). 
Hence, more massive radio BHNSs evolve faster with the closer companions assisting to speed up the evolution with rapid mass transfer episodes. 
The radio-alive BHNSs evolve $\approx$4\,Myr faster on average than the net ensemble. 
The mean birth time is biased to a larger value for radio BHNSs  
owing to the combination of drawing the birth times from an uniform distribution and the fact that 
the younger non-recycled pulsars are, the less likely they are to have spun down to the graveyard region of dead pulsars within the $P\dot{P}$ diagram. 

The NSBHs on the other hand show almost no difference between the radio and net population masses. 
The ZAMS primary and secondary mean masses are slightly lower for the radio NSBHs. 
The final NS mean mass is however larger for the radio systems. 
This apparent discrepancy can be explained by pulsar recycling. 
The radio NSBH population is dominated by recycled pulsars since the NS forms the primary. 
The spin up due to accretion is inversely proportional to the pulsar moment of inertia (see Eqn.~\ref{AngularMomentumAccretion}) and hence to the pulsar mass. 
However, due to recycling the mass transfer results in the pulsar gaining matter. 
The dominance of recycled pulsars in the population aligns the mean initially towards less massive NSs, which after mass accretion by the pulsars results in a slightly more massive radio NS mass distribution. 
The ZAMS separation for the radio NSBHs is approximately half that of the net NSBHs, where binaries with closer component stars experience faster and more efficient mass transfer in general. However, overall the radio NSBHs evolve $\approx$ 0.6 Myrs slower than the net population, due to the lower ZAMSs mass distribution. The birth time distribution of the radio NSBHs is less steep than radio BHNSs, as recycling of primary pulsars allows older pulsars to survive longer in the radio-alive phase. 

From the predicted rates of DNS and NS+BH radio observations 
in Table~\ref{tab:netNumbers}, we can start to segregate reasonable models from those that can be considered unfeasible. Observationally, there has been 15 confirmed Galactic DNSs (including one double pulsar system) discovered by previous pulsar surveys \citep[see][Table~1]{Chattopadhyay:DNS2019} and no detection of PSR+BHs so far \citep{Manchester:2005yCat}. The lack of observation of any PSR+BH serves as a constraint in itself and though the prediction of observing a small number stays within the uncertainty range, models predicting a large amount of current survey-observable PSR+BHs (e.g. the `Parkes' column in Tab.~\ref{tab:netNumbers}) can be ruled out.
The BHK-Z model clearly produces too many NS+BH systems to be reconciled with the observational constraint.
Similarly the BMF-FL model produces the most number of NS+BH systems and also  predicts DNS observations to be significantly higher than the actual detections. 
At the other end of the scale, the CE-Z model predicts only three DNS systems observed by Parkes which is significantly less than the current observed number. 
Hence the models BHK-Z, BMF-FL, CE-Z are judged to be unfeasible. 
From Tab.~\ref{tab:netNumbers} alone, the predicted upper-limit of SKA observed PSR+BHs is hence approximately 30. 
We modify this value to account for the initial binary parameters and Milky-Way stellar distribution or we modify the survey parameters (see Sec.~\ref{subsec:telescope_obs}).

We show the typical birth rates (in the Milky Way) and the mean radio lifetimes of the NSBH and BHNS binaries for our suite of models in Tab.~\ref{table:radio_lifetime}. 
We note that for all models the rate of NSBHs produced per BHNS in quite small. 
However, for most models the radio lifetime (due to pulsar recycling) of NSBHs is higher than BHNSs, assisting in the radio detectability of the former. 
The expected radio detection rates of the PSR+BH binaries (in Tab.~\ref{tab:netNumbers}) is a combined effect of their individual birth rates and radio lifespans.


\section{Results: Radio}
\label{sec:results_radio}

In this section we present results for radio observations of PSR+BH binaries. In Section~\ref{subsec:telescope_obs} we describe the results of our mock surveys with the Parkes, MeerKAT and SKA radio telescopes. In Section~\ref{subsec:radio_observables} we show the distributions of radio observable properties for PSR+BH binaries and describe how they vary with our suite of models. Finally, in Section~\ref{MSPs} we describe a subset of NSBHs which contain millisecond pulsars.

The section uses a population scaled to be the equivalent of the  Milky-Way to quote and predict NS+BH numbers, as in Table~\ref{tab:netNumbers}. However, for general analysis, such as calculating mean values and generating cumulative distribution functions (CDFs), we use a population equivalent to 10 Milky-Way systems. This is done to lower statistical noise, produce improved visual resolution in figures and give robust predictions of detection rates for the future radio-telescope surveys.

\subsection{Telescope Observations}
\label{subsec:telescope_obs}

\begin{figure}
\includegraphics[width=0.99\columnwidth]{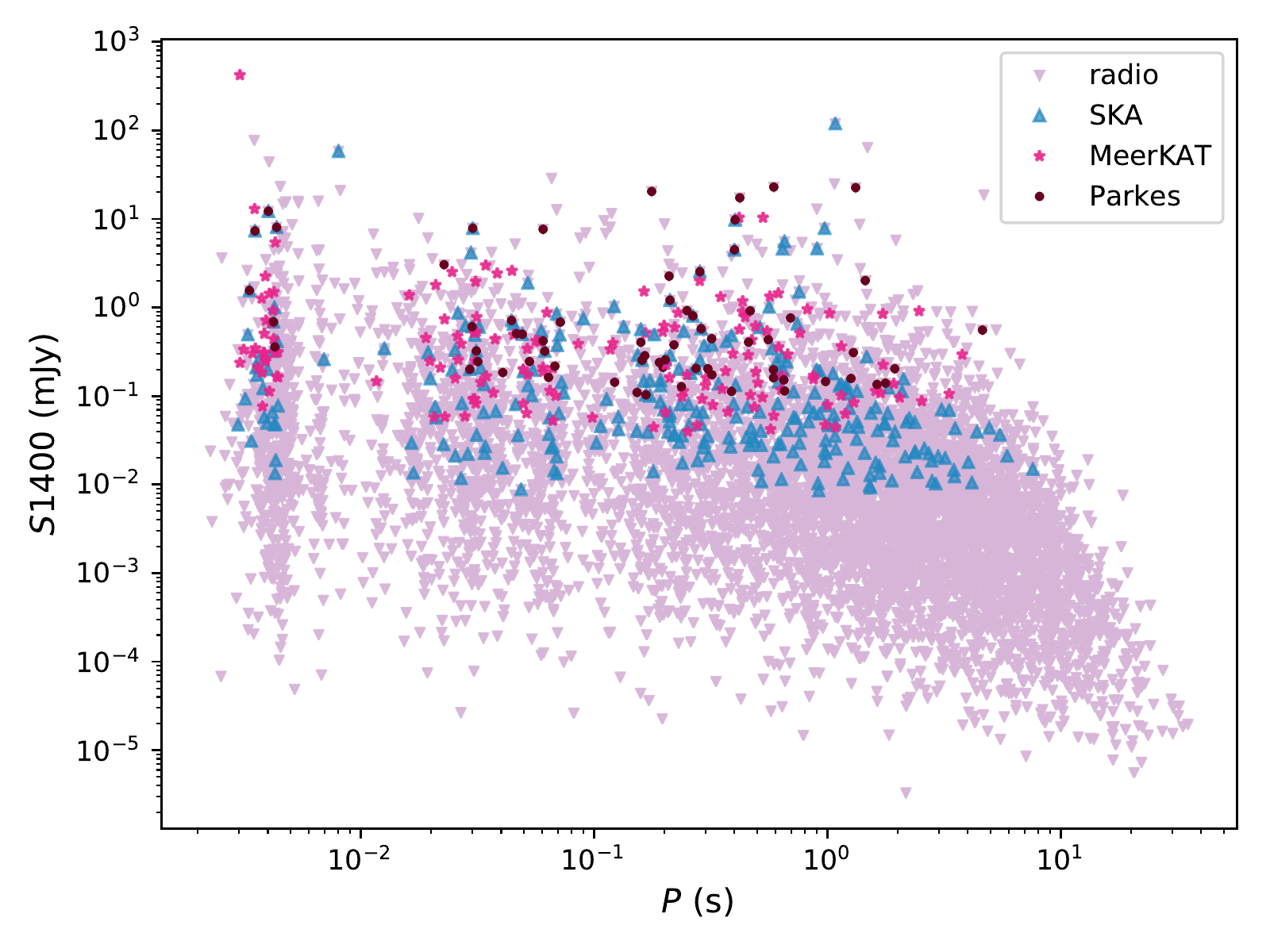}
\caption{Radio flux at a frequency of 1400\,MHz $S_{1400}$ as a function of pulsar spin period $P$ for PSR+BH binaries (both BHNS and NSBH) for 10 MW systems. 
The tail on the right side of the figure is created by the BHNS population.
The radio-alive pulsars are plotted in lavender with inverted triangles, while blue upright triangles, pink stars and maroon dots show the survey-observable pulsars for SKA, MeerKAT and Parkes respectively. 
It can be seen that SKA and MeerKAT have fluxes of around 0.01\,mJy, and 0.05\,mJy. SKA and MeerKAT are both capable of detecting fainter pulsars than current surveys such as Parkes. 
Many pulsars will remain too faint to be observable even with the SKA.}
\label{fig:SvP}
\end{figure}

We explore the quantitative pulsar observations from NS+BH systems for the telescope surveys by Parkes, SKA and MeerKAT, as well as the qualitative behaviour of independent observable parameters such as the spin period $P$, spin down rate $\dot{P}$, orbital period $P_{\mathrm{orb}}$, eccentricity $e$, scale height $|Z|$ and derived parameters such as the surface magnetic field $B$, pulsar mass $m_{\mathrm{psr}}$ and companion mass $m_{\mathrm{cmp}}$. For all the observed pulsar numbers we quote, it must be remembered that these are cumulative quantities rather than new detections. 

Fig.~\ref{fig:SvP} shows the radio flux at a frequency of 1400\,MHz against the pulsar spin period for the Fiducial model. 
We see that the next-generation of radio telescopes such as the SKA will enable observations of pulsars with an order of magnitude lower flux than Parkes, though many pulsars will remain too faint to be observed even with the SKA.
We find that MeerKAT and the SKA observe $\sim$ 1.5--2 and $\sim$ 2.5--3.5 times more PSR+BH binaries respectively than Parkes (see Table~\ref{tab:netNumbers}). Hence the lower limit of new detections from MeerKAT is $(1.5-1)=0.5$ times that from the Parkes observations while the higher limit of new detections from the SKA is $(3.5-1)=2.5$ times the Parkes observations. Thus with future surveys we expect a  0.5--2.5 fold increase in the known pulsar dataset. SKA observations are at least $\sim$2 times more than MeerKAT. The subpopulation of BHNSs create a tail of high $P$ and low $S_{1400}$ pulsars in Fig.~\ref{fig:SvP}. 
This is because we determine whether a pulsar is radio-alive using a combination of its radio efficiency and death-lines (see Section~\ref{subsec:pulsarPhysics} and Fig.~\ref{fig:PPdot}). 
Hence, it depends not only on the luminosity distribution but also on $P$ and $\dot{P}$. 
BHNSs, which contain non-recycled pulsars, are biased towards high $P$.

\input{MeerKAT}

The number of PSR+BH binaries observed in our mock surveys depends on our assumptions regarding the survey parameters, in particular the area of the sky surveyed and the integration time (see Table~\ref{table:telescope_specs} for details 
As an example, we demonstrate how our results change for different choices of survey parameters for our mock survey with MeerKAT in Table~\ref{tab:MeerKAT_rates}. 
Changing the integration time (MeerKAT$_\mathrm{T}$), sky-coverage (MeerKAT$_\mathrm{G}$) or both (MeerKAT$_\mathrm{GT}$) changes the number of observed PSR+BH binaries by factors of 1.6, 1.4 and 1.2 respectively (see Tab.~\ref{tab:MeerKAT_rates}). MeerKAT$_\mathrm{F}$ doubles the observations from MeerKAT.
The factor $\mathcal{F}$ gives the ratio of 
PSR+BH observations for each model version of the survey with respect to MeerKAT, but does not change the qualitative distributions of the populations. Since the change in the number of detections is model and sub-population independent, we do not quote them separately for each model. 

The number of observations of NSBH and BHNS binaries for each of our mock surveys (Parkes, the SKA and MeerKAT) are given in Table~\ref{tab:netNumbers} for each of our models. 
Some of our models may be disqualified based solely on the predicted number of DNS or NS+BH observations.
Model BHK-Z produces an excess of NS+BHs, predicting an observation rate $\mathcal{R}=$ 1.48 PSR+BH per DNS observed till now. Model BMF-FL creates far too many DNS and NS+BH binaries while CE-Z produces too few DNSs compared to the observed sample of Galactic DNSs \citep{Chattopadhyay:DNS2019}. We obtain a lower limit to the expected number of PSR+BH detections from an SKA all-sky survey from the model with the lowest number of predicted detections amongst feasible models; model CE-P predicts 6 PSR+BH observations with the SKA. 
Similarly, the upper limit of 40 detections comes from our RM-R model. 
For the lower limit, we further note that an SKA survey with lower integration time and covering only the Galactic plane (as for MeerKAT$_\mathrm{GT}$, Table~\ref{tab:MeerKAT_rates}) further decreases the lower limit to 3.
The spread in these numbers represents the uncertainties in the pulsar and binary evolution parameters we have varied. 
There are additional uncertainties associated with the initial distribution of binary parameters and the re-scaling of our simulation to the Milky Way (see sec.~\ref{sec:discussion}). 
We introduce an additional factor of 2 uncertainty to the predicted numbers of detections from Table~\ref{tab:netNumbers} to account for these.
This uncertainty is folded in by halving the lower limit and doubling the upper-limits. 
Hence our final prediction for the number of PSR+BH binaries observed by future pulsar surveys with the SKA is 1--80. For MeerKAT this number is halved to 0--40. 

The columns marked radio in Table~\ref{tab:netNumbers} give the radio-alive population of the net NSBH and BHNS binaries and are hence PSR+BHs. 
For this pre-radio selection effects radio data-set, the model producing the highest number of PSR+BHs is the RM-R model (1000),whilst model BHK-F produces around 100 and our
Fiducial model (650) having a value in-between. 
Hence, accounting for the uncertainties in our modelling as described above, we predict approximately 50--2000 PSR+BHs in the Milky Way field. 
Table~\ref{tab:netNumbers} shows that under our mock survey assumptions, MeerKAT and the SKA detect approximately equal numbers of PSR+BHs assuming the same sky coverage and integration time. 

Eqn.~\ref{eq:Radiometer} shows that the telescope-dependent radio selection effects primarily depend on the pulsar spin period $P$ explicitly, as well as implicitly through pulse width $W_e$.
Since $\dot{P}$ explicitly depends on $P$ as well, both of these two quantities are expected to show survey dependant behaviour.
Fig.~\ref{fig:FiducialPPdotB_distribution_for_telescopes} shows the CDF of $P$ and $\dot{P}$ for the three pulsar-surveys for our Fiducial model.
We see $P$ to be more strongly survey-dependant. 
Our detailed calculations allow $B$ to be obtained directly from the models, compared to pulsar-survey data, where the surface magnetic field strength is derived as $B=3.2\times10^{19}(P\dot{P})^{(1/2)}$ (in Gauss). 
Hence $B$ distributions from our simulations remain telescopic observation independent though for real observation it will be a survey dependant parameter through $P$ and $\dot{P}$.

The pulsar spin period $P$ shapes the radio selection effect through the beaming fraction $f_\mathrm{beaming}$ as well, with a lower value of $P$ giving a higher $f_\mathrm{beaming}$ (see the left panel of Fig.10 from \citealp{Tauris:1998}). This means that faster spinning pulsars have larger beaming angles and are more likely to be detected by a radio telescope.
Although, the beaming fraction selection effect is independent of the telescope survey, it reinforces the importance of $P$ in pulsar radio selection effect. 

We have also included the effect of eccentricity and orbital period in determining the observability of pulsars (see Section~\ref{subsubsec:eccentric_binaries}). 
The inter-dependency of $e$ and $P_\mathrm{orb}$ is to be noted, since though a highly eccentric orbit produces high acceleration, a longer orbital period may allow the pulsar to spend most of its time in a lower acceleration region, and hence still detectable with comparative ease compared to a binary pulsar of lower $e$ but also shorter $P_\mathrm{orb}$ spending more time in the accelerated part of the orbit. 
It is however highlighted that specifically designed acceleration searches can actually discover these pulsars. 
This selection effect is also telescope-independent.


\subsection{Radio observables}
\label{subsec:radio_observables}

Here we study the radio observable properties of the PSR+BH binaries. This includes the pulsar spin period $P$, spin down rate $\dot{P}$, surface magnetic field magnitude $B$, pulsar $m_\mathrm{psr}$ and companion $m_\mathrm{comp}$ masses, binary orbital period $P_\mathrm{orb}$, eccentricity $e$ and scale height $|Z|$. 
The mean values of these radio-observable properties for the radio population and SKA-observed population are presented in Tables~\ref{tab:radioObservablesMean:Radio} and~\ref{tab:radioObservablesMean:SKA} respectively. 
These form a reference point to compare the effects of model-wise initial parameter changes on the radio population, 
as discussed below, 
as well as the consequences of the radio selection effects modifying the observable population. 
Again to minimise statistical noise, we have used weighted sampling in an ensemble of NS+BH systems equivalent to 10 Milky-Way populations per model.

\subsubsection{$P\dot{P}-B$}
\label{subsubsec:radio_obs_PPdot}

\begin{figure}
\includegraphics[width=0.99\columnwidth]{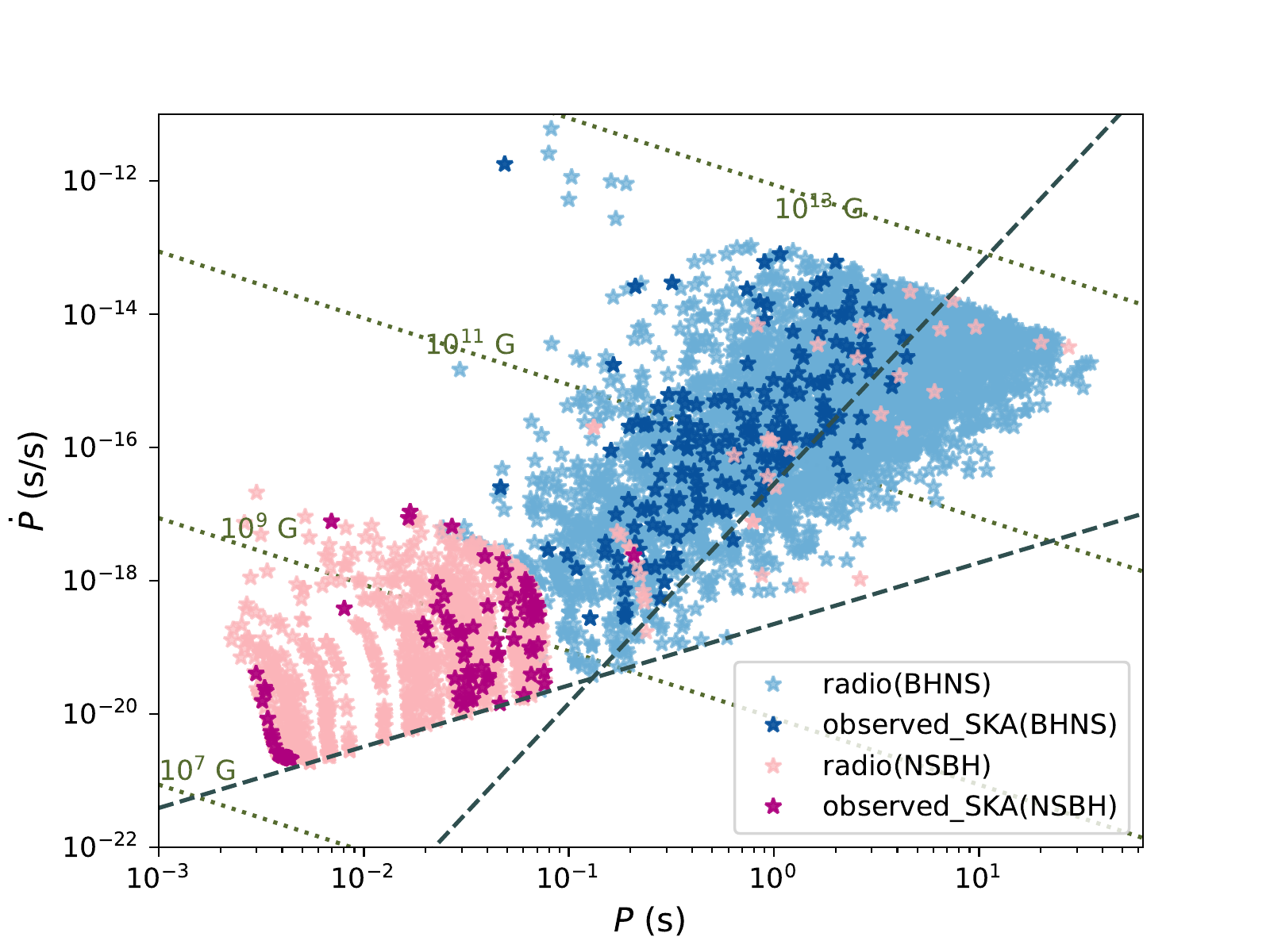}
\caption{$P \dot{P}$ diagram showing the PSR+BH binaries observable with the SKA in our Fiducial model for a population of stars 10 times larger than the Milky Way. The light coloured points show the radio-alive pulsars (pink for NSBHs and blue for BHNSs) while the darker colours show the survey-detected pulsars. 
The two dashed lines in black are the death lines from \citet{RudakRitter1994}, whilst the olive green dotted diagonal lines are the lines of constant magnetic fields (individual values noted on the figure in Gauss). 
}
\label{fig:PPdot}
\end{figure}

\begin{figure*}
\includegraphics[width=0.33\textwidth]{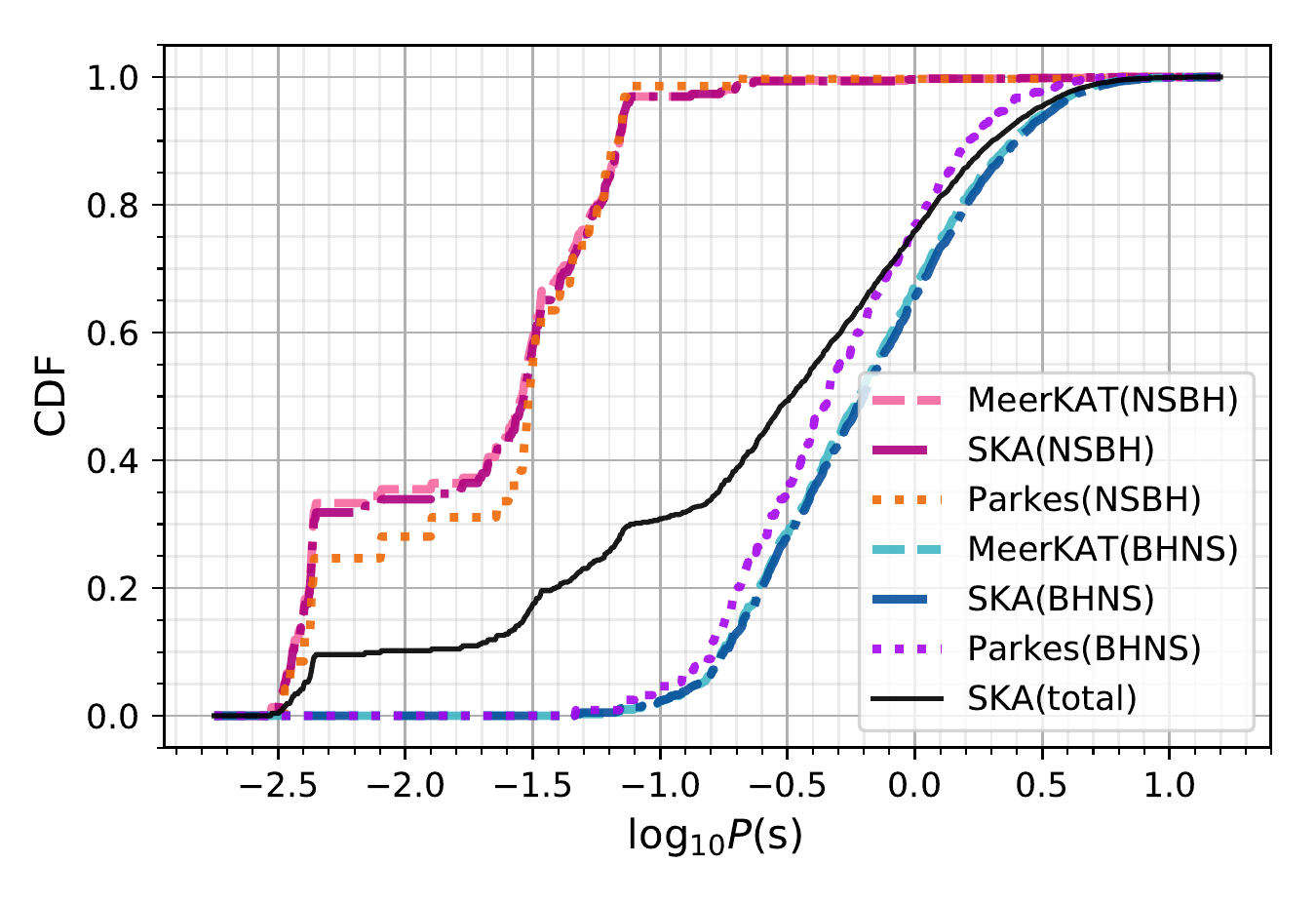}
\includegraphics[width=0.33\textwidth]{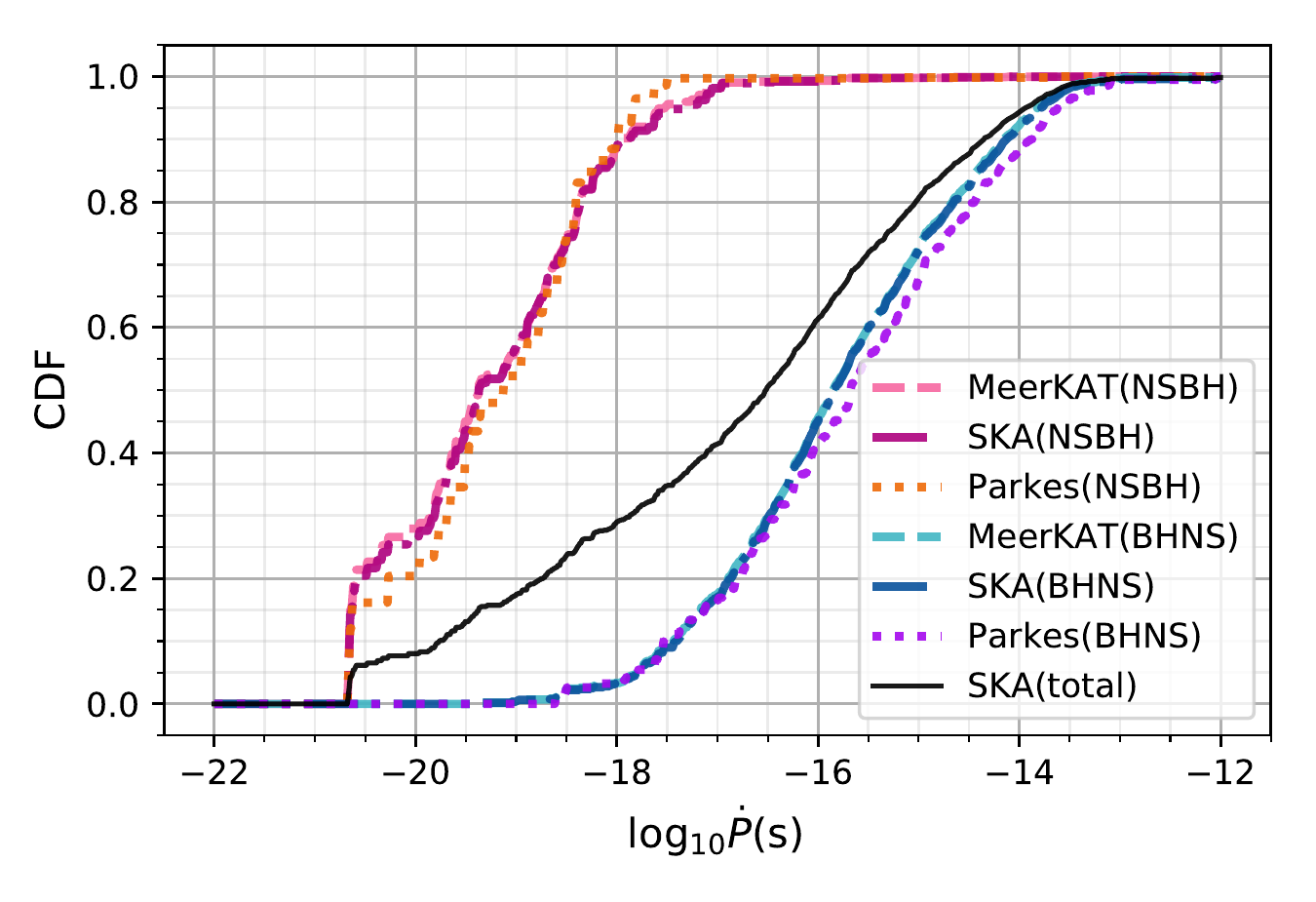}
\includegraphics[width=0.33\textwidth]{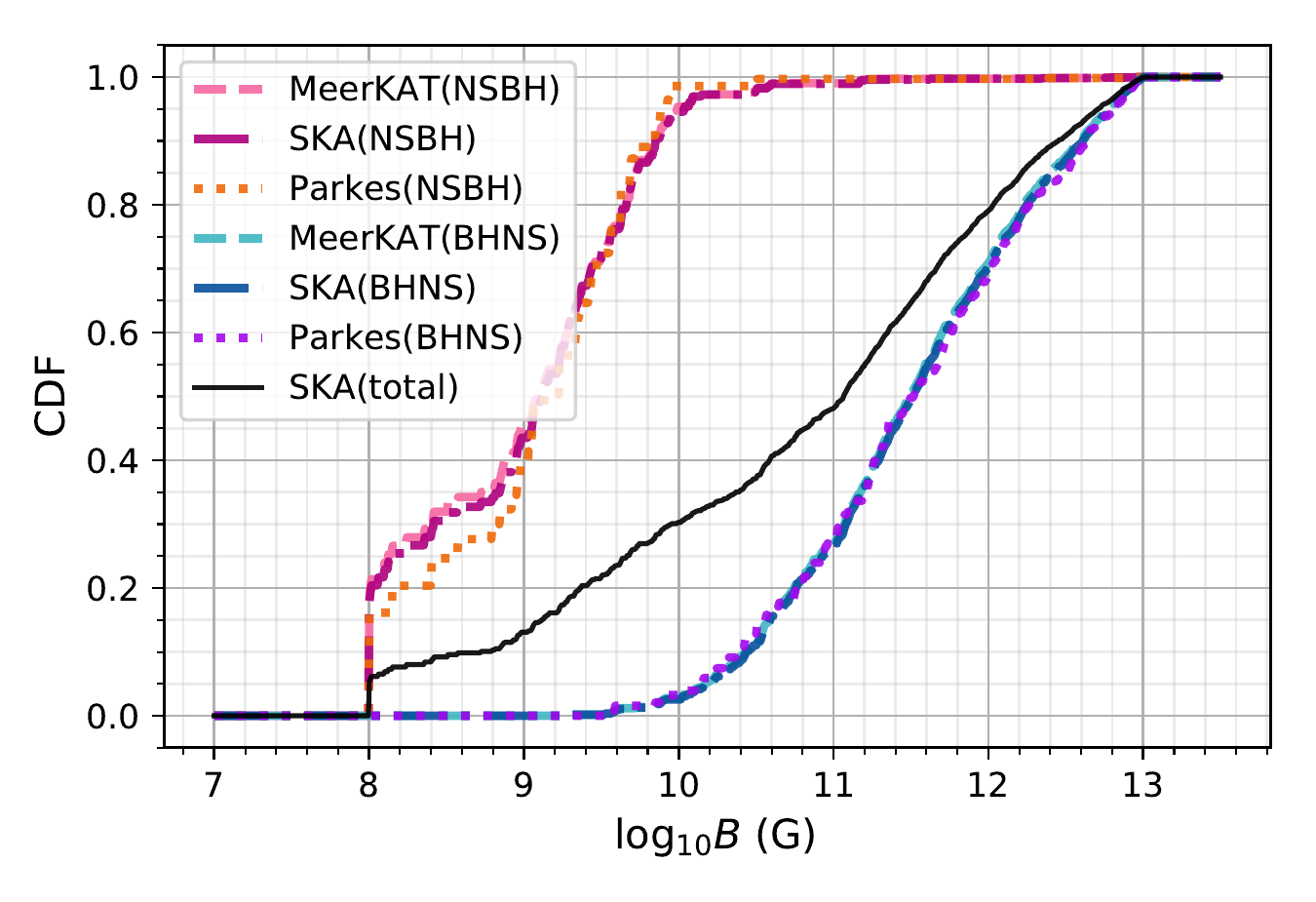}
\caption{Cumulative distributions of pulsar spin period $P$ (left), spin down rate $\dot{P}$ (middle) and magnetic field strength $B$ (right) for PSR+BH binaries in our Fiducial model. 
Simulated pulsar surveys for Parkes, MeerKAT and the SKA are shown separately. 
The jagged appearance of the Parkes curves reflects the lack of observed pulsars even for 10 Milky-Way worth systems. 
Further, each data-set is sub-grouped into NSBHs and BHNSs - showing distinct characteristics.
$P$ shows a slight dependency on the telescope type, which is propagated onto $\dot{P}$. 
Since we compute the surface magnetic field instead of deriving from the spin and spin down rate, $B$ distribution appears survey independent. 
The black solid line denotes the total SKA-observed population which is slightly biased towards BHNSs, due to the latter being a little more dominant quantitatively in the data-set. }
\label{fig:FiducialPPdotB_distribution_for_telescopes}
\end{figure*}

Observationally, $P$ vs $\dot{P}$ plots, often termed as `$P\dot{P}$' diagrams are used to characterise pulsars. 
Not only do these plots represent the spin and the spin down rate of the pulsar but they also show the pulsar surface magnetic field strength identified by the diagonal lines of constant magnetic field. 
Fig~\ref{fig:PPdot} shows the $P\dot{P}$ diagram for the radio and SKA-observed pulsars in NSBH and BHNS systems. 

We assume pulsar death occurs when the radio efficiency parameter \citep{Szary:2014dia} exceeds a threshold value, rather than assuming an abrupt cutoff beyond the deathlines (see Sec.2.3 of \citealp{Chattopadhyay:DNS2019} for more details).
In our models we therefore occasionally find pulsars beyond the deathline in the `graveyard' region of the $P\dot{P}$ diagram. 
On average, pulsars in NSBHs have lower surface $B$ fields, shorter spin periods $P$ and lower spin down rates $\dot{P}$ compared to pulsars in BHNSs due to recycled pulsars being present amongst the former. 
NSBH pulsars that are non-recycled typically have larger surface magnetic field $B$ and spin period $P$, occupying similar region of the $P\dot{P}$ parameter space as the BHNSs. 
In comparison to the recycled DNSs, recycled NSBHs occupy a lower surface magnetic field and faster spin period region of the parameter space because of the latter accreting more matter than the former (see Sec~\ref{MSPs} for details). 
This difference in NSBH and BHNS pulsar parameters suggests that measurements of $P\dot{P}$ would ordinarily allow a PSR+BH binary to be identified as either a recycled NSBH or a BHNS. A non-recycled NSBH, though having a considerably lower detection probability (about 1\% of the SKA NSBH population for Fiducial) than the recycled NSBH population may still be distinguished by its typically lower mass distribution (see section~\ref{subsec:masses}) or larger scale height (see section~\ref{subsec:Z}). 
However, the uncertainties in initial metallicity and supernova kicks may not make it always possible to distinguish between non-recycled pulsars and BHNSs. 
This distinction in properties between NSBHs and BHNSs arises because NSBHs contain recycled pulsars unlike BHNSs, where all pulsars are non-recycled. 
As mass accretion spins the pulsar up as well as buries the surface $B$ (Eqn.~\ref{AngularMomentumAccretion},~\ref{MagneticFieldAccretion}), NSBH pulsars show a higher mean $P$ and a lower $B$, as reflected in their positions in the $P\dot{P}$ diagram. 
The $P$, $\dot{P}$ and $B$ of the pulsar is qualitatively model dependant, and are primarily determined by magnetic field decay time-scale $\tau_d$, mass-scale $\Delta M_d$ and the CE mass accretion assumption. 
These are explored by models FDT-500, FDM-20 and CE-Z respectively. 
Lower values of $\tau_{d}$ cause the pulsar magnetic field to decay on a shorter timescale. For example, in model FDT-500 ($\tau_{d}=500$\,Myr) pulsars die faster than in the Fiducial model ($\tau_{d}=1000$\,Myr).
Hence, in this model there are  fewer radio pulsars for both BHNSs and NSBHs. 

Given that all pulsars in BHNSs are unrecycled they are therefore unaffected by the value of the parameter $\Delta M_d$. 
Increasing this parameter (as in model FDM-20) does not affect BHNSs.
The recycled pulsars, forming the bulk of the NSBH sub-population are affected as determined by Eqn.~\ref{MagneticFieldAccretion}.
The magnetic fields of pulsars in model FDM-20 (with $\Delta M_d=0.02$\,M$_\odot$) are buried less by accretion than in the Fiducial model ($M_d=0.015$\,M$_\odot$), leading to a higher mean $B$ for pulsars in NSBH binaries.
Though accretion induced spin up is independent of $\Delta M_d$, spin down is dependant on the surface magnetic field by Eqn.~\ref{SpinDownIsolated}. 
The higher mean $B$ of recycled pulsars causes higher spin down rate for FDM-20, also resulting in quantitatively fewer NSBH radio pulsars than in the Fiducial model. 

The birth distribution of the magnetic field also affects the $P$, $\dot{P}$ and $B$ distributions, as explored by model BMF-FL. 
The distribution of magnetic field strengths in model BMF-FL is shifted to lower values than the Fiducial model (see Fig.8 of \citealp{Chattopadhyay:DNS2019}). 
The smaller average value of $B$ also causes the net population to have a slower spin down rate (smaller $\dot{\Omega}$), resulting in more rapidly spinning pulsars 
and increasing their radio lifetime.
The latter causes the number of radio NS+BH systems to be higher in BMF-FL than in the Fiducial model (see Table~\ref{tab:netNumbers}, columns 3 and 8). 
Hence for BMF-FL both NSBHs and BHNSs have lower mean values of $P$, $\dot{P}$ and $B$ than the Fiducial model.

The CDFs of $P$, $\dot{P}$ and $B$ for model Fiducial are shown in Fig.~\ref{fig:FiducialPPdotB_distribution_for_telescopes}, comparing the survey-detectable populations across Parkes, MeerKAT and SKA. 

In model CE-Z we do not allow mass accretion during CE events. 
This means that in this model, pulsars are not spun up during CE leading to a drastic lowering of the number of recycled pulsars.
This is reflected in the resultant population having a higher mean value of $P$ (i.e. the pulsars spin more slowly) causing a lower number of radio NSBHs than in the Fiducial model. 
This is also reflected in the higher $P$, $\dot{P}$ and $B$ mean values of NSBHs for CE-Z relative to Fiducial. 

\begin{figure}
\includegraphics[width=0.99\columnwidth]{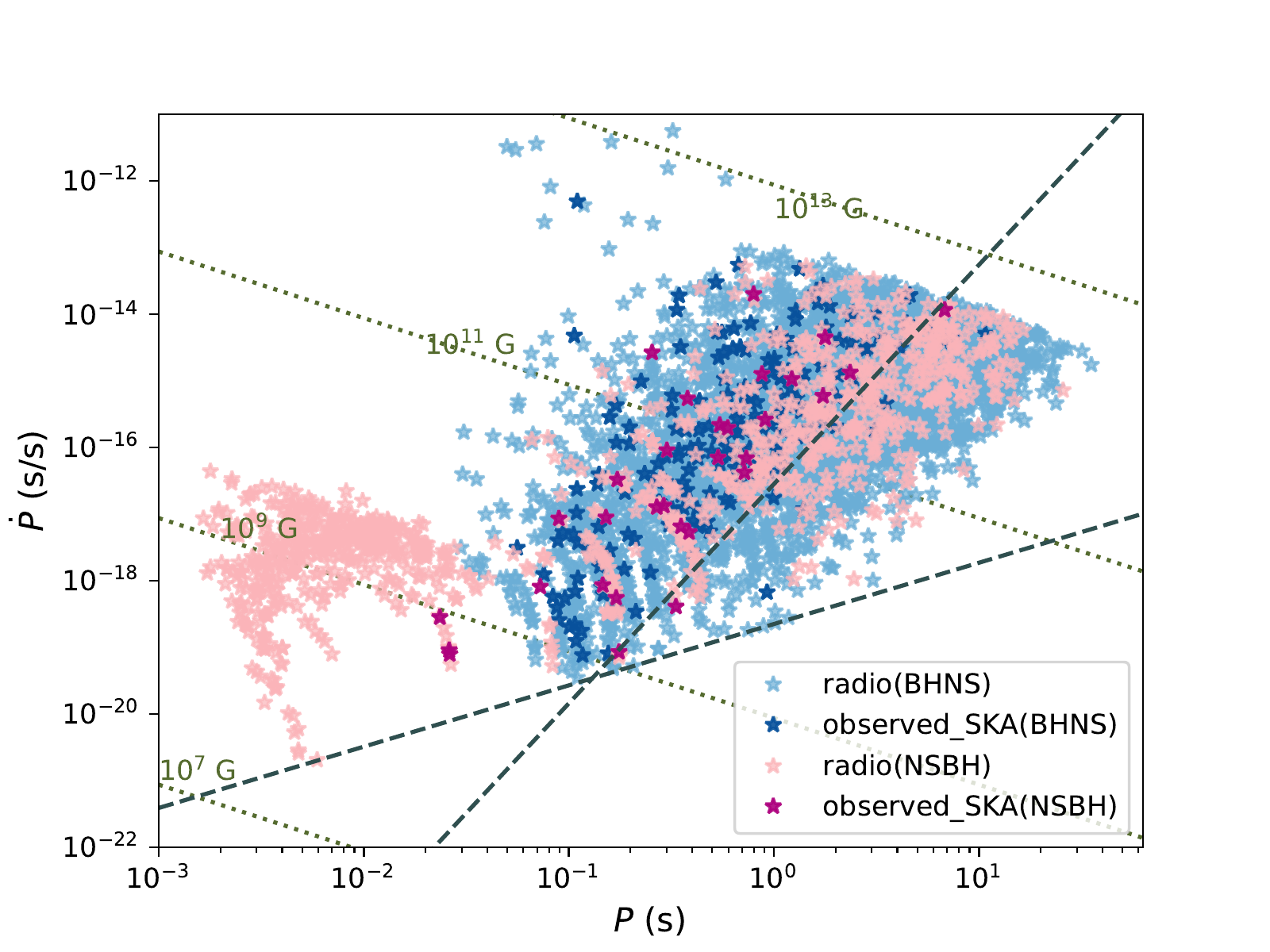}
\caption{$P \dot{P}$ diagram showing the SKA-observable pulsars for model ZM-001. Symbols have the same meaning as in Figure~\ref{fig:PPdot}. Compared to the Fiducial model (Figure~\ref{fig:PPdot}), there is a lack of recycled NSBH pulsars in the bottom left corner of the plot. This is created by the rapid merger of the radio NSBHs (light pink). See Figure~\ref{fig:Z001_merger} for more details.
}
\label{fig:ZM001_PdotP}
\end{figure}

In model ZM-001, at a low metallicity of $Z=0.001$, 
we see from Tables~\ref{tab:radioObservablesMean:Radio} 
that the mean values of $P$, $\dot{P}$ and $B$ are at least an order of magnitude larger for the NSBH population than in the Fiducial model.
This is due to the presence of a larger proportion of non-recycled pulsars in NSBH binaries.
Decreased wind mass-loss at lower metallicity creates more massive He stars, which tend to expand less than their less massive counterparts \citep{Hurley:2000pk}. 
The lower expansion rate reduces the occurrence of mass transfer that is the essential phenomenon to spin-up pulsars. 
Moreover, increased formation of double-core CE without case-BB mass transfer at lower metallicity further aids in decreasing pulsar recycling and only $\approx$ 36\% of pulsars of the radio NSBH population are recycled \citep{Broekgaarden:2020NSBH}.
The lower fraction of recycled pulsars reduces the number of radio NSBH pulsars by 1/3 in ZM-001 compared to the Fiducial model. 
The $P\dot{P}$ diagram for ZM-001 is shown in Fig.~\ref{fig:ZM001_PdotP}. 
The recycled NSBHs have shorter delay time ($t_\mathrm{m}$, the time difference between double compact object formation and merger) compared to those in the Fiducial model and merge faster than the timescale for magnetic field decay. 
This effect is more apparent in the recycled pulsar BH population of ZM-001. 
We discuss this more in section ~\ref{subsec:masses}.

\subsubsection{$P_\mathrm{orb}$--$e$}
\label{subsubsec:orbital_period_eccentricity}

\begin{figure}
\includegraphics[width=8cm, height=6.5cm]{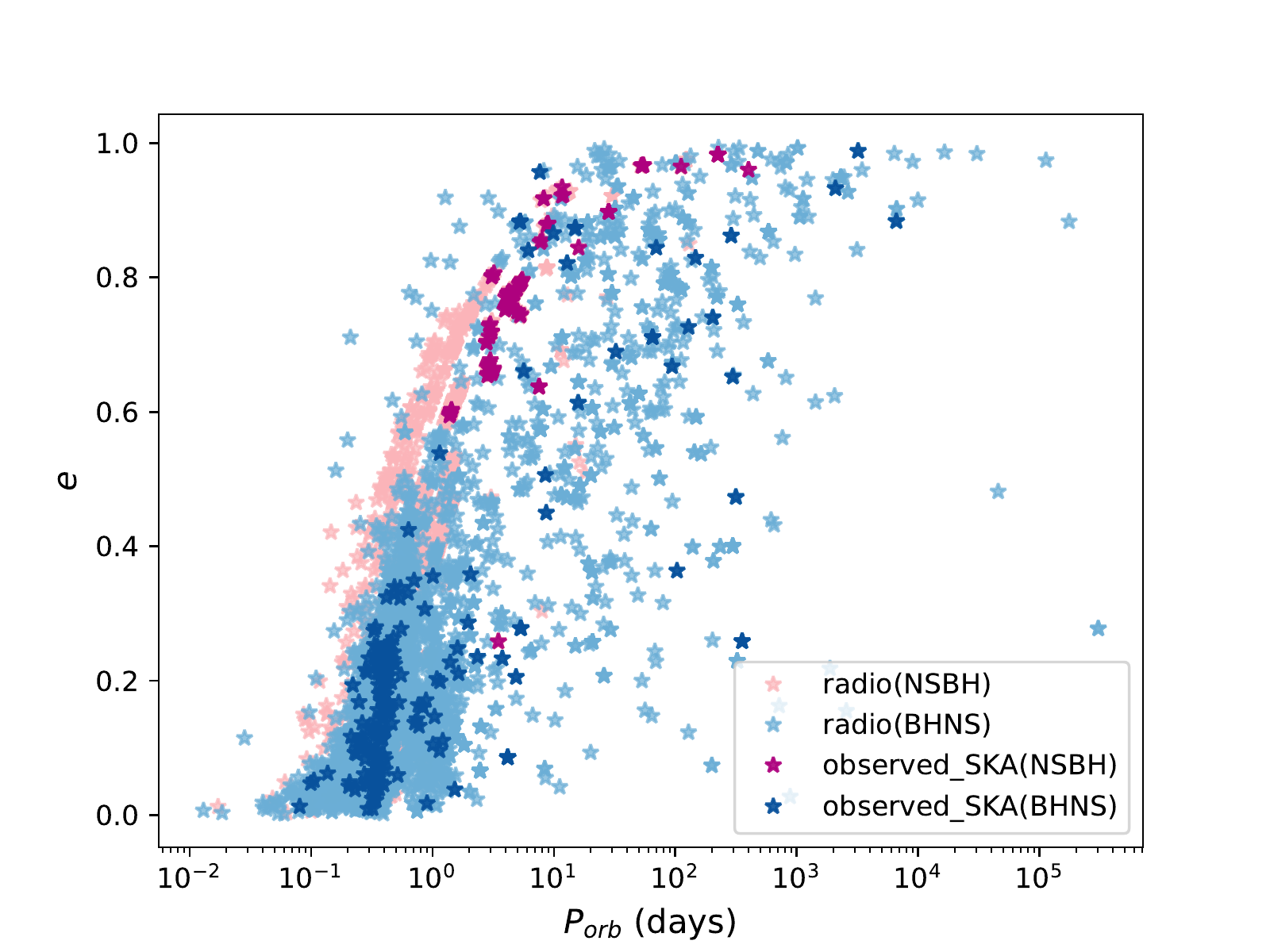}
\caption{$P_{\mathrm{orb}}$ vs. $e$ scatter plot for radio-alive and SKA observable pulsar+BH binaries from our Fiducial model (size $\sim$10 Milky-Way). 
The NSBHs are plotted in pink and BHNSs in blue. 
The radio-alive population is lighter in colour, the SKA-observed sub-population is in a darker shade. Distinctive segregation of NSBHs and BHNSs is noted in the SKA population, with BHNSs at a lower $e$ compared to NSBHs. }
\label{fig:PorbVSe_SKA}
\end{figure}

\begin{figure}
\includegraphics[width=6.3cm, height=5cm]{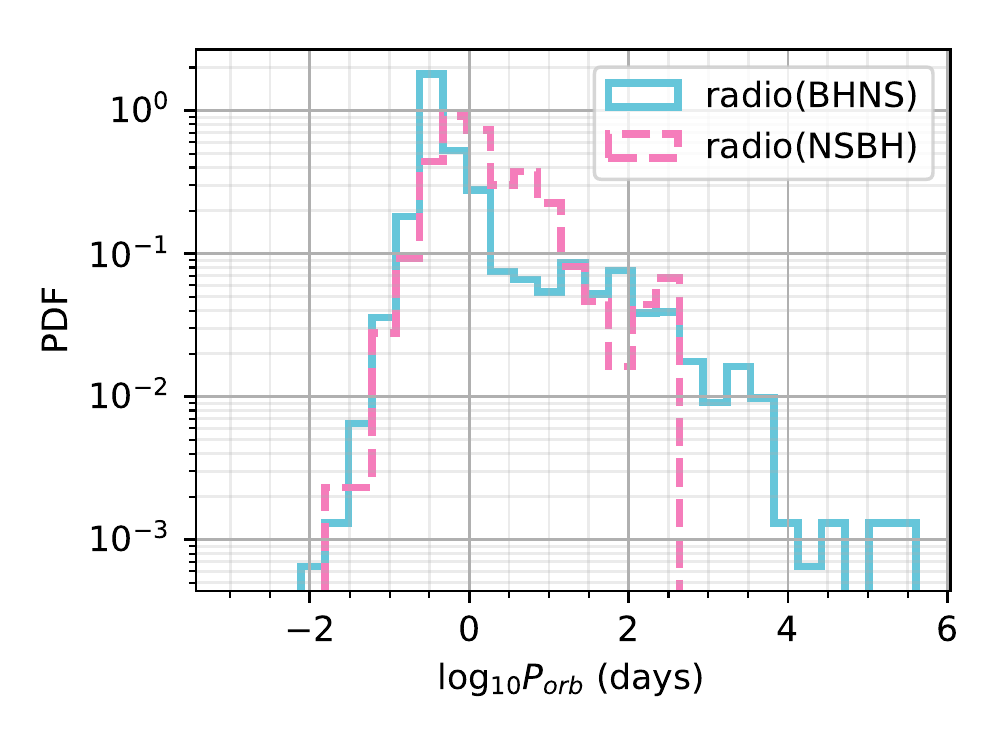}
\includegraphics[width=6.3cm, height=5cm]{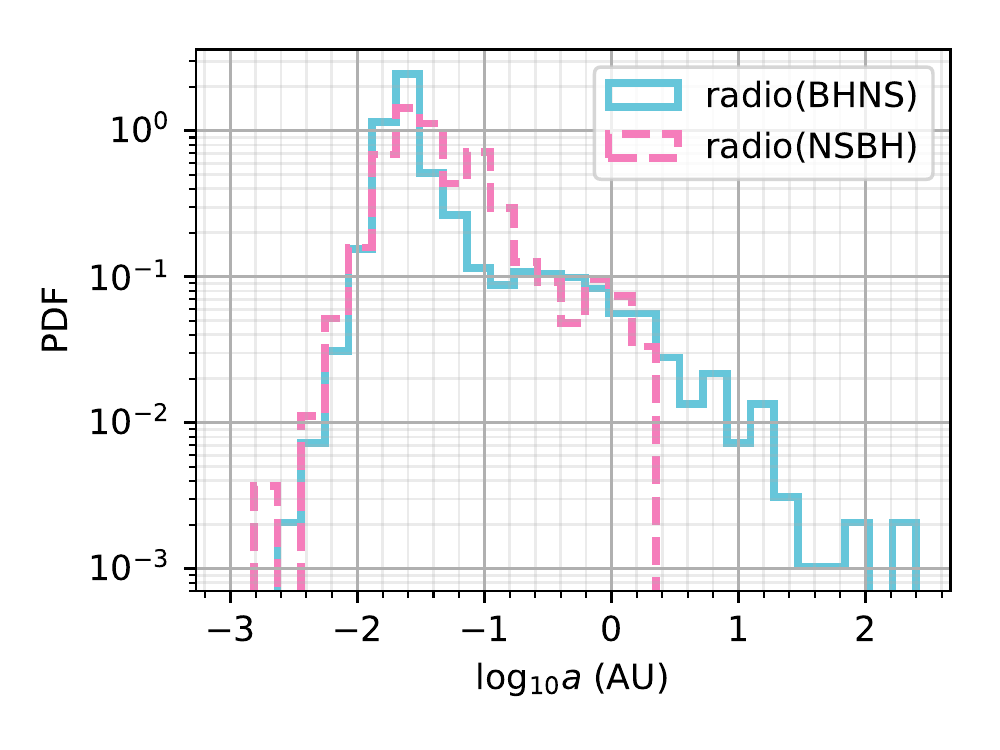}
\caption{
Histograms of logarithms of orbital period $P_{\mathrm{orb}}$ (top) and binary separation $a$ (bottom) showing radio-alive populations for the Fiducial model. The BHNSs are shown as blue solid lines and the NSBHs as pink dashed lines. Both SNs that form the BHNS systems are dominantly USSNe, causing the systems to have a lower average SN kick than NSBHs primarily formed through CCSNe. The loosely bound systems with higher orbital period and separation hence survive more often for BHNSs,  accounting for the tail at the upper end of the distributions. Lower mean SNe kicks for BHNSs also causes less increase in post-SNe binary $a$ and $P_{\mathrm{orb}}$ compared to higher kick magnitudes for NSBHs. The BHNSs hence show a slightly lower peak log$_{10}P_{\mathrm{orb}}$ and log$_{10}a$ than NSBHs.}
\label{fig:porb_a_pdf}
\end{figure}

\begin{figure}
\includegraphics[width=0.43\textwidth]{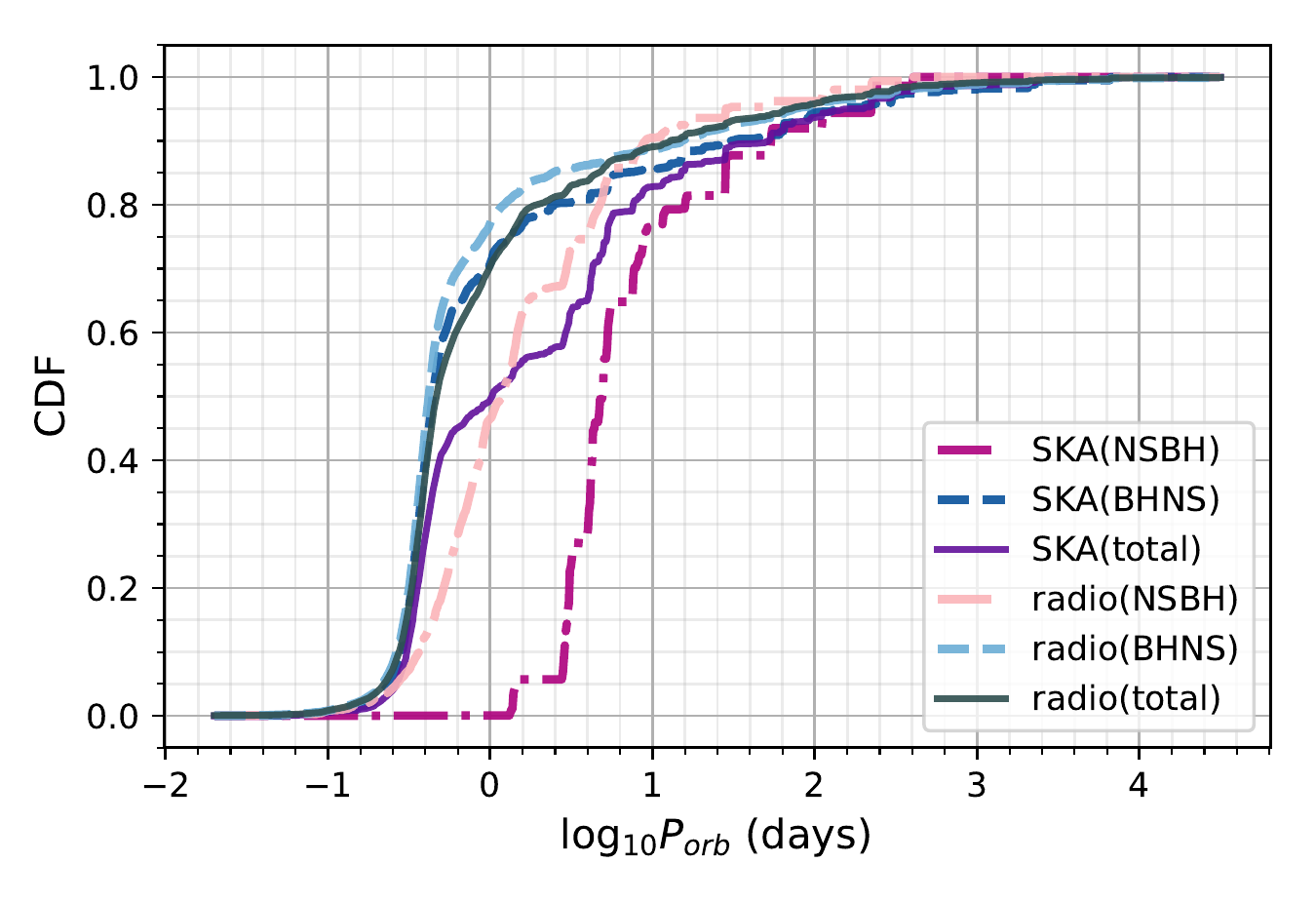}
\includegraphics[width=0.43\textwidth]{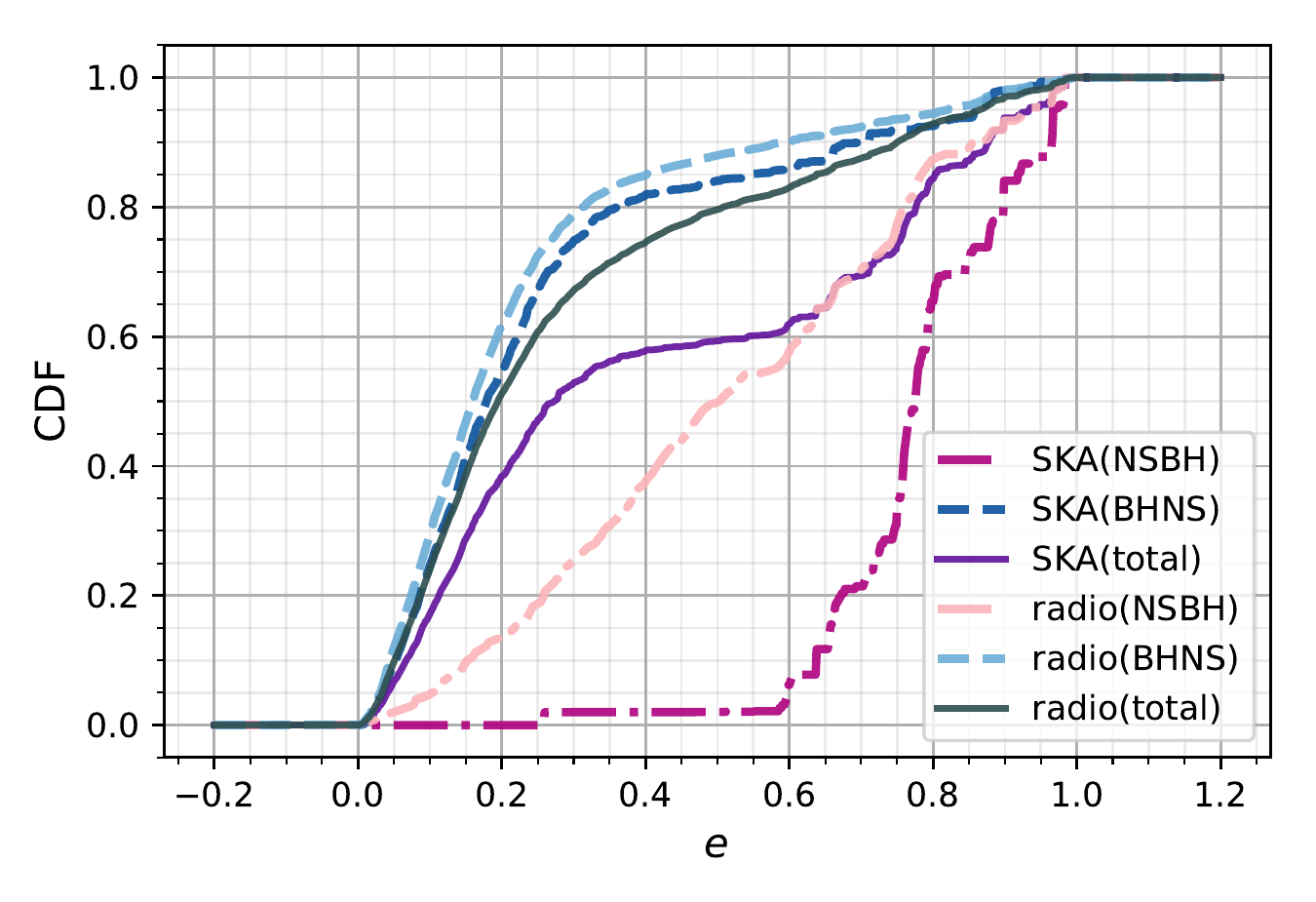}
\caption{
Cumulative distributions of orbital periods $P_{\mathrm{orb}}$ (top) and eccentricity $e$ (bottom) for the radio-alive and SKA-observed populations for the Fiducial model. 
The total population (NSBH and BHNS) is shown with a solid line, and the NSBH and BHNS sub-populations in dot-dashed and dashed lines,  respectively. 
The shift of the observed distribution towards longer orbital periods and higher eccentricity (larger values of both are dominated by NSBHs) is due to radio-selection effects described in Sec.~\ref{subsubsec:eccentric_binaries}}
\label{fig:porb_and_e_distribution}
\end{figure}

The orbital period $P_\mathrm{orb}$ and eccentricity $e$ of NS+BH binaries depend on the order in which the compact objects form (i.e. NSBH vs BHNS), our assumptions about massive binary evolution and the radio selection effects. 
For $P_{\mathrm{orb}}$, the efficiency of CE ejection ($\alpha_{\mathrm{CE}}$) also plays a key role \citep[e.g.][]{Dominik:2012,Giacobbo:2018etu}, as it determines the separation of the binary after ejection of the envelope.
We have assumed $\alpha_{\mathrm{CE}} = 1$ in all of our models.
Other factors that determine $P_{\mathrm{orb}}$ include case BB mass transfer and the supernova natal kick. 
The orbital eccentricity $e$, on the other hand, is primarily dependent on the SN kick where 
higher magnitude, asymmetric kicks create binaries with higher eccentricity.
The $P_\mathrm{orb}-e$ scatter-plot for the Fiducial model is shown in Fig.~\ref{fig:PorbVSe_SKA}.

In our Fiducial model, over 90\% of NSs in radio BHNSs are born in USSNe, whereas 100\% of BHs in NSBHs are born in CCSNe. 
The latter receive a second SN natal kick which is nearly a factor of 4 higher than the former.
Higher natal kicks for NSBHs will disrupt the wider binaries, leaving only the compact systems behind, whereas the the lower second SN birth kick for BHNSs allow the existence of broader binaries. 
Thus, 
as we see in Tables~\ref{tab:radioObservablesMean:Radio}, 
the Fiducial model NSBHs show a lower mean $P_{\mathrm{orb}}$ than BHNSs.

Models ZM-001 and ZM-02 with respectively lower and higher metallicity than the Fiducial model experience different SNe and hence natal kick distributions than the Fiducial model. 
In the ZM-001 radio BHNS population, $\approx$ 62\% of the NSs experience USSNe, $\approx$ 3\% ECSNe and the remaining 35 \% are CCSNe remnants. 
This higher proportion of NSs formed in CCSNe results in radio-BHNSs having a mean second SN kick magnitude of around 100\,km/s, about twice that of the Fiducial model.
The BHs of radio NSBHs in ZM-001 are all formed from CCSNe as in the Fiducial model. 
However, lower metallicity facilitates the formation of more massive BH remnants due to reduced stellar winds (the mean radio-NSBH BH mass for ZM-001 is about 5$\times$ the Fiducial) and therefore increased fallback mass. 
Since BH kicks are scaled down by this fallback mass (the fallback mass fraction for BHs of radio NSBHs in model ZM-001 is about 0.88 compared to 0.37 in the Fiducial model), the mean second supernova kick for the NSBHs of ZM-001 is around 50 km/s, almost one-fourth of that in Fiducial. 
As a result the mean $P_{\mathrm{orb}}$ values for the NSBH and BHNS sub-populations in ZM-001 are lower and higher than Fiducial respectively. 
For the higher metallicity model ZM-02, the reverse of the explained effect occurs, rendering the mean BHNS NS kick to be 60\,km/s and NSBH BH kick to be around 230\,km/s. 
Lower and higher mean $e$ values for the NSBHs of the ZM$-$001 and ZM$-$02 models relative to the Fiducial model can be explained by their lower and higher mean kick magnitudes as described above, which also explains the reason for the higher BHNS mean $e$ values for both models.

Model FDT-500 shows much larger mean $P_{\mathrm{orb}}$ values for both radio NSBHs and radio BHNSs than Fiducial. 
This can be understood by the orbital separation of the binaries.
The radio NSBHs of FDT-500 show a mean separation of 0.639\,AU compared to Fiducial's 0.097\,AU.
Since lowering $\tau_{d}$ decreases the radio lifetime of the binaries, FDT-500's radio population is considerably younger than Fiducial's. 
Younger NS+BH binaries tend to have a larger separation and hence larger orbital period, as the binary does not get sufficient time to become compact by emitting gravitational radiation. 

Model BHK-Z shows orders of magnitude higher $P_{\mathrm{orb}}$ compared to the Fiducial model for both NSBHs and BHNSs. 
Since the BHs receive no natal kick at birth in this model, systems with loosely bound orbits, with larger separation and higher values of $P_{\mathrm{orb}}$ that would not have survived in the Fiducial model survive for BHK-Z. 

In the Fiducial model, the mass distribution of the BHNS population is much higher than the NSBHs (see Table~\ref{tab:ZAMS_radioVSnet}). A larger ZAMS mass distribution results in a larger fallback mass and hence greater fallback mass scaling for the BH progenitors \citep{Fryer:2012}. BHs in radio BHNSs have a mean natal kick velocity of $\approx$85\,km/s due to having a mean fallback-mass scaling factor of about 0.72, compared to BHs of NSBHs that have a mean natal kick of around 200\,km/s because of the fallback mass scaling being around 0.37. The lower natal kick allows loosely bound binaries in the BHNS population to survive more frequently, whereas for NSBH systems, only the binaries with a tightly bound orbit (and thus higher binding energies) survive the SN explosion of the BH progenitor. If BHs receive no kicks at formation, more NSBH systems would survive, while a full, un-scaled kick would disrupt more BHNS systems than NSBHs. In model BHK-Z there are around 300 times more NSBHs than the Fiducial model, and around 60 times more BHNSs. More NSBHs survive a full, non-scaled BH natal kick compared to BHNSs, as the former tend to have tighter orbits. For model BHK-F, where the BHs obtain a full, un-scaled birth kick from the SN event, only 16\% of the net BHNSs survive compared to 39\% of the net NSBHs in the Fiducial model. 

The radio selection effects for $P_{\mathrm{orb}}$ and $e$ are  inter-correlated and pulsar search algorithm dependant (see Sec~\ref{subsubsec:eccentric_binaries}). 
In general, this shifts the mean eccentricity and orbital period for NSBHs and BHNSs towards higher values, if other selection effects remain constant. 
For the Fiducial model, the $P_{\mathrm{orb}}-e$ diagram in Fig.~\ref{fig:PorbVSe_SKA} shows the background radio population and the SKA-observed distribution. 
The observed NSBH and BHNS pulsars appear as two separate populations in the higher and lower eccentricity regions. 

Orbits with high $e$ and large separation can have low binding energy. 
Model BHK-Z (which imparts no kick to BHs at formation) allows such systems to survive (they are easily disrupted by BH natal kicks in the Fiducial model). 
Such loosely bound wide systems tend to completely avoid CE evolution, unlike closer binaries where the CE phase circularizes the systems. 
The presence of these wide systems (with low binding energies) in model BHK-Z causes its $e$ distribution to have a significantly higher mean compared to the Fiducial model.

Fig.~\ref{fig:porb_a_pdf} shows the orbital period and separation distributions for Fiducial radio BHNSs and NSBHs represented through their probability density function (PDF).
Both of the SNe that form the BHNS systems are dominantly USSNe, causing the systems to have a lower average SN kick than NSBHs primarily formed through CCSNe. 
The loosely bound systems with higher orbital period and separation hence survives for BHNSs accounting for the tail at the upper end of the distributions. 
Lower mean SN kicks for BHNSs also causes less increase in the post-SNe binary $a$ and $P_{\mathrm{orb}}$ values compared to higher kick magnitudes for NSBHs. 
The BHNSs hence show a slightly lower peak log$_{10}P_{\mathrm{orb}}$ and log$_{10}a$ than NSBHs.

Fig.~\ref{fig:porb_and_e_distribution} shows the CDFs of $P_\mathrm{orb}$  and $e$ of PSR+BH binaries from the Fiducial model. 
We find that PSR+BHs have orbital periods $P_\mathrm{orb}$ in the range $0.1$--$100$\,days. 
BHNS binaries tend to be more compact with lower $P_\mathrm{orb}$  and $e$ than NSBH binaries due to USSNe kicks. 
This means they will be more accelerated, making them harder to observe in fast Fourier transform based pulsar searches, as discussed in Sec.~\ref{subsubsec:eccentric_binaries}. 
The post-radio selection effects distribution is hence shifted to larger values of orbital period.

The dominant formation channel of radio NSBHs consists of binaries with high ZAMS mass-ratios ($q>$0.7, where $q=m_\mathrm{2}/m_\mathrm{1}$ and $m_\mathrm{1}>m_\mathrm{2}$) within a mass range of about $20$--$25$\, M$_\odot$ for the primary.
However, the `rapid' prescription (model RM-R) -- which  enforces the existence of a mass gap between NSs and BHs -- creates lower mass-ratio NSBH binaries. 
Thus, RM-R NSBHs have ZAMS $q\approx 0.5$--$0.6$, and the primary mass peaking around $25$--$30$\,M$_\odot$. 
The complete suppression of the formation channel of more equal-mass ZAMS binaries reduces the net number of NSBHs in model RM-R by about 40\% compared to the Fiducial model. 
Such massive binaries with highly unequal masses require a huge amount of mass-loss and mass-transfer from the primary to create a NS first. 
Typically the ZAMS separation of NSBH RM-R binaries are only slightly lower than in the Fiducial model.
However, more gravitational attraction and hence progressively more stable mass transfer causes a rapid reduction of orbital separation in the RM-R case. 
Following the formation of the NS, the dominant channels of both Fiducial and RM-R NSBHs experience unstable mass transfer and CE. 
The lower orbital separations and more massive BH-progenitor for RM-R causes them to become ultra-stripped. 
Thus the second SN is predominantly an USSN for RM-R NSBHs. 
The lower natal kick distribution of USSNe together with increased fallback for more massive BH-progenitors makes the average second SN kick for RM-R radio NSBHs about $4 \, {\rm km} \, {\rm s}^{-1}$, an order-of-magnitude lower than that of the Fiducial population.
Increased mass transfer, decreased orbital separation and lowered second SN kick also reduces the final $e$ of the binaries, making them more circular than Fiducial.


\subsubsection{$|Z|$}
\label{subsec:Z}

\begin{figure}
\includegraphics[width=9cm, height=2cm]{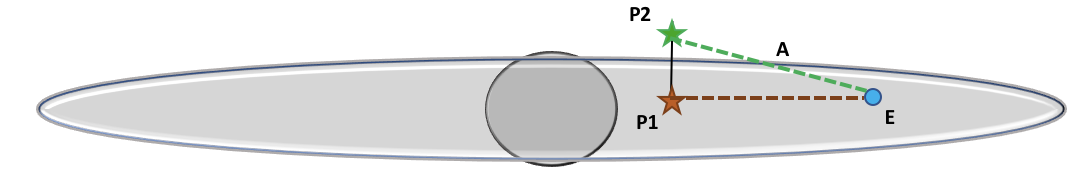}
\caption{Schematic diagram to explain the variance of mean $|Z|$ for the radio and survey observed populations of NS+BHs. The grey disc is assumed to be the disc of the Milky Way and the Earth is denoted by the blue circle E. P1 (orange) and P2 (green) are two pulsars located on the same vertical line, hence P2 having the same X and Y but higher Z in the Galacto-centric Cartesian co-ordinate system. A radio beam from P1 travels completely through the Galactic disc (orange line P1E) and hence suffers more dispersion compared to beam P2E (green line, A being the point where the beam intersects the Galactic plane). 
As long as AE$<$P1E, P2 will be easier to detect than P1, provided all other conditions are the same. However, if P2 is too distant, the beam P2E will lose its intensity and become harder to detect.}
\label{fig:MilkyWay}
\end{figure}

\begin{figure}
\includegraphics[width=0.99\columnwidth]{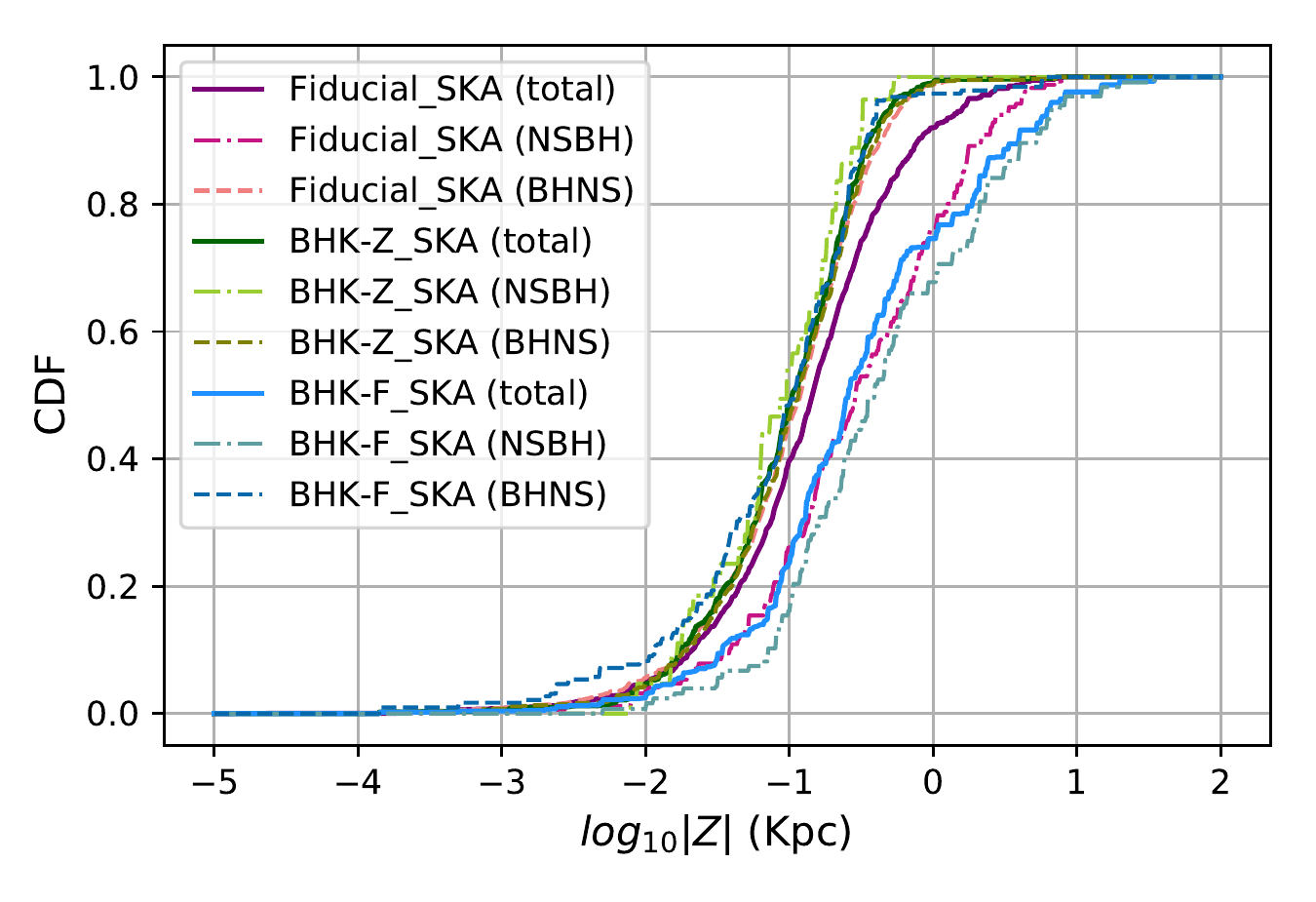}
\caption{Cumulative distribution of scale heights $\log_{10}|Z / \mathrm{kpc}|$ for SKA-observed PSR+BH binaries. We show predictions from the Fiducial model, as well as a model with larger BH natal kicks (BHK-F) and no BH natal kicks (BHK-Z). 
Solid lines denote the total population, while dash-dotted and dashed lines shown the NSBH and BHNS sub-populations. 
The Fiducial models are shown in shades of violet, BHK-Z in blue and BHK-F in green. 
The abundance of systems due to the lack of BH kicks in model BHK-Z is evident in the smoothness of the associated CDFs. 
BHK-F shows the most bias towards larger values of $|Z|$ due to receiving the highest BH kicks followed by Fiducial where BH kicks are scaled by fall-back mass. 
BHK-Z has much smaller $|Z|$ values, as the BHs receive no kick. 
Real observations by future pulsar-surveys may hence aide in constraining BH natal kicks.}
\label{fig:Z}
\end{figure}

The scale height $|Z|$ of the NS+BH binaries is primarily determined by the birth locations of massive binary stars, the mass distribution of the binaries, the Galactic potential and supernova kicks. 
Since all the presented models use the same Milky-Way-like galactic potential and NS natal velocity distributions, ordinarily the BH birth kicks determine the $|Z|$ distribution for each model. However, secondary effects from changing the mass distribution (for instance, by changing the metallicity range) and hence the fallback mass kick scaling factor also affects the $|Z|$ distribution. 
In our Fiducial model, BH supernova kicks are scaled down by the fallback masses (see Section~\ref{subsubsec:neutronStarBlackHoleKicks}). 
The BH kick prescriptions are varied in models BHK-Z and BHK-F. 
For the former, the BHs receive no velocity at birth, while for the latter, BH natal kicks are the exact same as for NSs, without any fallback scale-down. 
Thus, BHK-Z and BHK-F explore the two extremities of BH birth kicks, while the Fiducial represents the parameter space in-between. 
We show the distributions of $Z$ observable by the SKA for models BHK-Z, BHK-F and the Fiducial model in Figure~\ref{fig:Z}.

The $|Z|$ distributions are decided by the natal kicks of NSs for BHNSs and BHs for NSBHs, because the second supernova plays the deterministic role for the $|Z|$ distribution. 
As shown in Tables~\ref{tab:radioObservablesMean:Radio}, 
for the Fiducial model, BHNSs have a lower mean $|Z|$ (0.16\,kpc) than NSBHs (0.38\,kpc). 
Though it appears counter-intuitive since BHs kicks are scaled by the fallback mass it can be explained as the following. 
The BHs of radio-NSBHs are always created by CCSNe for Fiducial.
More than 90\% of NSs in radio-BHNSs in this model are created from USSNe, $\approx$ 0.2\% from ECSNe and the rest from CCSNe. 
Since the USSNe birth kick distribution is of considerably lower magnitude than CCSNe (Section~\ref{subsubsec:neutronStarBlackHoleKicks}), even with fallback mass birth velocity scale-down, the BHs of NSBHs experience a higher natal kick than BHNS NSs. Moreover BHNS systems are typically more massive (about twice, for Fiducial)
than NSBHs. Hence for the same natal kick, the BHNSs adjust to a lower centre of mass velocity -- thus lower scale height, compared to NSBHs. In our Fiducial model, the mean second supernova birth kick for radio-NSBHs is $\approx$ 200 km/s and for radio-BHNSs is $\approx$ 50 km/s. 

At sub-solar metallicity (as in model ZM-001), reduced wind mass-loss leads to more massive stars at core collapse \citep[e.g.][]{2010ApJ...714.1217B}. 
Hence more massive BH progenitors have larger fallback mass which reduces the natal kicks. 
Most of the potential NSBHs in the Fiducial model are disrupted due to the CCSNe BH progenitor natal kick. 
However, lower kicks due to higher amounts of fallback mass strongly affects the net number of ZM-001 NSBHs, also accounting for the larger value of mean $|Z|$. The higher metallicity model ZM-02 shows the opposite effect.  

Higher BH birth kicks causes more disrupted binaries, lowering the final number of NS+BHs as seen in BHK-F, while no BH birth kick results in an increased quantity of the binaries in the resultant population as seen in BHK-Z compared to model Fiducial.
In our models, the second supernova determines the final position of the double compact object in the Galaxy when it is observed and thus NSBH scale heights are affected by the change of BH kick prescription, reflecting also on the overall $|Z|$ distribution of the NS+BHs. 
This is reflected by BHK-Z NSBHs having lower mean $|Z|$ and BHK-F NSBHs have higher mean $|Z|$ than Fiducial NSBHs. 
An observed population of Galactic NSBHs with their scale heights measured will hence be able to constrain the BH supernova kicks \citep[see also][]{2005ApJ...628..343P,Kiel:2010MNRAS}. 

Since the rapid SNe prescription causes the more energetic explosions, the radio BHNSs of model RM-R experience a slightly higher second SN kick ($\approx 66 \, {\rm km} \, {\rm s}^{-1}$) than for the Fiducial radio BHNS. 
By the time of the second SN the binary is usually hardened (having a higher binding energy) due to mass transfer onto the first formed compact object and thus slightly higher kicks from the second SN explosion do not cause more binary disruptions to affect the BHNS numbers. 
The higher NS SN kick of radio BHNSs, in addition to the fact that the total system mass is slightly reduced (see Tab.~\ref{tab:radioObservablesMean:Radio} and sec.~\ref{subsec:masses}) causes the $|Z|$ distribution of the RM-R radio BHNSs to be higher than in our Fiducial model. 
The $\approx 10 \times$ lowering of the second SN of RM-R radio NSBHs is reflected in the $|Z|$ mean, which is about 1/3rd of the Fiducial.

Also, the average $|Z|$ for the SKA-observed population is slightly higher than the radio population across all models.
This is because pulsars located a little above the Galactic plane can be easier to observe due to lower stellar density and hence lower sky temperature. 
This is illustrated in Fig.~\ref{fig:MilkyWay}, where two pulsars with the same X and Y Galactro-centric coordinates but with different Z coordinates are compared. 
Radio waves from P2, located above the plane of the Galaxy, travel through a less dense environment en-route to Earth (and hence suffer less dispersion) than those from P1 located in the plane of the Galaxy (see Eqn.s 28 and 29 of \citealp{Chattopadhyay:DNS2019} for details). However, a beam that is too far off from the Galactic plane will become fainter and harder to detect -- as happens for radio NSBHs of model BHK-F. The mean $|Z|$ of the radio NSBHs is 0.27\,Kpc higher than those observed by the SKA (see Tab.~\ref{tab:radioObservablesMean:Radio} and \ref{tab:radioObservablesMean:SKA}).
We note that the $|Z|$ distribution for survey-observable binaries is dependant on the sky-coverage by the telescope.
A pulsar survey restricted to the Galactic plane (such as MeerKAT, MeerKAT$_\mathrm{G}$,MeerKAT$_\mathrm{GT}$ of Table~\ref{table:telescope_specs}) will naturally have lower $|Z|$ than quoted above assuming an all-sky survey. 

\begin{figure}
\includegraphics[width=1.1\columnwidth]{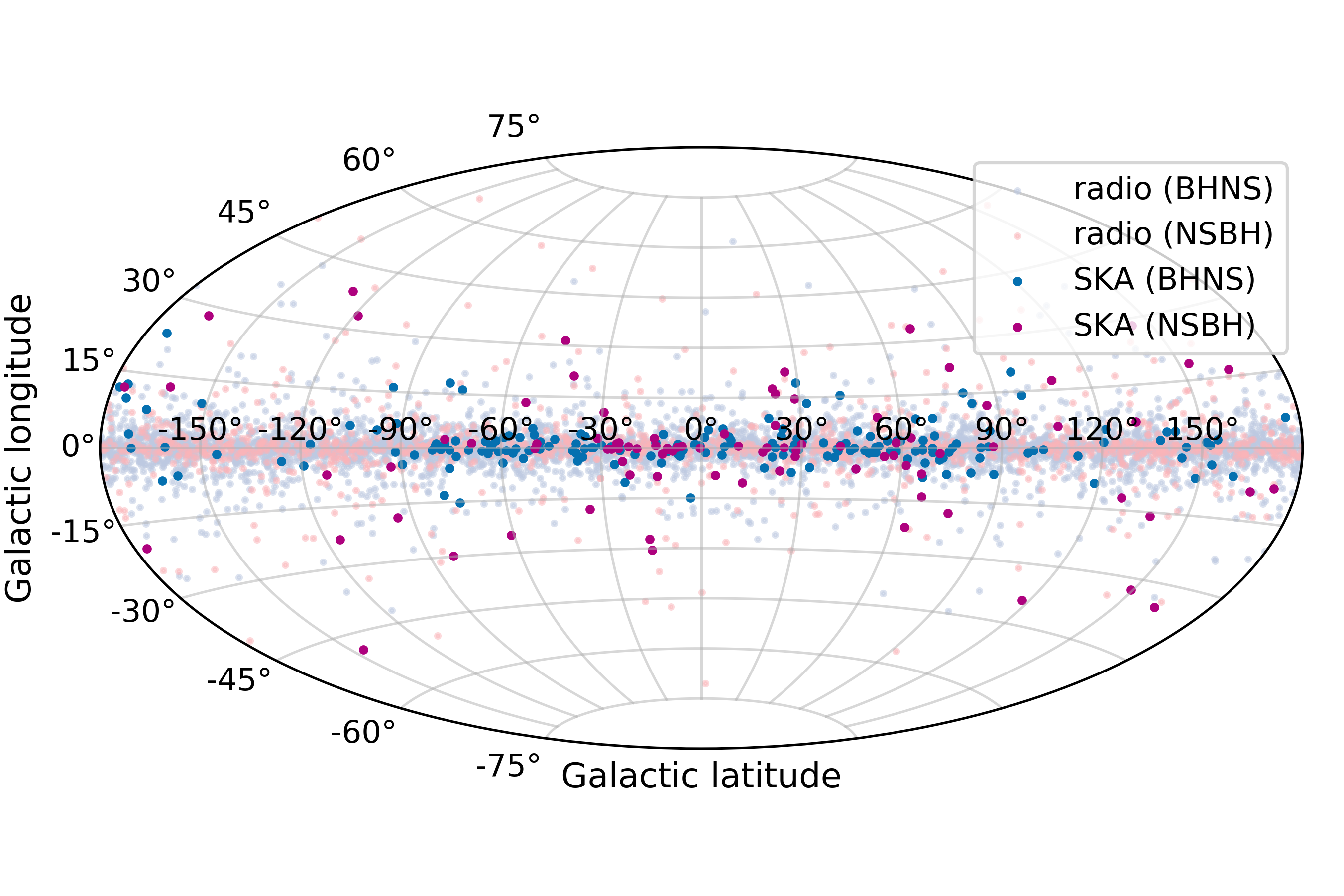}
\caption{The skymap projection of the observations by survey SKA specified in Tab.~\ref{table:telescope_specs}. The radio NSBH and BHNS populations are plotted in light pink and light blue respectively, while the SKA-observable NSBHs and BHNSs are represented in dark pink and dark blue respectively. The visualization represents about 10 times the size of Milky-Way systems.}
\label{fig:skymap}
\end{figure}

We show an Aitoff  projection of the spatial distribution of PSR+BHs observed by the SKA in Fig.~\ref{fig:skymap}. The trend of NSBHs to have a higher $|Z|$ than BHNSs, as well as the survey detectable pulsars being off the Galactic plane, can be readily seen. 

\input{meanValues_Observables}
\input{meanValuesSKA}


\subsubsection{$m_{\mathrm{psr}}$-$m_{\mathrm{cmp}}$}
\label{subsubsec:pulsar_companion_masses}

\begin{figure}
\includegraphics[width=0.8\columnwidth]{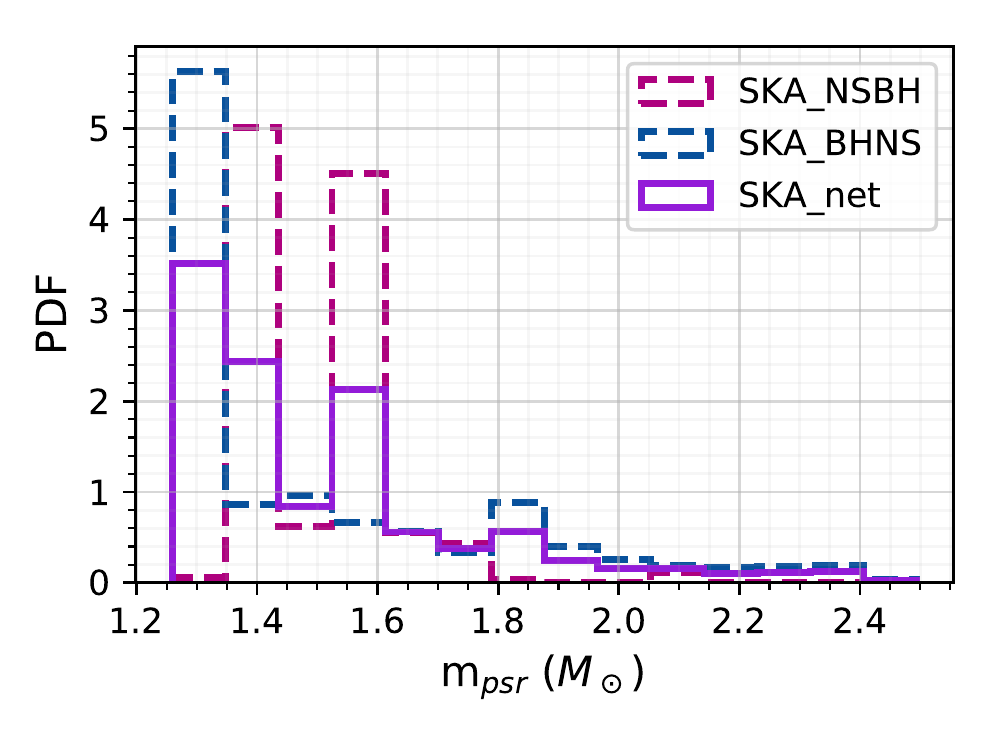}
\includegraphics[width=0.8\columnwidth]{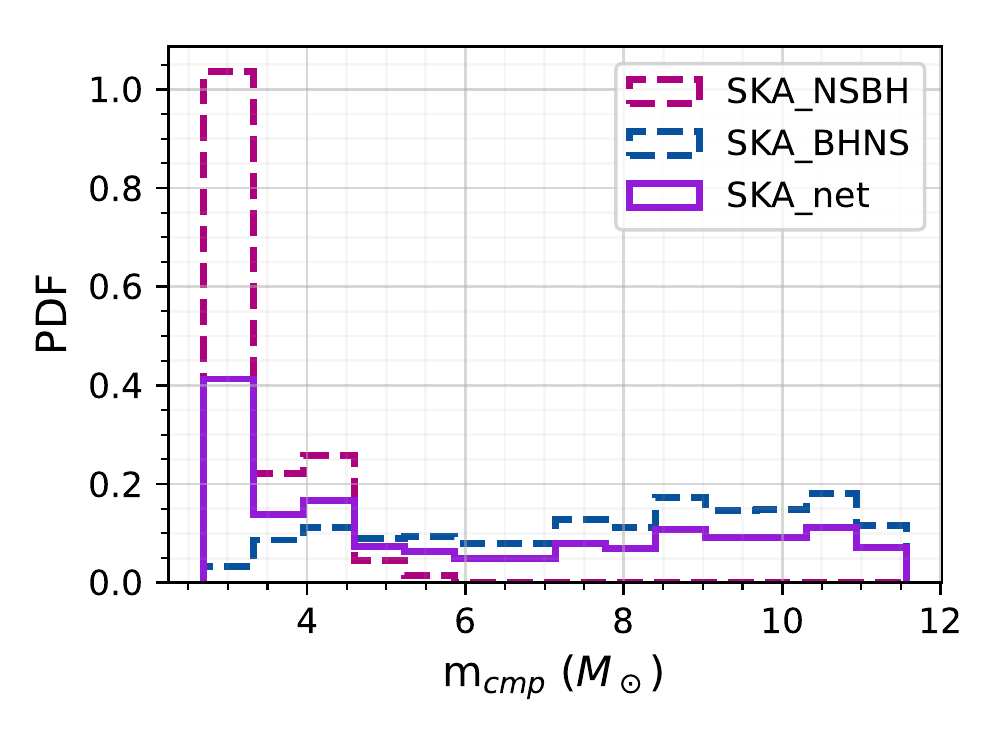}
\caption{Pulsar (top) and companion (bottom) mass distributions for our Fiducial model. 
The NSBH and BHNS sub-populations are shown with broken lines and the net population is shown with the solid line. BHNS and NSBH observable by SKA have a bias to pulsar masses  $<1.3$\,M$_\odot$ and companion masses  $<6$\, M$_\odot$, respectively.
}
\label{fig:M_pulsarM_comp}
\end{figure}

\begin{figure}
\includegraphics[width=0.8\columnwidth]{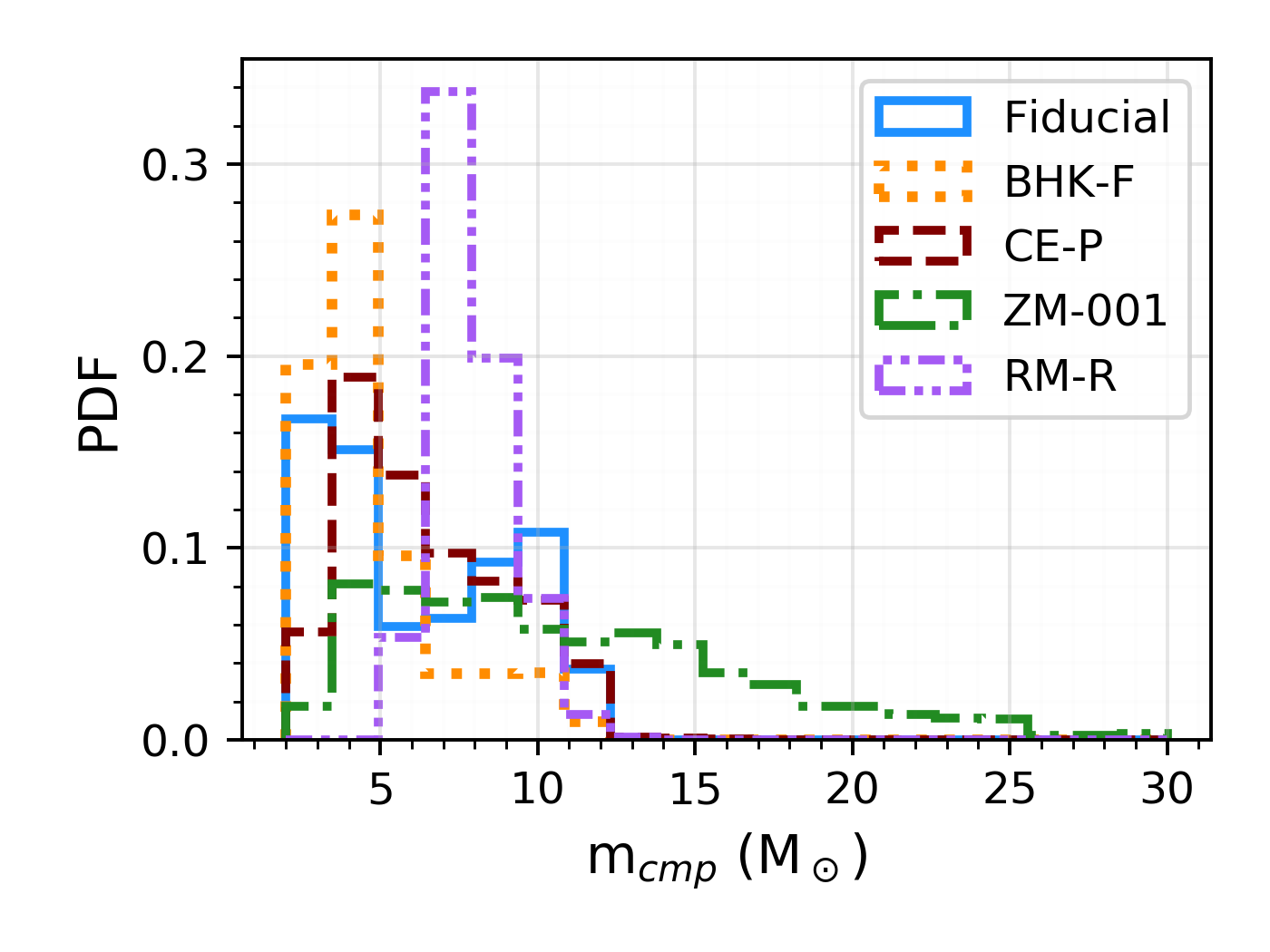}
\includegraphics[width=0.8\columnwidth]{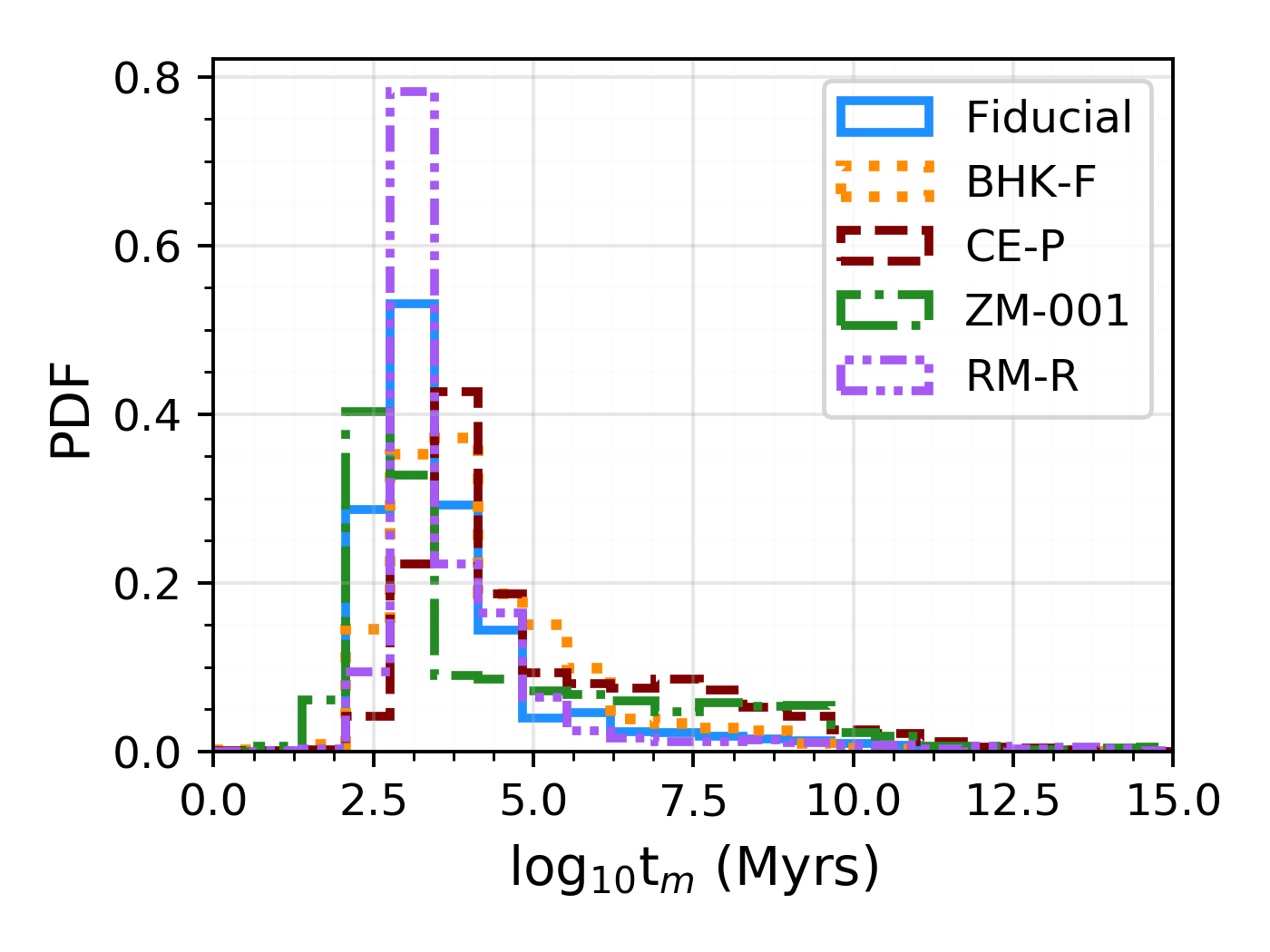}
\caption{The companion BH mass distribution of all radio pulsars (radio NS+BH binaries) for models Fiducial, BHK-F, CE-P, ZM-001, RM-R are shown in the top panel. RM-R $m_\mathrm{cmp}$ peaks around 8 M$_\odot$ while Fiducial shows a much broader distribution between about 2.5 - 12 M$_\odot$ and the ZM-001 $m_\mathrm{cmp}$ distribution extends upto about 30 M$_\odot$.
The delay time distribution of the same radio NS+BH binaries for the same models are shown in the lower panel. CE-P and BHK-F has the largest $t_\mathrm{m}$ peaks, while ZM-001 has the lowest.}
\label{fig:M_compt_delay}
\end{figure}

The pulsar and the companion masses ($m_{\mathrm{psr}}$ and $m_{\mathrm{cmp}}$, respectively)  may also be derived from radio observations \citep[e.g.][]{Ferdman:2020huz}. 
The pulsar and companion mass distributions for survey-detectable SKA population are shown in Fig.~\ref{fig:M_pulsarM_comp}. 
We choose the SKA population because it has the largest sample-size and the mass distribution is survey independent according to our radio selection prescription. 
For $m_{\mathrm{psr}}$ the NSBH population is biased towards heavier masses than the NSs of the total NSBH population (see Tab.~\ref{tab:ZAMS_radioVSnet}) due to the dominant presence of recycled pulsars that accrete more matter from the companion. Since more massive non-recycled pulsars have higher moment of inertia $I$ which decreases the spin deceleration $\dot{\Omega}$ (see Sec.~\ref{subsec:formation_channels}), the BHNS sub-population, though lacking mass transfer, also constitutes heavy pulsars ($m_{\mathrm{psr}}$ $>$ 1.8 M$_\odot$). However, BHNSs still peak at low masses because the \citet{Kroupa:2000iv} initial mass function favours the birth of low mass stars which form lower mass pulsars.
The companion mass distribution is also shown in  Fig.~\ref{fig:M_pulsarM_comp}, and is overall flatter in nature than for $m_{\mathrm{psr}}$. 

The BHNS $m_{\mathrm{cmp}}$ are dominated by more massive BHs (mean value of around 8\,M$_\odot$ for the Fiducial model) since the originating ZAMS population is comprised of such systems (this is due to the preference of non-recycled radio pulsars being heavier, and hence partial towards having massive companions). 
The BHs from the radio-detected NSBHs however are less massive (mean value of about 3.3\,M$_\odot$ for the Fiducial model). 
This trend is apparent even in the SKA-net curve, and hence being able to infer the companion mass of a population of NS+BH binaries can constrain and help to segregate the BHNS and NSBH populations. 

In general, given all other factors remaining constant, the masses of recycled pulsars are affected by several factors --- i) higher mass and hence larger $I$ which reduces the spin deceleration $\dot{\Omega}$ and increases the radio lifetime, ii) lower mass and hence smaller $I$ which increases the spin acceleration due to accretion $\dot{\Omega}_\mathrm{acc}$ and also increases the radio lifetime, and iii) larger mass accretion rate $\dot{M}_\mathrm{NS}$ which leads to higher $\dot{\Omega}_\mathrm{acc}$. In the context of PSR+BH systems, the particular formation channel also plays an important role in determining the pulsar mass distribution. Comparing the mean m$_\mathrm{psr}$ for the pre-selection effect radio population (see Tab.~\ref{tab:radioObservablesMean:Radio}) of all models other than Fiducial, we notice that NSBHs tend to have larger values (ranging from slightly to significantly, as discussed below). This is because ordinarily the formation channel for NSBHs requires relatively more massive NS progenitors than for BHNSs. Higher mass main-sequence stars form heavier carbon-oxygen cores and hence heavier NSs. For the Fiducial model the mean m$_\mathrm{psr}$ is only slightly heavier than for BHNSs and is due to the $\dot{\Omega}_\mathrm{acc}$ effect, more evident in the top histogram of Fig.~\ref{fig:M_pulsarM_comp}. Selection effects due to radio telescope surveys (pulse width $W_\mathrm{e}$, beaming fraction $f_\mathrm{beaming}$ -- see equation~\ref{eq:Radiometer} and \ref{equ:fBeaming}) that typically bias the detected pulsars towards smaller $P$ (i.e. fast spinning) also indirectly affect the $m_\mathrm{psr}$ distribution of the binaries. Hence the observed pulsar mass distribution for binary systems remain a complicated function of binary evolution, pulsar evolution and radio selection effects. 

The CE-Z model effects the radio NSBH $m_{\mathrm{psr}}$ and $m_{\mathrm{cmp}}$ since the change in CE mass accretion prescription only affects the recycled pulsars. 
Since no mass is accreted by the pulsar during CE, they are only recycled through RLOF. 
Hence, heavier pulsars that have accreted more mass are spun up more and that aids them in sustaining their radio-lifetime longer by lowering the spin deceleration through having a higher moment of inertia. 
Since the Fiducial model allows pulsar recycling through both RLOF and CE, with mass accretion during the latter being typically higher \citep[see section 2.2.1 and 2.2.2 of][]{Chattopadhyay:DNS2019}, this suppressed effect due to $I$ becomes apparent for CE-Z.
The heavier companion mass is also noted since the companion progenitors with higher mass are biased towards transferring more mass during RLOF, provided all other conditions remain constant.

Model CE-P shows a strong bias towards more massive pulsar and companion masses for the NSBH sub-population, while BHNS companions tend to be slightly less massive than in the Fiducial model. 
The latter phenomenon can be explained by the fact that more massive companions typically experience a larger radial expansion during the HG phase and thus tend to engage in CE. 
In model CE-P such systems that involve a HG donor during CE \citep{Dominik:2012} end in a merger and hence this shifts the $m_\mathrm{cmp}$ towards a smaller mean value. 
The affinity of the CE-P NSBHs towards heavier masses is due to the presence of nearly equal-mass progenitors that evolve in a similar time-span creating double core CE \citep{1995ApJ...440..270B,1998ApJ...506..780B}. 
See formation channel IV of \citet{Broekgaarden:2020NSBH} for further details. 
We note that in model CE-P, the NSBH population is extremely rare and suffers from biases of small number statistics.

\begin{figure}
\includegraphics[width=0.7\columnwidth]{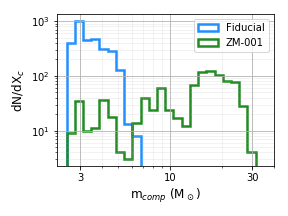}
\includegraphics[width=0.7\columnwidth]{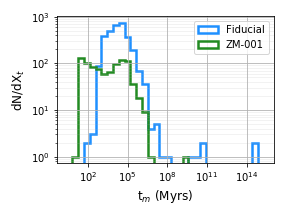}
\caption{Companion mass (top panel) and delay time (bottom panel) distributions for radio NSBHs in the Fiducial model (blue) and in the low metallicity model ZM-001 (green). X$_\mathrm{c}$=log$_{10}$m$_\mathrm{comp}$ and X$_\mathrm{t}$=log$_{10}$t$_\mathrm{m}$. dN/dX$_\mathrm{c}$ and dN/dX$_\mathrm{t}$ signify the number of systems per log$_{10}$m$_\mathrm{comp}$ and per log$_{10}$t$_\mathrm{m}$ bins respectively. Log-spaced bins are used for easier visual identifiability.
dN/dm$_\mathrm{comp}$ and dN/dt$_\mathrm{m}$ signify the number of systems per m$_\mathrm{comp}$ and per t$_\mathrm{m}$ bins respectively. The more massive m$_\mathrm{comp}$ distribution causes the radio NSBHs of ZM-001 to have a lower t$_\mathrm{m}$ distribution.}
\label{fig:Z001_merger}
\end{figure}

Model ZM-001 results in more massive $m_{\mathrm{psr}}$ and $m_{\mathrm{cmp}}$ on average for both NSBHs and BHNSs due to reduced wind mass loss of the respective progenitors at lower metallicity. 
The decreased wind mass-loss leads to larger carbon-oxygen cores for lower metallicity \citep[e.g.][]{WoosleyHegerWeaver:2002} and hence biases the pulsar-companion mass distribution towards higher values.
The model with higher metallicity, ZM-02, shows the opposite behaviour due to increased wind mass-loss. 
The increase in the typical companion mass for radio NSBHs in ZM-001 compared to the Fiducial model results in shorter delay times \citep{Peters:1964}. It is important to note that Fiducial NSBHs typically have slightly higher orbital separation (about 0.01 AU) and lower eccentricity (less than half, see Tab.~\ref{tab:radioObservablesMean:Radio}) at the observation time than those in ZM-001. Even though higher eccentricity and smaller orbital separation decreases the inspiral time, given the masses remain constant, here the significantly lower mass distribution (by more than 4$\times$ for the companion BH mass than in ZM-001, see Tab.~\ref{tab:radioObservablesMean:Radio}) is causing Fiducial NSBHs to have a larger delay time.

For the radio NSBHs of model ZM-001 the average delay time is $t_\mathrm{m}\approx9\times10^4$\,Myr, while for the Fiducial model it is $t_\mathrm{m}\approx2\times10^6$\,Myr (see \citealp{Dominik:2012} for metallicity and delay times). 
The $m_{\mathrm{cmp}}$ and $t_\mathrm{m}$ distributions are compared in Fig~\ref{fig:Z001_merger}. The shorter binary lifetime also negatively affects the radio lifetime of the ZM-001 binaries, reducing them to about 7\% of Fiducial radio NSBHs (see Tab.~\ref{table:radio_lifetime}).

Model RM-R, due to the constructed mass gap by the rapid SNe prescription creates slightly less massive NSs \citep{Fryer:2012}, as apparent in the mean NS masses of radio BHNSs and NSBHs (see Table~\ref{tab:radioObservablesMean:Radio}). 
The BH masses of the BHNS population are slightly increased due to higher fallback mass, which decreases the first SN kick for radio BHNSs to $\approx 30 \, {\rm km} \, {\rm s}^{-1}$, which is only about 35\% of the Fiducial case. 
While this increases the net BHNS numbers by $\approx 2.8\times$ from our Fiducial model, the increase in radio BHNSs is only $\approx 1.9\times$. 
Since the masses of the NSs are decreased from Fiducial, the reduction of the moment of inertia causes non-recycled pulsars (hence the full radio BHNS population) to show a slightly reduced radio lifetime (Tab.~\ref{table:radio_lifetime}).  
Even with a lower $e$, RM-R radio NSBHs have increased total mass (by $\approx$1.9 from Fiducial) and decreased orbital separation ($\approx 0.02\,$AU at observation time, compared to $\approx 0.04\,$AU for Fiducial), lowering their delay time to $\approx 0.8 \times$ the Fiducial population.
Such accelerated mergers also lowers the radio lifetime of the binaries (see Tab.~\ref{table:radio_lifetime}).

We show the $m_\mathrm{cmp}$ and $t_\mathrm{m}$ distributions for radio NS+BHs, i.e. pulsar BH binaries, for the models Fiducial, BHK-F, CE-P, ZM-001 and RM-R that show the most distinct differences in BH masses in Fig.~\ref{fig:M_compt_delay}.


\subsubsection{Radio population vs. observations}
\label{radioVSobs}

\begin{figure}
\includegraphics[width=.8\columnwidth]{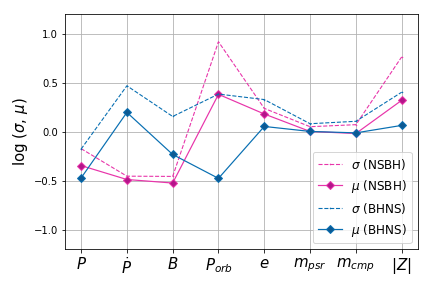}
\caption{
Impact of radio selection effects on the observable parameters of PSR+BH binaries in our Fiducial model. 
We distinguish NSBH (pink) and BHNS (blue) systems. For each parameter we show the ratio of the mean of the distribution observable by the SKA to the the mean of the intrinsic radio population as $\mu=$(SKA mean value)/(radio mean value) as solid lines with diamond markers.
See Tables~\ref{tab:radioObservablesMean:Radio} and \ref{tab:radioObservablesMean:SKA} for more information. 
The dotted lines show the width of the distribution in the same units as $\sigma=$(SKA mean value + SKA standard deviation)/(radio mean value), i.e. $\sigma$ shows an estimate of the spread of the distribution and hence the outliers. 
}
\label{fig:radio_vs_SKA_mean}
\end{figure}

We have highlighted the importance of radio selection effects in determining the observational properties of the PSR+BH population. The underlying radio population --- which is again a subset of the net population --- may have a different distribution from the observations. As we see from Tables~\ref{tab:radioObservablesMean:Radio} and \ref{tab:radioObservablesMean:SKA}, which respectively show the mean values of the radio and the SKA observed parameters, the selection effects do change the nature of the detected population compared to the base radio population. 
Fig.~\ref{fig:radio_vs_SKA_mean} further illustrates these differences by showing the ratio of the mean values of the population observational parameters with and without selection effects for the Fiducial model.  
The observable parameter primarily involved in determining the radio selection effects is the pulsar spin $P$ (see equations~\ref{eq:Radiometer} and \ref{equ:fBeaming}) so it is no surprise to see substantial variation in the measured mean value of this parameter. 
Similarly, $\dot{P}$ being an explicit function of $P$ is also affected. 
The surface magnetic field $B$, although explicitly independent of selection effects, also shows variation. 
This is more pronounced for NSBHs where during pulsar recycling, a larger amount of mass accretion causes more spin-up and a hence smaller $P$, while $B$ gets buried. 
For non-recycled pulsars (hence the entire BHNS population) a larger value of B makes the magnitude of $\dot{\Omega}$ higher (see eqn.~\ref{SpinDownIsolated}), causing more pronounced spin deceleration and thus a decreased $P$ which feeds into the selection effects calculation. 
Hence both observed data-sets of NSBHs and BHNSs tend to have a slightly lower mean $B$ than the base radio population. 
$P_\mathrm{orb}$ and $e$ are both affected due to the orbital selection effects discussed in Sec.~\ref{subsubsec:eccentric_binaries}. 
While there is no direct selection effect dependence on the masses $m_\mathrm{psr}$ and $m_\mathrm{cmp}$, we do see slight variations in the mean masses because the pulsar mass is related to the moment of inertia which in turn is related to $P$ which is part of the selection effects calculation as already discussed. 
Finally, the scale height $|Z|$ changes for the radio observations as this is 
a factor in the selection effect calculations (Fig.~\ref{fig:MilkyWay}). 

\subsubsection{Birth times and ages}
\label{subsubsec:birth_times}

\begin{figure}
\includegraphics[width=0.8\columnwidth]{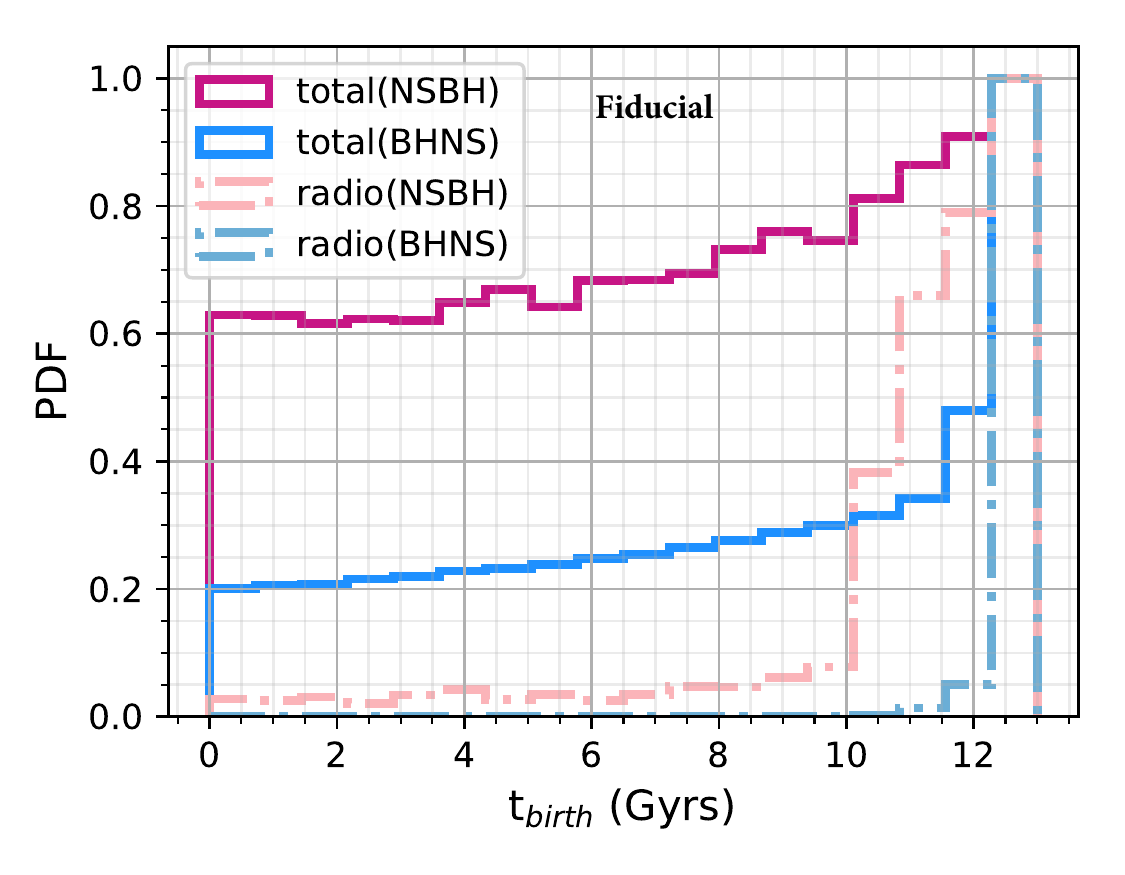}
\includegraphics[width=0.8\columnwidth]{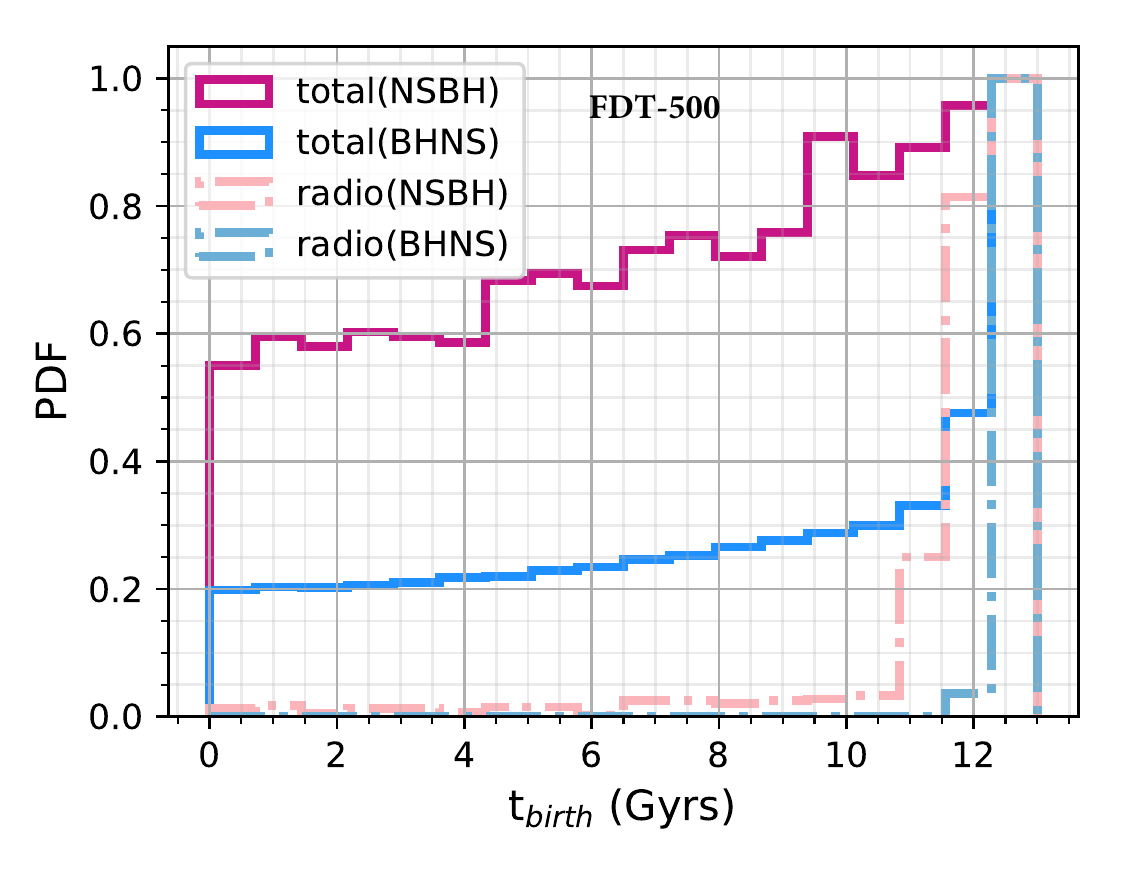}
\caption{Birth times of NSBH (blue) and BHNS (pink) binaries existing in the Milky Way at the present time. The total populations are shown with solid lines, whilst the populations observable in the radio are shown with dotted lines. The top panel shows the Fiducial model and the bottom panel shows model FDT-500 (see Table~\ref{tab:NSBH} for details). Each population is normalised to a maximum of 1; the absolute values of the populations are given in Table~\ref{tab:netNumbers}. 
}
\label{fig:birth_times}
\end{figure}

We show the birth times $t_\mathrm{birth}$ of NSBH and BHNS binaries existing in the Milky Way at the current time in Figure~\ref{fig:birth_times}. 
Both types of binaries are assigned birth times that span  the history of the Milky Way. 
Due to the preference for short time delays between double compact object formation and merger \citep[e.g.][]{Dominik:2012,Neijssel:2019}, there is a slight preference for NS+BH binaries born recently. 
This preference is stronger than for DNSs due to the higher masses in NS+BH binaries. 
The population of NS+BH binaries observed in the radio in our model are all born within the last $\sim 2$\,Gyr. 
This is because the lifetime of recycled pulsars in our model is a few $\tau_{d}$; models with smaller $\tau_{d}$ show a stronger preference for NS+BH binaries being born recently.

Since PSR+BH binaries are all born in the last $\sim 2$\,Gyr, they are most likely formed from stars born in high metallicity environments (via the age-metallicity relation for Milky Way disk field stars, for a review see \citealp{2002ARA&A..40..487F}).
This is a notable difference to the population of NS+BH mergers that will be observed in gravitational-waves, which may form predominately in low metallicity environments, due to the long time delays and higher total masses \citep{Broekgaarden:2020NSBH}.

\begin{figure}
\includegraphics[width=0.85\columnwidth]{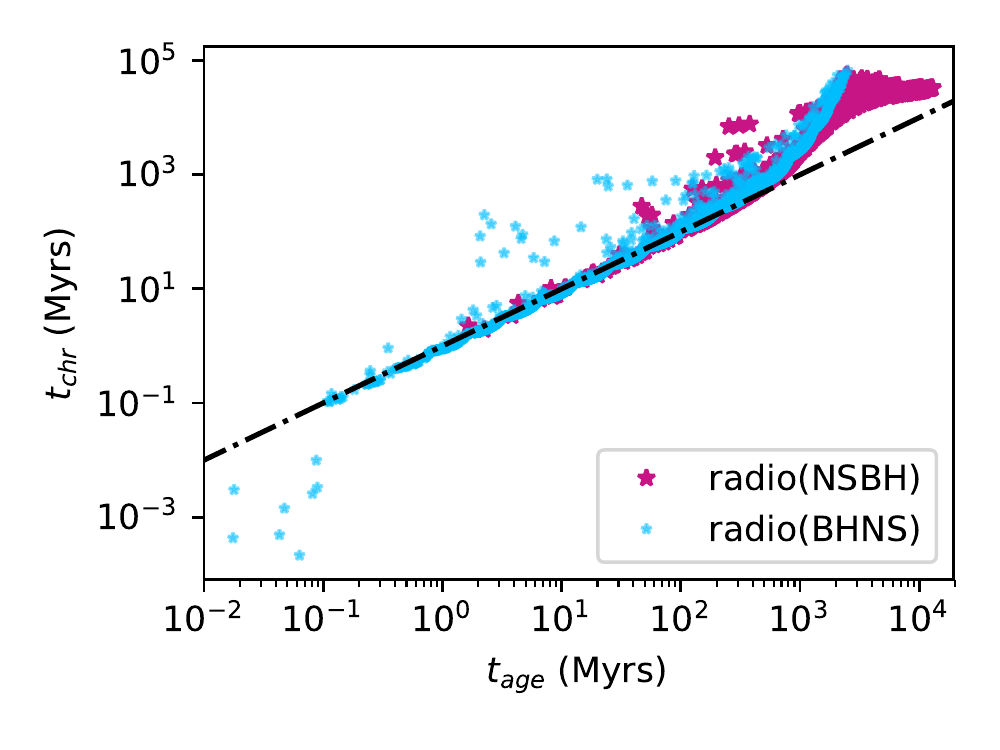}
\caption{Characteristic age of a pulsar $t_{\mathrm{chr}}(= P/2\dot{P})$ plotted against the real age $t_{\mathrm{age}}$ of pulsars in our Fiducial model; the pulsars of BHNSs are in blue and NSBHs in magenta. The NSBHs, due to harbouring recycled pulsars in the data-set have higher values of $t_{\mathrm{age}}$. 
The black line signifies $t_{\mathrm{chr}}=t_{\mathrm{age}}$. 
The fanned out structure of the magenta points are due to different amounts of spin-up of the recycled pulsars.}
\label{fig:T_chr}
\end{figure}

The spin down age or characteristic age of a pulsar is a derived parameter expressed as $t_{\mathrm{chr}}= P/2\dot{P}$. 
Though $t_\mathrm{chr}$ is not the exact real age of the pulsar, it can act as an approximation, especially for younger, non-recycled pulsars. 
Fig.~\ref{fig:T_chr} shows the $t_{\mathrm{chr}}$ compared to the real age $t_{\mathrm{age}}$ for radio pulsars in our Fiducial model.
Interpreting $t_{\mathrm{chr}}$ as the pulsar age may wrongfully give the impression of older or recycled pulsars to be older than they actually are (see top right corner of Fig.~\ref{fig:T_chr}). 
For very young pulsars, $t_{\mathrm{chr}}$ can make them appear younger than the true age.  
 
\subsection{Millisecond Pulsars}
\label{MSPs}

Millisecond pulsars (MSPs) are pulsars with spin period of the order of milliseconds. 
In this paper, we define any pulsar with a spin period $\leq$ 30\,ms as a MSP. 
MSPs---being very fast rotating and extremely accurate in their motion---can act as clocks with atomic-clock order precision, and are hence used in pulsar timing arrays. 

\begin{figure}
    \centering
    \includegraphics[width=0.95\columnwidth]{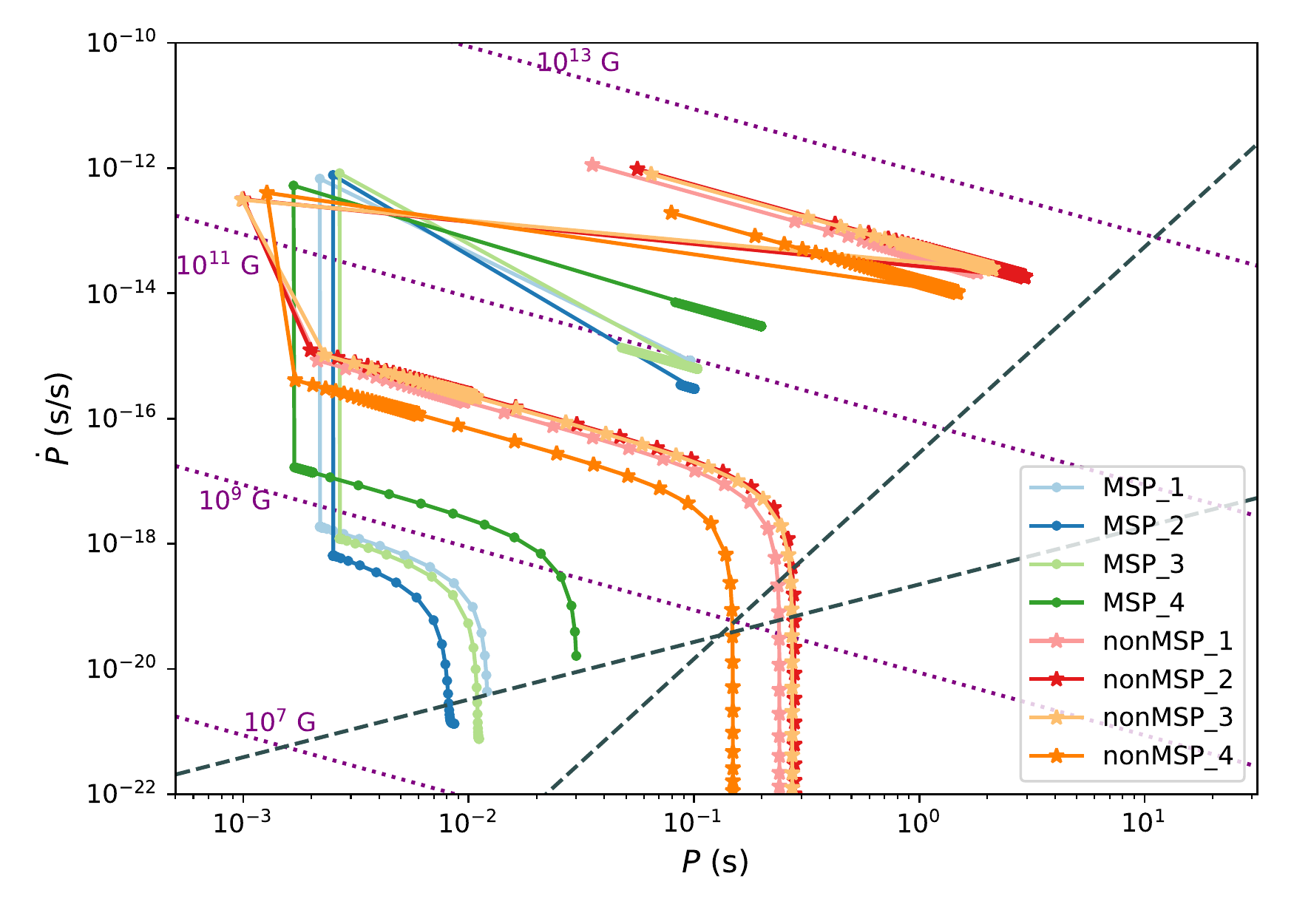}
    \caption{The P$\dot{P}$ diagram during the entire evolution of four sample millisecond (orange/red stars) and non-millisecond (green/blue dots) pulsars from the Fiducial model. For NSBHs, MSP progenitors typically have a lower birth magnetic field strength and experience mass transfer earlier after birth than their non-MSP counterparts. During observation, after accounting for the radio selection effects, we pick only one snapshot from the pulsar's entire trajectory corresponding to the current time.}
    \label{fig:MSP_NSBH_Fiducial}
\end{figure}

In our model, a pulsar may be born with a spin period 10--100\,ms. The fastest of these pulsars (with $P < 30$\,ms) are thus born as MSPs. Such pulsars are expected to rapidly spin down to spin periods of 0.1--1\,s in about 0.1--5\,Myr ($\tau_d$ plays a key role in determining the decay time-scale). 
Hence the possibility of detecting a young, non-recycled pulsar as a MSP is very low.
Observationally, the MSPs are detected to typically have lower surface magnetic fields than other pulsars \citep{2005AJ....129.1993M}.
It is hence assumed that recycling due to mass transfer creates MSPs by imparting angular momentum and the accreted matter buries the surface magnetic field ($\Delta M_d$ quantifying the burial).
For NS+BH systems, the NSBHs provide the only channel that has the possibility of producing MSPs.

Equations~\ref{AngularMomentumAccretion} \&~\ref{MagneticFieldAccretion} show that MSP formation, like pulsar recycling, depends on the rate of mass accretion ($\dot{M}_\mathrm{NS}$), the mass (${M_\mathrm{NS}}$), the moment of inertia ($I$), the spin ($\Omega$) and magnetic field ($B$) of the pulsar. 
Since we assume a fixed radius for NSs and they follow the same equation of state in all our models, the dependence on $I$ reduces to $M_\mathrm{NS}$ only. 

The formation channels of recycled pulsars in binaries with NSs (DNSs) or BHs (NSBHs) are very similar. 
After the MSP progenitor (the NS) is formed through the first SN, the companion evolves to become a Hertzsprung gap (HG) star. 
Unstable mass transfer from the HG star to the NS leads to CE, recycling the pulsar. 
If the binary survives the CE, the expulsion of the envelope leads to the companion evolving to become a stripped (helium) star. 
For DNS progenitors, the stripped star may once again fill its Roche lobe as a Helium HG star and case BB mass transfer can occur, further recycling the pulsar.
For NSBH progenitors, the BH is formed through the collapse of the He HG star and there is no further mass accretion onto the NS.

We find that for our Fiducial model, around 10\% of radio-DNSs contain an MSP. 
For the radio-NSBHs, however, about 27\% contain MSPs at the time of observation. 
The numbers are further reduced after accounting for the radio selection effects and becomes null for ~1 Milky Way DNSs systems.
Though both DNS and NSBH MSPs are created in a similar formation channel as described above, the MSP excess in NSBHs can be explained by two key reasons - i) the amount of mass accreted by the NS, and ii) the time required for the main-sequence companion to reach CE phase.
Furthermore, the mass accretion during CE is dependant on the mass and radius of the companion star. 
On average for MSP companions, in NSBHs the BH progenitor HG star is $\approx$ 1.5 times heavier than the NS progenitor HG star in DNSs. 
The radius of the former is $\approx$ 1.3 times larger than the latter. This results in $\approx$ 10 \% less mass accretion onto the NS accretor in the case of DNSs. 

The time span of the companion evolution is of particular importance because a fast spinning younger pulsar requires less angular momentum to spin up and become a MSP compared to an older, slower pulsar. 
Hence it is crucial that the companion star reaches the CE phase as quickly as possible. 
Since heavier stars evolve more quickly, the BH progenitor star evolves to the naked Helium main sequence star phase in $\approx$ 12\,Myr while the NS progenitor takes $\approx$ 22\,Myr. 
Thus a heavier companion benefits the pulsar by providing more mass to accrete and on a faster timescale. 

The birth magnetic field is of utmost importance for the formation of MSPs through its effect on the angular frequency decay (Eqn.~\ref{SpinDownIsolated}). 
The pulsars with birth surface magnetic field $\leq$ 10$^{12}$ Gauss have more possibility of forming an MSP than other pulsars born with a higher surface magnetic field, if all other conditions remain comparable. 
On the other hand, the birth spin period does not have a significant  impact on MSP formation. 
The choice of birth magnetic field range and distribution hence strongly affects MSP formation. 
NSs with heavier donors have more chances of forming MSPs. 
Lighter pulsars are spun up more with the same torque than heavier pulsars, and hence pre-MSP phase pulsars tend to be less massive on average. 
However, more mass accretion imparts more angular momentum and hence the resultant MSP after formation tends to be of nearly the same mass or heavier than its non-MSP counterparts. 
MSPs of DNSs are $\approx$ 0.07\,M$_\odot$ heavier than the net primary pulsars. 
The MSPs of NSBHs however, are $<$ 0.01\,M$_\odot$ less massive than the net pulsar population. 
This is because all NSBHs have such massive BH progenitors that almost all pulsars experience similar amounts of mass accretion ($\approx$ 0.1\,M$_\odot$) and hence the other variables define MSP formation rather than the amount of mass transfer. 
For DNSs, the non-MSP pulsars have companion HG stars with radii $\approx$ 0.8 times of that of the MSP counterparts, resulting in less mass accretion. 

MSP formation in DNSs and NSBHs is thus a varied and complex process, rather than a separate unique formation channel. Fig.~\ref{fig:MSP_NSBH_Fiducial} shows the evolution of four distinct MSPs contrasted with four distinct non-MSPs in binaries with BHs from model Fiducial in the $P\dot{P}$ diagram. 
The upper-limit for the number of MSP-BH binaries in the Galactic field is estimated from our Fiducial model as being about 40 (27\% of the  145 radio NSBH systems) and the lower limit is zero which comes from the CE-P model (see Table ~\ref{tab:netNumbers}). 
Accounting for the factor of two uncertainty (see section~\ref{sec:discussion}), we thus estimate that approximately 0-80 MSP-BH binaries exist in the Galactic field.
The detection rate for the SKA is 0--2 from models CE-P and FDM-20 after accounting for the radio selection effects which becomes 0--4 after accounting for the factor of two uncertainty.
The number of SKA-observable MSP-NS binaries is estimated to be 6 (8\% of the 78 SKA-observed DNSs) for our Fiducial model. 


\section{Results: Gravitational Waves}
\label{sec:gravitational_waves}

We present predictions for gravitational-wave observable mass and spin distributions for our Fiducial model, both for the full populations and the radio-alive subsets. 
The primary objective for this section is the qualitative analysis marking the unique formation channels of NS+BH binaries distinguishable through GW observables.

As described earlier, the NS+BH binaries can be segregated into two sub-populations: NSBHs and BHNSs depending on whether the NS or the BH forms first, respectively. 
We denote the systems within each sub-population that contain a pulsar as the `radio' population 
and 
refer to the combined radio and non-radio populations as a whole for each of these sub-populations as `total'. 
Moreover, to denote the NSBHs and BHNSs together we use the term `net'.
We now further subdivide these sub-populations into three different categories:  
\begin{enumerate}
    \item LIGO: The systems observable by current ground based gravitational wave observatories such as LIGO,Virgo and KAGRA.
    For this category, we select only those binaries which merge due to gravitational wave emission within the age of the Universe (assumed 13 Gyrs in this paper), which we determine using the formulas for inspiral time described in  \citep{Peters:1964}. 
    We also account for gravitational-wave selection effects which favour more massive binaries in terms of effective volume $V_{\mathrm{eff}} \propto (m_1m_2)^{3/2}(m_1+m_2)^{1/2}$ \citep[][]{Stevenson:2015bqa,Chattopadhyay:DNS2019}. 
    It must be noted that the quantitative number count of binaries in this category does not have a physical meaning in this paper since we are not accounting for cosmological modelling of Milky Way-like galaxies observable by LIGO. 
    However, the qualitative analysis remains unaffected by this.
    We do not attempt to calculate detection rates for LIGO/Virgo/KAGRA; for a detailed calculation of rate predictions see \citet{Broekgaarden:2020NSBH}.
    \item LISA: The systems observable by the future space based gravitational wave observatory LISA.  
    The lowest gravitational-wave frequency LISA is sensitive to is around $10^{-5}$\,Hz, approximately corresponding to a $\sim$ 1 day orbital period \citep{2017arXiv170200786A}. 
    We therefore selectively include binaries with orbital periods $\leq 1$\,day in this category.
    For a more thorough investigation of NS+BHs observable by LISA, see Wagg et al. in prep and \citet{Breivik:2019lmt}.
    \item radio: The systems comprising a black-hole with a pulsar companion, hence potentially observable by pulsar surveys of current and future radio telescopes.
\end{enumerate}

Rather than calculating signal-to-noise ratios to determine which binaries in our models will be detectable for each telescope (e.g. for LIGO, LISA and the SKA), we instead select the underlying population for each telescope. 

\subsection{Mass distribution}
\label{subsec:masses}

\begin{figure}
\includegraphics[width=0.8\columnwidth]{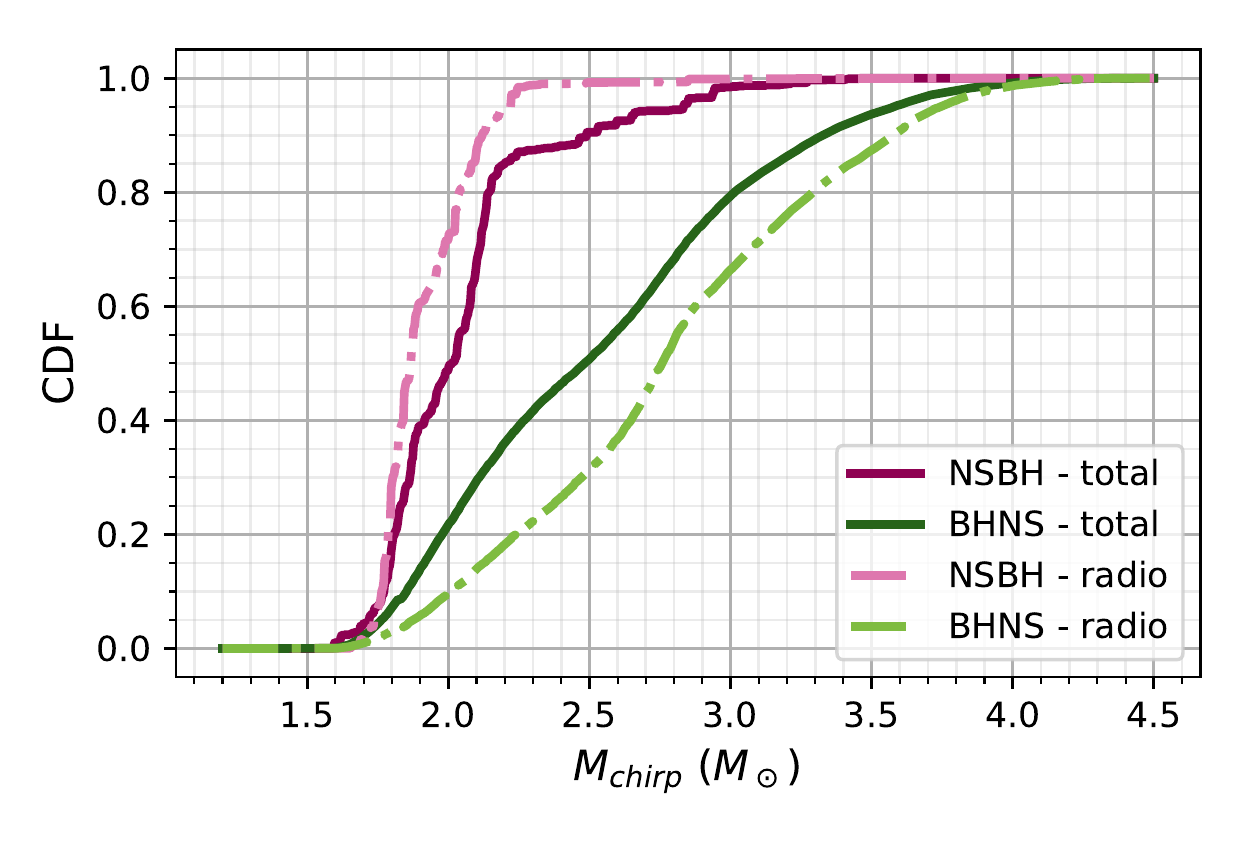}
\caption{Predicted chirp mass distributions of NSBH (magenta) and BHNS (green) binaries in the Fiducial model. The total populations are shown with solid, dark lines, whilst the radio sub-populations are shown with dotted, lighter-coloured line. }
\label{fig:chirp_mass_radio_total_NSBH_BHNS}
\end{figure}

In Figure~\ref{fig:chirp_mass_radio_total_NSBH_BHNS} we show the mass distribution of NSBH and BHNS binaries. 
We choose to quote the chirp mass $\mathcal{M} = (m_1 m_2)^{3/5} (m_1 + m_2)^{-1/5}$, which is the parameter measured with the highest precision with gravitational waves. 
We find that the radio NSBHs are less massive and radio BHNSs are more massive than the total population. 
This is explained in Sec.~\ref{subsec:formation_channels}, which shows the slight bias of NSBH pulsars to less massive binaries as lower moment of inertia allows faster spin-up of dominant recycled pulsars in the sub-population while for the non-recycled BHNS, more massive and thus larger moment of inertia allows slower spin-down rate and hence a longer radio-lifetime.

\begin{figure}
\includegraphics[width=0.99\columnwidth]{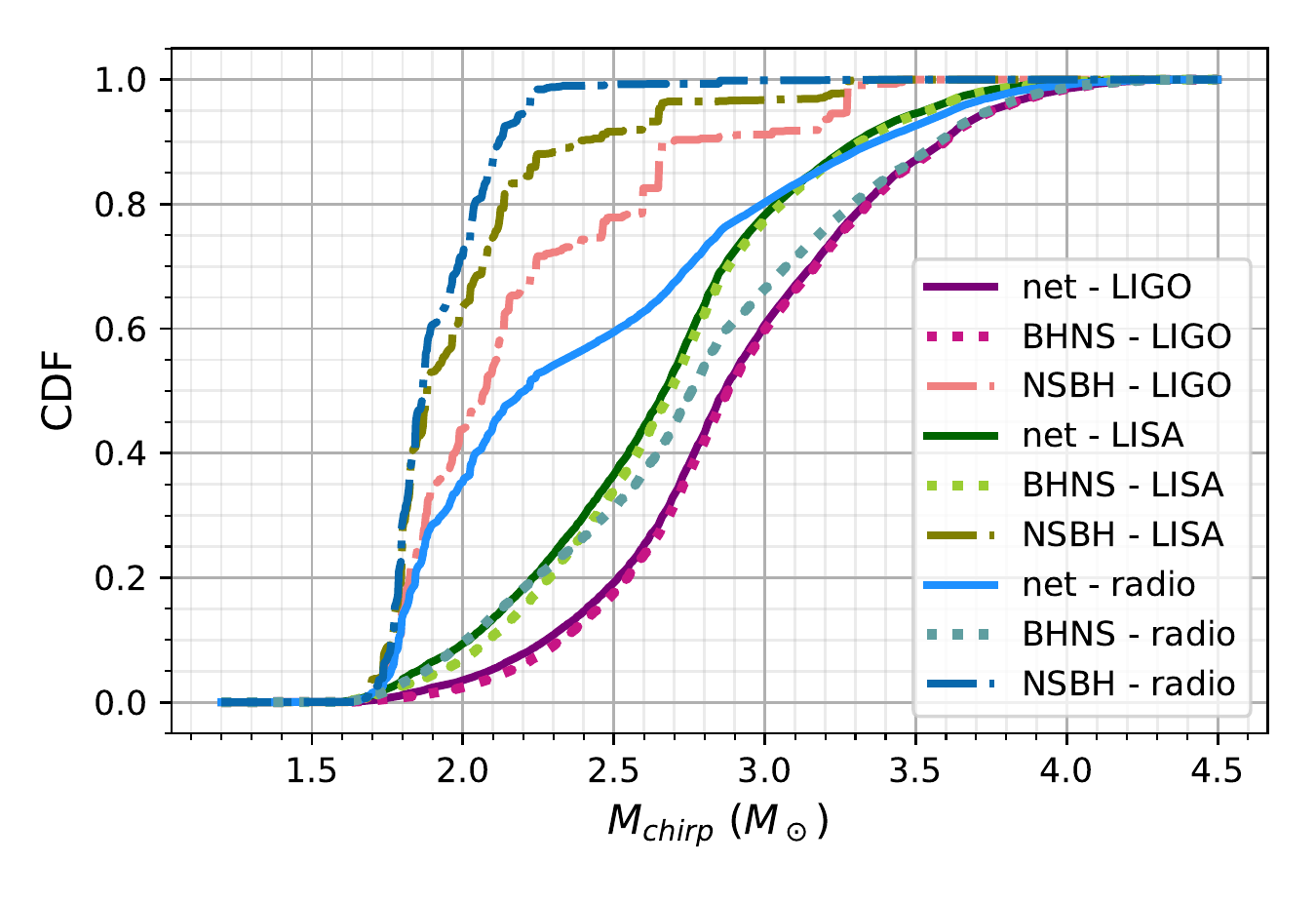}
\caption{Predicted chirp mass distributions of NS+BH binaries detectable in GWs by LIGO and LISA and in the radio in our Fiducial model. 
The solid line shows the net (NSBH + BHNS) population, the dotted lines show the BHNS sub-population and the dot-dashed lines show the NSBH sub-population. 
The net distributions for LIGO (purple) and LISA (green) are dominated by the BHNS populations, while around 20\% of the radio population (blue) population are NSBHs, skewing the chirp mass distribution to lower mass.}
\label{fig:chirp_mass_LIGO_LISA_SKA}
\end{figure}

Fig.~\ref{fig:chirp_mass_LIGO_LISA_SKA} shows the CDF of chirp mass distribution of LIGO, LISA and SKA population. 
Each net population is segregated into the constituting BHNS and NSBH sub populations. 
BHNS make up $\sim$ 97 $\%$ of net-LIGO and $\sim$ 98 $\%$ of the LISA population. 
This dominance of BHNSs biases the net - LIGO and net - LISA curves towards BHNS - LIGO and BHNS - LISA curves respectively.
Since the LIGO population is selectively biased towards heavier chirp masses, the CDFs for LIGO have a smaller slope than LISA and radio. 
The radio population is also dominated by BHNSs ($\sim$ 77 $\%$). However, the relative abundance of NSBHs in radio compared to LIGO or LISA populations effects the net chirp mass distribution of the former, specifically at the low mass end of the spectrum. 

\subsection{Effective spin}
\label{subsec:effective_spin}

\begin{figure}
\includegraphics[width=0.7\columnwidth]{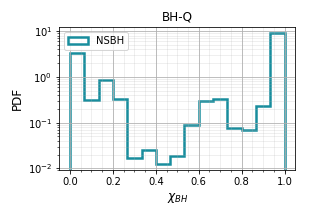}
\caption{The spin distribution of BHs for model BH-Q using Eqn.~\ref{eq:aBHfitQin}. Since BHs in BHNS binaries are assumed to have a spin of 0, only NSBHs are shown.}
\label{fig:chi_BH}
\end{figure}

\begin{figure}
\includegraphics[width=0.7\columnwidth]{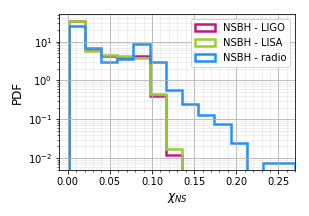}
\caption{The spin distribution of NSs of NSBH systems for populations LIGO(magenta), LISA(green) and radio (blue) calculated through detailed pulsar evolution as discussed in section~\ref{subsec:pulsarPhysics}.}
\label{fig:chi_NS}
\end{figure}

We characterise the spin of NS+BH binaries through the effective inspiral spin parameter $\chi_\mathrm{eff}$ which can be expressed as \citep{Cutler:1992tc,Ajith:2009bn,Ng:2018neg}
\begin{equation}
    \chi_\mathrm{eff} = \frac{m_1 \chi_1 + m_2 \chi_2}{m_1 + m_2}
    \label{eq:chi_eff}
\end{equation}
where $m_{1,2}$ are the component binary masses and $\chi_{1,2} = a_{1,2} \cos\theta_{1,2}$ are the dimensionless projections of the component spins along the direction of orbital angular momentum, where the spin magnitudes $a_{1,2} = 2 \pi c I / G P m_{1,2}^2 $, $I$ is the moment of inertia and $P$ is the rotational spin period.
For simplicity, we assume that the compact object spins are aligned with the orbital angular momentum, so that $\chi_{1,2} \in [0,1]$. There remains an observational bias for positive $\chi_\mathrm{eff}$ systems which produces comparatively long-lived, high frequency signals than its negative counterparts \citep{Zhu:2017znf, Ng:2018lkc}. The selection effect has been shown to be $< 10$\% by \citet{Ng:2018lkc} and is ignored in our analysis. 

Hence, the effective spin includes contributions from both the NS and the BH. To calculate the spin of the BH, we consider two separate models: 
\begin{itemize}
    \item \textbf{BH spin model 1: BH-Z} In this model, we assume that all BHs are effectively non-spinning ($a_{1,2} = 0$). This is motivated by models in which angular momentum is efficiently transferred from the stellar core to the envelope \citep{Fuller:2019sxi}, which is subsequently removed by stellar winds and mass transfer, resulting in a core with little angular momentum, forming a slowly rotating BH.
    \item \textbf{BH spin model 2: BH-Q} In this model, we assume that for BHNS binaries the BH spin is 0, as in model BH-Z. For NSBH binaries, where the immediate progenitor is a pulsar-helium star binary, the spin of the second born compact object (the BH in this case) is determined by a combination of wind mass loss (spinning the helium star down) and tides (spinning the helium star up) \citep{Zaldarriaga:2017qkw,Qin:2018nuz}.  
\end{itemize}

For model BH-Q, we fit the black hole spin $\chi_{\mathrm{BH}}$ as a function of the orbital period $P_\mathrm{orb}$ (in days) prior to the second supernova using the top middle panel of Figure 6 of \citet{Qin:2018nuz} as
\begin{equation}
    \chi_{\mathrm{BH}} = \left\{\begin{array}{ll}
                    0, & \text{for } \log_{10}{P_\mathrm{orb}} >  0.3 \\
                    1, & \text{for } \log_{10}{P_\mathrm{orb}} < -0.3 \\
                    m \log_{10} P_\mathrm{orb} + c, & \text{for } -0.3 < \log_{10}{P_\mathrm{orb}} < 0.3
                    \end{array}\right\}
    \label{eq:aBHfitQin}
\end{equation}
where $m = -5/3$ and $c = 0.5$. \citet{Bavera:2019} recently applied a similar method for binary BHs. 
Model BH-Q gives predictions for pulsar timing observations of PSR+BH binaries, which will be able to determine the properties of the companion BH, possibly measuring its spin to 1\% \citep{Wex:1998wt,Liu:2014uka}. 
The obtained BH spin distribution is shown in Fig.~\ref{fig:chi_BH}.

\begin{figure}
\includegraphics[width=8cm, height=5.5cm]{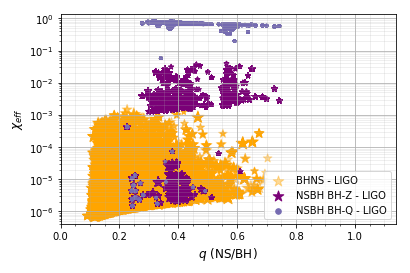}
\includegraphics[width=8cm, height=5.5cm]{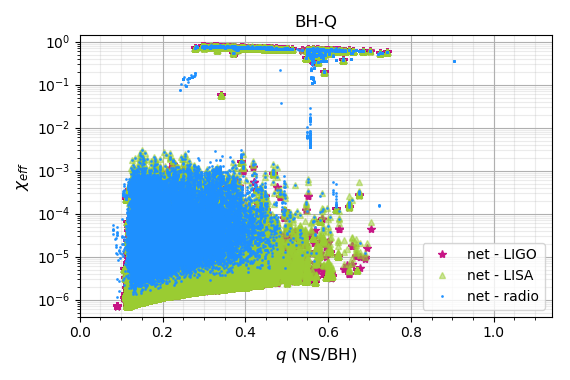}
\caption{Effective spin $\chi_\mathrm{eff}$ vs. mass ratio $q$ plot of NS+BH binaries. The top panel shows the BHNS (in orange star) and NSBH (spin model BH-Z in dark purple star and spin model BH-Q in violet circular symbols) system of the LIGO population. The marker size is proportional to weight $V_\mathrm{eff}$. The bottom panel shows the same distribution of the net systems of LIGO (pink star), LISA (green triangle) and radio (blue dot) population for the spin model BH-Q.}
\label{fig:chieffq}
\end{figure}

We are computing the NS spin through pulsar evolution in detail (c.f. Eqn.~\ref{SpinDownIsolated}) and obtaining the dimensionless spin parameter $\chi_{\mathrm{NS}}$. The $\chi_{\mathrm{NS}}$ distributions for NSBHs of the LIGO, LISA and SKA populations are presented in Fig.~\ref{fig:chi_NS}. The relative abundance of recycled pulsars in the radio population causes the NS spin distribution to be more weighted towards higher spins ($\chi_{\mathrm{NS}}>0.15$) than LIGO and LISA populations. This is also due to the fact that while the radio population observes the binaries at the current time, LIGO population detects the same at the merger time and LISA population observes it at P$_\mathrm{orb}\leq 1$day. Hence the LIGO and LISA populations have spun-down, slower NSs when compared to the radio population with typically smaller $P$.
Comparing the horizontal axes of Fig.~\ref{fig:chi_BH} and Fig.~\ref{fig:chi_NS}, we note that even for radio NSBHs the maximum value of $\chi_{\mathrm{NS}}$ is approximately 0.3 times $\chi_{\mathrm{BH}}$. Compared to a BH with same spin, even a fast spinning recycled pulsar will contribute less to $\chi_{\mathrm{eff}}$ of the system, NSs ordinarily being the less massive compact object of the binary. We show the distribution of $\chi_\mathrm{eff}$ and mass ratio $q$ in Figure~\ref{fig:chieffq}. 
The top panel shows the distribution of $\chi_\mathrm{eff}$ observable by LIGO, using both models (BH-Z and BH-Q) for the spin of the BH in NSBH binaries. 

\begin{figure}
\includegraphics[width=0.5\columnwidth]{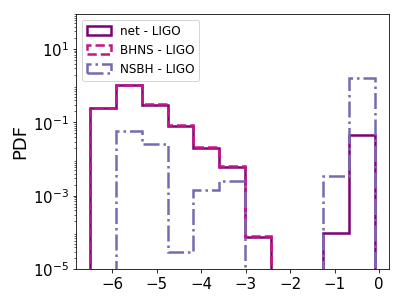}
\includegraphics[width=0.5\columnwidth]{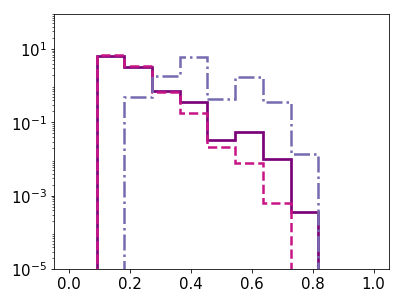}
\includegraphics[width=0.5\columnwidth]{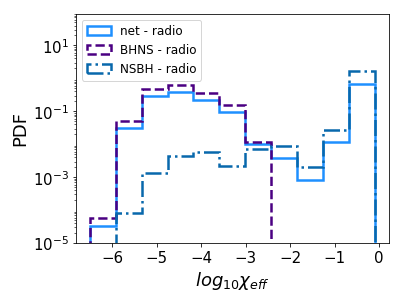}
\includegraphics[width=0.5\columnwidth]{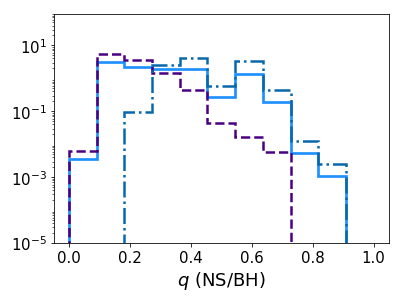}
\caption{Histograms of the effective spin parameter $\chi_\mathrm{eff}$ and mass ratio $q$ for the LIGO (top, plotted in shades of magenta) and radio (bottom, plotted in shades of blue) NS+BH populations.}
\label{fig:chi_q_pdf}
\end{figure}

\begin{figure}
\includegraphics[width=7cm, height=5cm]{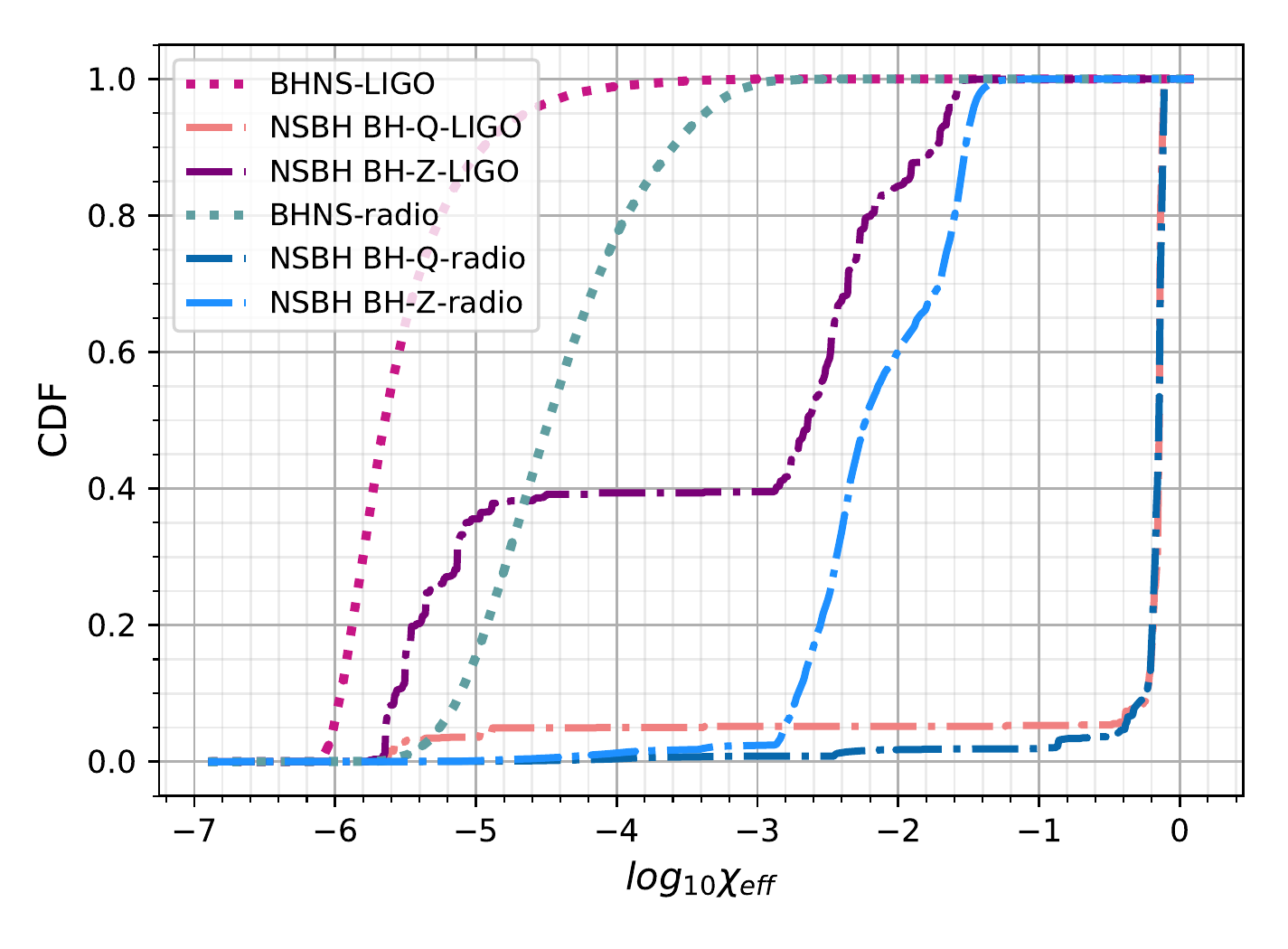}
\includegraphics[width=7cm, height=5cm]{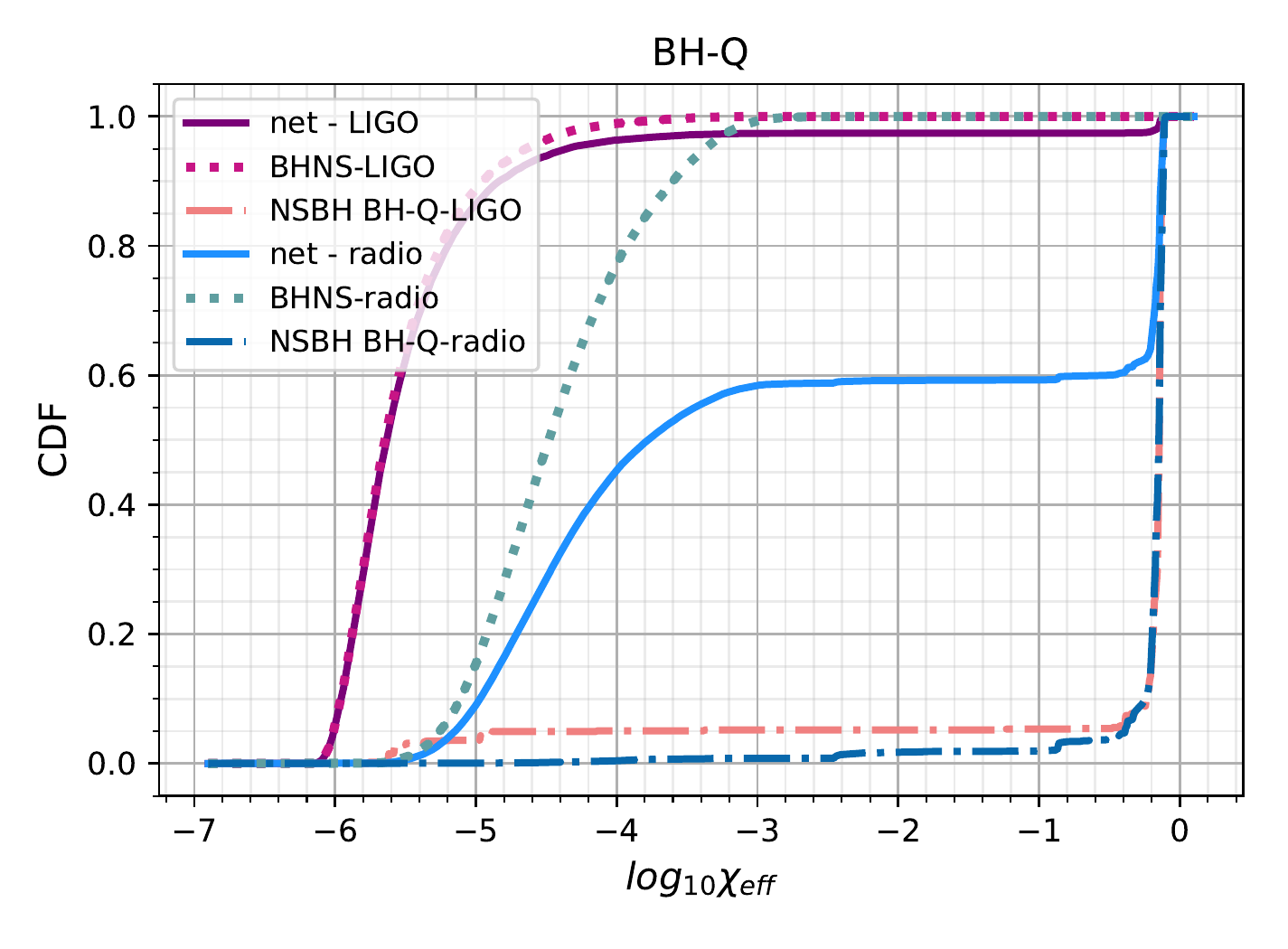}
\caption{CDF of log$_{10}\chi_\mathrm{eff}$ distribution for LIGO(purple) and radio(blue) populations are shown. Both BHNS and NSBH sub-populations for radio have higher effective spin than LIGO. 90\% LIGO BHNSs have log$_{10}\chi_{\mathrm{eff}}\leq -5$ and 80\% radio BHNSs have log$_{10}\chi_{\mathrm{eff}}\leq -4$. The NSBH BH-Q population shows log$_{10}\chi_{\mathrm{eff}}\leq -0.025$ for only 10\% of the population. For BH-Z model, 85\% LIGO NSBHs and 60\% radio NSBHs have log$_{10}\chi_{\mathrm{eff}}\leq -2$. }
\label{fig:chi_cdf}
\end{figure}

Our model results in a bi-modal $\chi_\mathrm{eff}$ distribution, with the BHNS population having low $\chi_\mathrm{eff} \lesssim 10^{-3}$, whilst NSBHs show two sub-populations with $\chi_\mathrm{eff} \gtrsim 10^{-3}$ and $\chi_\mathrm{eff} \lesssim 10^{-4}$. 
This is because the NS is second formed in BHNS binaries, and hence is always a non-recycled pulsar, making the average NS spin slower; resulting in small $\chi_{\rm{eff}}$. 
For NSBHs, the sub-group with higher ($10^{-3} < \chi_\mathrm{eff} < 10^{-1}$) effective spins contain recycled pulsars, while the lower sub-group contains non-recycled pulsars. 
The BH spin for BHNSs is assumed to be 0 for both BH-Z and BH-Q prescriptions \citep[c.f.][]{Fuller:2019sxi}. 
Hence there is no difference for the BHNS subpopulation between the two BH spin models. 
For NSBH, the NS may get recycled, making the contribution of $\chi_\mathrm{NS}$ higher due to spin up through mass accretion.
For model BH-Q the segregation occurs as described by Eqn.~\ref{eq:aBHfitQin}. 

The lower panel of Fig.~\ref{fig:chieffq} shows the  net $q$ vs $\chi_\mathrm{eff}$ distribution for LIGO, LISA and the radio for the BH-Q model. 
The $\chi_\mathrm{NS}$ for the LIGO population is computed using the NS spin $P$ at the time of merger, while for the LISA and the radio populations we use the $P$ at the current time. 
The apparent smearing of the LISA and radio data-points is because of our method of statistical recycling allowing us to re-use the binaries 100 times (see section~\ref{subsec:models}) and hence biased in selecting the same binary as radio-alive, at different points of its lifetime. 
While this changes the spin value, and hence $\chi_\mathrm{eff}$, the binary mass remains the same causing the degeneracy in $q$.
However, at merger, all such `pseudo-unique' binaries converge to a particular merger $P$, giving the LIGO points.  
We note that the BH will always be tidally spun up in NSBHs observed by LISA (c.f. Equation~\ref{eq:aBHfitQin}).
The lower limit of $\chi_\mathrm{eff}$ for the radio population appears to be greater than LISA or LIGO (Fig~\ref{fig:chieffq}, lower panel). 
This is because radio population selectively harbours faster spinning pulsars.

The PDFs of $\chi_\mathrm{eff}$ and $q$ for LIGO and SKA are shown in Fig~\ref{fig:chi_q_pdf}. 
The bi-modality of $\chi_\mathrm{eff}$ distribution due to BHNS and NSBH is more apparent in LIGO than SKA. 
This is because SKA population is dominated by NSBHs and the BHNSs that are spinning fast enough to be radio-alive. 
The preferential low $q$ for BHNSs is visible in both LIGO and SKA.

Fig.~\ref{fig:chi_cdf}
shows the $\chi_\mathrm{eff}$ CDFs for radio and LIGO individual sub-populations (for both BH-Q and BH-Z spin models) at the top panel and compares the net LIGO and radio distributions for spin model BH-Q at the bottom panel.


\subsubsection{Post-merger remnant spin}
\label{eq:post_merger_remnant}

The remnant of a NS+BH merger is expected to be a black hole \citep{2013PhRvD..88j4025P,Zappa:2019ntl}. 
The spin of the NS+BH post-merger remnant $\chi_\mathrm{rem}$ is expected to depend on the progenitor mass ratio, spins and NS compactness (see e.g. Eqn.1 of \citealp{Zappa:2019ntl}, which assumes pre-merger NS spin to be 0, for details).
We use polynomial fitting from \citet{Zappa:2019ntl} for the cases with pre-merger BH spin $\chi_\mathrm{BH} \geq 0.0$. 
We use approximate our assumptions by removing the dependency from $\Lambda$ (NS quadrupole tidal polarizability) that measures tidal deformability of the NS.
This can be justified by the fact that the parameter $\Lambda$ is a function of the NS compactness given by the mass-radius ratio of the NS and since all our models have constant NS radius, a degree of degeneracy is introduced.
Secondly, from the the bottom panel of Fig.1 of \citet{Zappa:2019ntl}, it can be seen that the dependency of the final BH spin on $\Lambda$ is less prominent for lower progenitor BH spins.
Since we only focus on giving a general estimate of the post-merger remnant spin, we hold the said assumptions valid for the scope of this paper. 
We show the results for the LIGO population, since only for LIGO $\chi_{\mathrm{rem}}$ remains an infer-able quantity. 

Defining $q_s = M_{\mathrm{BH}}/M_{\mathrm{NS}} \geq 1$ we have the symmetric mass ratio $\gamma = q_s/(1+q_s)^2$. Then, the remnant BH spin $\chi_{\mathrm{rem}}$ is calculated by the fourth order polynomial, 
\begin{equation}
  \chi_{\mathrm{rem}}= a_4\times\gamma^4 + a_3\times\gamma^3 + a_2\times\gamma^2 + a_1\times\gamma^1 + a_0 ,
  \label{equ:chi_remnant}
\end{equation}
where
\begin{equation}
\begin{array}{l}
  a_4 = -2310.4\times\chi_{\mathrm{BH}}^2 + 2088.4\times\chi_{\mathrm{BH}} - 400 , \\
  \\
  a_3 = 1582.08\times\chi_{\mathrm{BH}}^2 - 1417.68\times\chi_{\mathrm{BH}} + 253.33 , \\
  \\
  a_2 = -367.46\times\chi_{\mathrm{BH}}^2 + 325.37\times\chi_{\mathrm{BH}} - 54.99 , \\
  \\
  a_1 = 34.43\times\chi_{\mathrm{BH}}^2 - 32.83\times\chi_{\mathrm{BH}} + 7.56 , \\
  \\
  a_0 = -1.17\times\chi_{\mathrm{BH}}^2 + 1.95\times\chi_{\mathrm{BH}} - 0.1 ,
  \label{equ:chi_remnant_coefficients}
  \end{array}
\end{equation}
given a pre-merger progenitor BH spin $\chi_{\mathrm{BH}}$. 
For all BHNS, we have $\chi_{\mathrm{BH}}=0$, while for NSBHs in model BH-Q, we calculate  $\chi_{\mathrm{BH}}$ from Eqn.\ref{eq:aBHfitQin} and for model BH-Z NSBH BHs have $\chi_{\mathrm{BH}}=0$. 

\begin{figure}
\includegraphics[width=0.8\columnwidth]{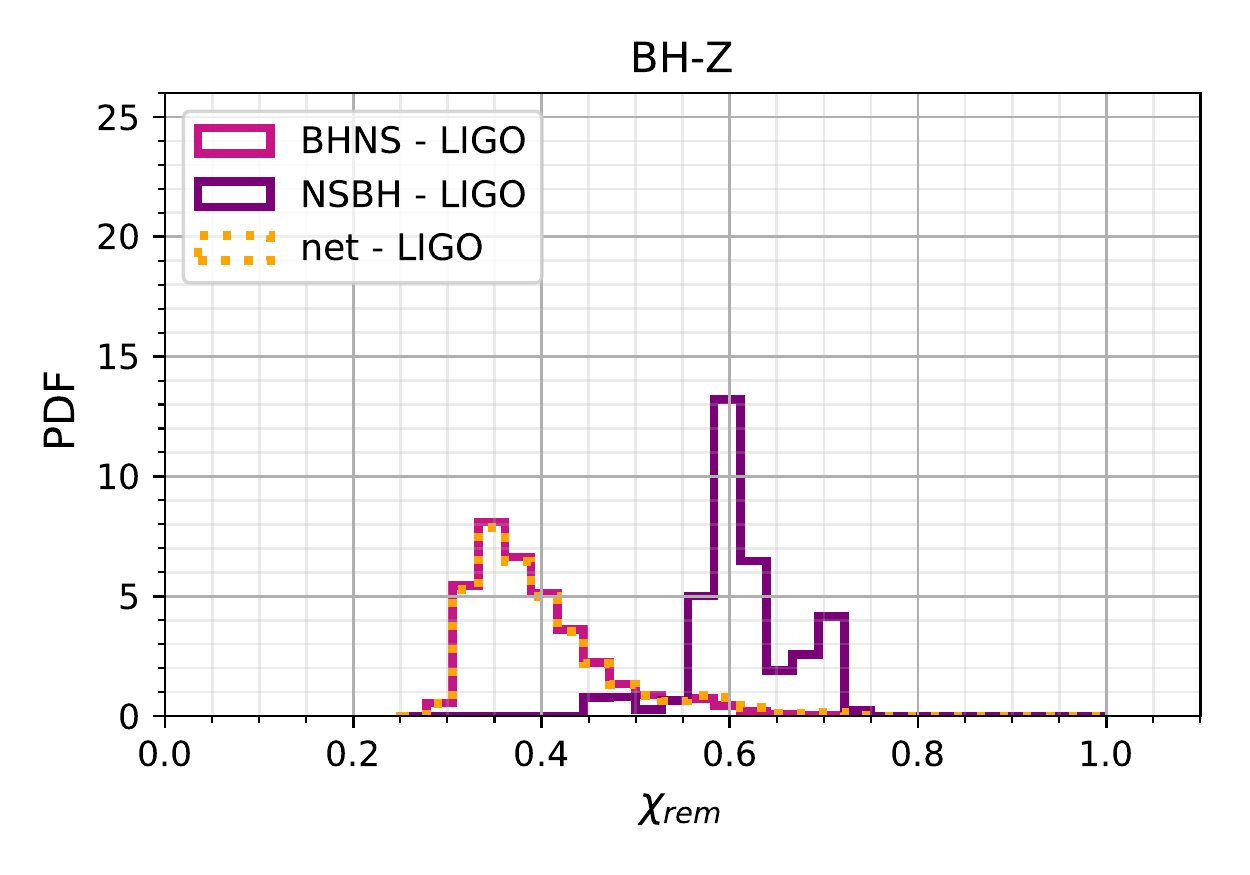}
\includegraphics[width=0.8\columnwidth]{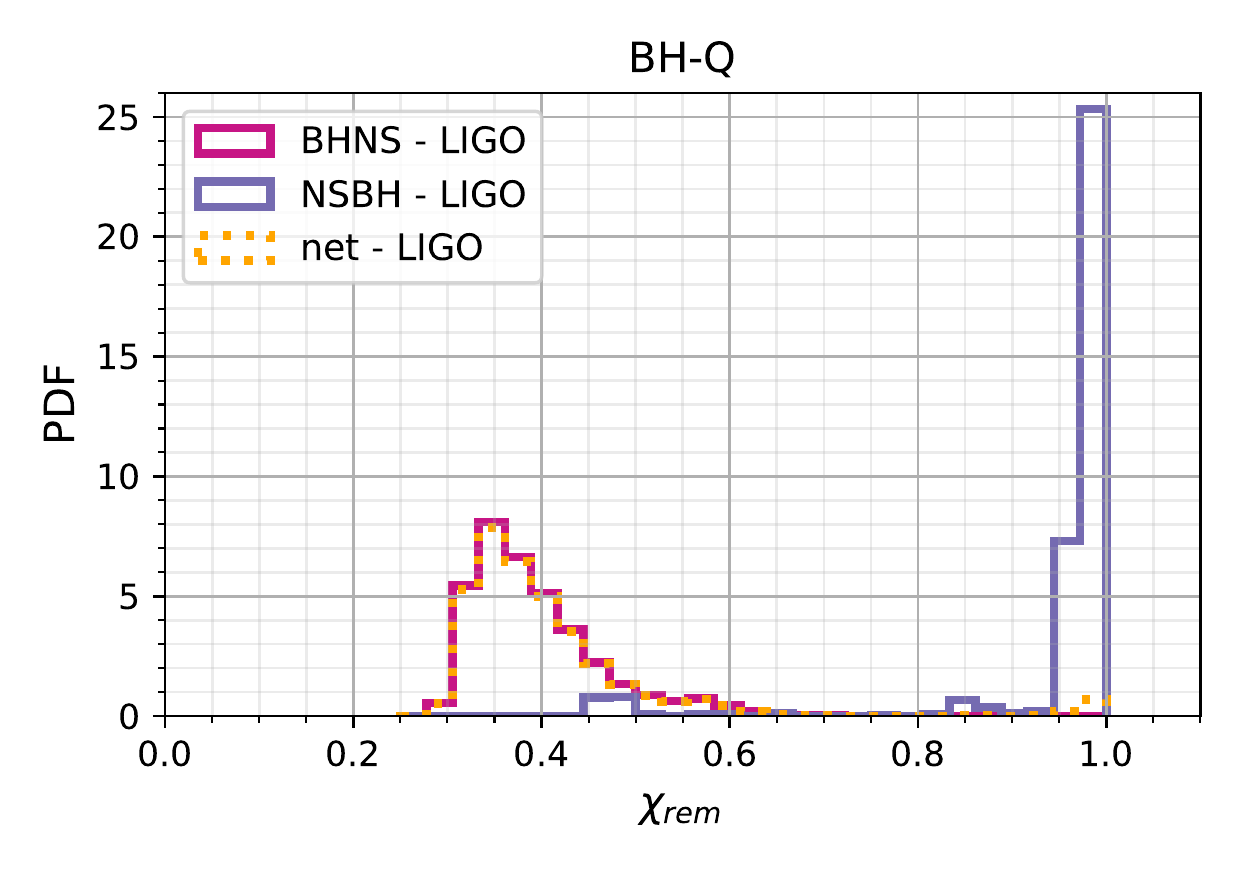}
\includegraphics[width=0.8\columnwidth]{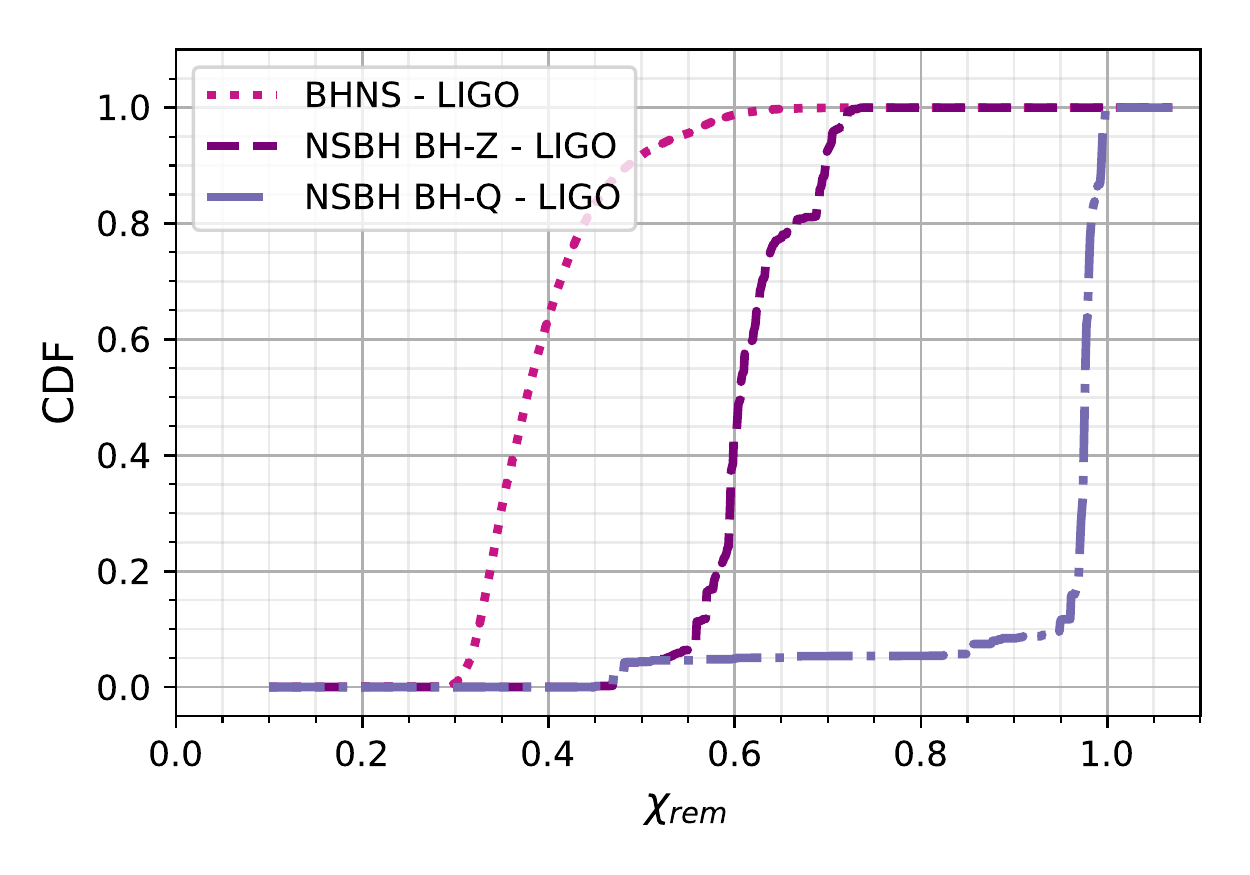}
\caption{Predicted remnant spin distributions of the LIGO population for the two progenitor BH spin models BH-Z (top) and BH-Q (bottom), under the assumption of pre-merger NS spin to be 0. Since it is the LIGO sub-population, the histograms are weighted by effective volume $V_{\mathrm{eff}}$ and are normalized so that the area under each histogram is 1. The magenta solid line, the violet solid line and the orange broken line show the BHNS, NSBH and the net populations respectively. Even when the progenitor BH is non-spinning, the remnant BH shows non-zero spin. Moreover, for both BH-Z and BH-Q models BHNS and NSBH populations are distinguishable from each other.}
\label{fig:chi_rem}
\end{figure}

Fig.~\ref{fig:chi_rem} shows the remnant spin distribution for the two progenitor spin models BH-Z and BH-Q. 
The BHNS remnant spin distribution remains constant across two models. 
Though for BHNSs, the progenitor BH spin is zero, the post-merger remnant has non-zero positive spin, depending on the symmetric mass ratio $\gamma$ of the binary, peaking around $\chi_{\mathrm{rem}}=0.35$. 
The NSBHs remnant spin distribution is dependant on the pre-merger BH spin assumptions, though the median $\chi_{\mathrm{rem}}$ is higher than BHNSs for both models (BH-Z around 0.6 and BH-Q around 1.0). Though NSBHs are comparatively rarer than BHNSs in Milky Way-like galaxies, this method of determining the post-merger remnant spin allows the NS+BH LIGO observations to be categorized to BHNSs or NSBHs .



\section{DISCUSSION}
\label{sec:discussion}

There are many uncertainties associated with each stage of our model, from the initial properties of massive binary stars, through to massive binary evolution (mainly regarding supernovae, CE evolution and stellar winds), to pulsar evolution, radio selection effects and the re-scaling of our simulation to a Milky Way equivalent population. 
In this section we discuss some of the most uncertain stages of our modelling and attempt to quantify the impact on our results.

The absolute numbers of PSR+BH binaries that our model predicts depends on the re-scaling of our simulation to a Milky Way equivalent population. Our assumptions for this re-scaling are described in Section~\ref{subsubsec:rescaling} \citep[see also][]{Chattopadhyay:DNS2019}. A number of the assumptions in this calculation, such as the number of stars in the Milky Way, the binary fraction and the initial mass function are uncertain. Based on this, we simply estimate a factor of two for the uncertainty in the number of PSR+BH binaries due to the uncertain star formation rate. The relative rates between DNS and NS+BH binaries that we quote in columns $\mathcal{R}$ and $\mathcal{R}_\mathrm{SKA}$ in Table~\ref{tab:netNumbers} should be robust to these uncertainties. 

The impact of uncertainties in massive binary evolution on the formation of NS+BH binaries has been studied by many authors. Uncertainties in the initial binary conditions such as the initial mass function and orbital period distribution lead to a factor of two  uncertainty in the yield of NS+BHs \citep[e.g.][]{deMink:2015yea,Klencki:2018zrz}, comparable to the variation due to uncertainties in binary evolution processes and massive star evolution \citep[e.g.][]{Dominik:2012,Vigna-Gomez:2018dza,Agrawal:2020}.

Uncertainties in massive binary evolution such as CE evolution and black hole kicks translate to large uncertainties in the properties of NS+BH binaries \citep[e.g.][]{Dominik:2012,2018ApJ...866..151A,Broekgaarden:2020NSBH} that affect the final distributions both quantitatively and qualitatively. 
For CE, we have explored the `pessimistic' approach for model CE-P (see Sec~\ref{subsec:models}). 
However, a more `realistic' estimate is expected to be somewhere in-between the `optimistic' (as for Fiducial) and pessimistic. 

It has been shown by \citet{Tauris:2017omb} that case BB mass transfer can allow enough accretion onto the first born NS and recycle the pulsar to match the Galactic observations. 
Unlike in COMPAS, \citet{Tauris:2017omb} assumed super Eddington mass transfer with the free parameter $X_\mathrm{Edd}$ determining the accretion efficiency and $X_\mathrm{Edd}\approx 2.0-3.0$ can explain observations. 
In COMPAS we assume case BB mass transfer to be Eddington limited \citep{Chattopadhyay:DNS2019, Vigna-Gomez:2018dza}, and hence require CE mass accretion by the NS \citep{MacLeod:2014yda} to explain the observation of recycled pulsars \citep{Chattopadhyay:DNS2019}. 
The dominant formation channels for both double NSs and NSBHs involves both case BB mass transfer and CE phases \citep{Vigna-Gomez:2018dza, Chattopadhyay:DNS2019, Broekgaarden:2020NSBH}, and the uncertainty remains which (or both) of the two mass transfer phases from the companion causes the pulsar recycling. 
We note that the in-feasibility of model CE-Z (no accretion onto NS during CE) is dependant on the uncertainty of the accretion channels.

The supernova kick distribution is a key uncertainty that determines the PSR+BH retention fraction and distribution in the host galaxy. 
Along with the Fiducial model where BH natal kicks are scaled by the fallback mass, we have explored two extreme cases for BH kicks. 
In model BHK-F, BHs receive the same distribution of kicks as for NSs, while in model BHK-Z BHs receive no kick at birth. In the Fiducial model, supernovae occurring in NSBHs on average impart a higher kick velocity than for supernovae occurring in BHNSs. For radio NSBHs, the supernova events that create both remnants are most likely to be CCSNe: the mean kick velocity for the first SN (forming the NS) is approximately 180\,km s$^{-1}$ while for the second SN (forming the BH, with fallback scaling factor of 0.37) it is about 200\,km s$^{-1}$. In comparison, for radio BHNSs, the first SN is predominantly a CCSN (forming the BH, with fallback scaling factor of  0.72) and the second SN is mainly an USSN (forming the NS): mean kick velocities of about 85 and 50\,km s$^{-1}$, respectively. The BH natal kick, which is scaled by the fallback mass for the Fiducial model, is lower for more massive binaries and hence for the BHNS population. This unequal SNe kick distribution allows loosely bound BHNSs to survive more often, while only NSBH binaries with a very high orbital binding energy survive. The models that directly change the BH kick distributions (i.e. BHK-F and BHK-Z) thus affect the relative observation rates of NSBHs and BHNSs. 
We find that detections of NSBH binaries by future radio surveys will aid in constraining BH kicks with measurements of their scale heights and orbital eccentricities.

The formation rate of NS+BHs is also affected by the metallicity \citep[e.g.][]{Neijssel:2019,Broekgaarden:2020NSBH}. 
This is explored in models ZM-001 and ZM-02 as shown in Table~\ref{tab:netNumbers}. 
The wind mass-loss is decreased at lower metallicity, causing a larger fraction of the collapsing star to fallback onto the proto-compact object, suppressing the natal kicks and disrupting fewer binaries \citep[c.f.][]{Fryer:2012}. Reducing the kick velocity in this manner affects NSBHs more, compared to BHNSs where the kick velocity is already low. The fallback scaling factor for BHs of radio NSBHs in model ZM-001 is 0.88, allowing more NSBHs to survive the SN than in the Fiducial model. At low metallicity (model ZM-001), the formation efficiency of  NSBHs increases by approximately 10 times compared to the Fiducial model, while around twice as many BHNSs are formed. 
At high metallicity (model ZM-02), the formation efficiency of both NSBHs and BHNSs decreases by about a factor of 2 compared to the Fiducial model.  
The Galactic metallicity evolution \citep{Panter:2002ed, Mackereth:2018} hence plays an important role in determining the star formation rate and evolution \citep{2019A&A...624A..19B}, thus affecting the NS+BH numbers.  
More thorough analysis integrating over the cosmic history of the Universe finds that many NS+BH binaries observed by LIGO might have formed at solar-like metallicities \citep[e.g.][]{Neijssel:2019,Broekgaarden:2020NSBH}, suggesting that using solar metallicity to examine the properties of NS+BH mergers observable with LIGO is a reasonable approximation. 

Pulsar luminosity and its possible correlations to the pulsar and binary parameters introduces an uncertainty that affects the post-radio selection effect results. 
We assumed the luminosity distribution from \citet{Szary:2014dia} which is independent of the pulsar parameters, and determine radio-death of pulsars from a hybrid approach of computing the radio efficiency and death-line cut-off \citep[see sections 2.3 and 2.7.2 of][]{Chattopadhyay:DNS2019}. 
With additional insight on the luminosity function of radio pulsars in the future, the modelling of the selection effects will become more accurate. 

There are two kinds of uncertainties associated with our models: systematics associated with the modelling of pulsar and binary evolution processes (as shown in Tables~\ref{tab:netNumbers},~\ref{tab:radioObservablesMean:Radio} and \ref{tab:radioObservablesMean:SKA}) and statistical uncertainties.
The latter are introduced due to our use of Monte Carlo sampling from the birth distributions of massive binaries and pulsars, for determining the pulsar luminosity, for selecting compact object kicks and for distributing binaries within the Milky Way. 
We estimate the size of statistical errors by performing multiple, otherwise identical simulations of our Fiducial model, using different random number seeds. 
We find that the net number of NS+BHs changes by $<$10\% , while for the radio population the change is limited to $<$ 15\%. 
These cause variations of two (five) in the number of NSBHs (BHNSs) observed by the SKA.
We conclude that uncertainty in our modelling is dominated by systematic uncertainty associated with the modelling of massive binaries.

A radio luminous PSR+BH with an orbital period $<$ 1 day can theoretically be observed both in radio and in gravitational waves with a space based detector. 
The possibility of observing the same PSR+BH binary with both LISA and the SKA is intriguing \citep[c.f.][]{Kyutoku:2018aai,Thrane:2019lwv}. 
We have shown that both the LISA and SKA populations are dominated by BHNSs. Accounting for radio selection effects, about 70\% of the population of PSR+BHs observed by the SKA in our Fiducial model are BHNSs (see Table~\ref{tab:netNumbers} for other models).
This suggests that if LISA and the SKA observe the same PSR+BH binary, it is more likely to be a BHNS than a NSBH, despite the fact that the pulsars in BHNSs are always non-recycled and have shorter radio lifetimes.
For the Fiducial model we find $\approx$ 70\% of radio alive NS+BHs have P$_\mathrm{orb}<$1\,day ($\approx$77\% of radio-BHNS and $\approx$47\% of radio-NSBH, total estimate about 464 such binaries per MW), making them fit for such multi-messenger detection.
For model BHK-F, with the lowest number of NS+BHs, we find 47 ($\approx$ 49\% of PSR+BHs) radio-alive NS+BHs have P$_\mathrm{orb}<$1\,day. For model RM-R we find about 18 radio NSBHs and 680 radio BHNSs have orbital period of less than a day.
Rounding the numbers and accounting for a factor of two uncertainty, we obtain about 25--1400 PSR+BHs present in the Milky-Way at the current time, potentially to be observable by LISA. 
We note that properly accounting for both radio and gravitational-wave selection effects will reduce the number of multi-messenger candidates considerably (see Sec.~\ref{subsec:telescope_obs}).



\section{Conclusions}
\label{sec:conclusions}

We have predicted the population of Galactic NS+BH binaries using the rapid binary population synthesis suite COMPAS \citep{Stevenson:2017tfq,Vigna-Gomez:2018dza,Neijssel:2019,Chattopadhyay:DNS2019} in the context of the next generation radio telescopes including MeerKAT, SKA and FAST, as well as current ground and future space-based gravitational wave detectors LIGO-Virgo and LISA respectively. 
Our key findings are summed up as: 
\begin{enumerate}
    \item Future pulsar surveys with the SKA and MeerKAT are expected to discover 1--80 Galactic field PSR+BH systems from our feasible models (see Table~\ref{tab:netNumbers}). 
    The uncertainty stems from uncertainties in massive binary evolution, pulsar evolution and in the re-scaling of our simulation to a Milky Way equivalent population.
    \item 
    Our models also predict that the SKA will observe $\mathcal{O}(100)$ Galactic DNSs. 
    We find that the ratio $\mathcal{R}_\mathrm{SKA}$ of observable PSR+BH systems per radio DNS system observed by the SKA is independent of our assumptions on the Galactic star formation history, metallicity and initial pulsar parameter distribution uncertainties. 
    We find $\mathcal{R}_\mathrm{SKA}=0.13 - 0.53$ from our feasible suite of models, with CE-P and ZM-001 giving the lower and upper limits respectively. 
    \item The formation order of the NS and BH in the radio-observable PSR+BH systems can be characterised by pulsar and binary parameters. 
    In our Fiducial model, NSBHs are expected to have spins $P \sim$ $\mathcal{O}(10^{-2})$\,s and spin down rate $\dot{P}\sim$ $\mathcal{O}(10^{-17})$ s/s, while pulsars in BHNS systems have $P \sim$ 1\,s and $\dot{P} \sim \mathcal{O}(10^{-15})$\,s/s. 
    The surface magnetic field $B$ of NSBHs is expected to be two orders of magnitude lower than BHNS pulsars with $B \sim \mathcal{O}(10^{12})$\,G. NSBH binaries have a higher orbital eccentricity $e \ge 0.5$, compared to BHNSs with $e \leq 0.4$.
    \item The scale height of radio NSBH binaries is shown to have a higher mean value ($|Z| \sim 0.38$\,kpc) compared to the BHNS mean ($|Z| \sim 0.16$\,kpc: see the Fiducial model in Table~\ref{tab:radioObservablesMean:Radio}). 
    This is because the NSs of the radio BHNS population are created by ultra-stripped supernovae for $\sim$ 93\% of the cases, making the second supernova kick for the systems an order of magnitude lower compared to core-collapse supernova natal kicks. 
    The BHs of radio NSBHs are formed through core-collapse supernova and are scaled down by the BH fall-back mass for the Fiducial model. 
    Therefore, NSBH observations can give insights into the formation channel of the binary as well as constrain the magnitude of BH natal kicks.
    \item We estimate that of all Galactic field radio NSBH systems, around 30\% contain millisecond pulsars ($P<30$\,ms), in contrast to only $\lesssim$10\% of all radio DNS systems formed through isolated evolution. 
    We expect 0--80 Galactic MSPs in binaries with BHs while radio-selection effects reduce the number observed with the SKA to 0--4 systems (see Sec.~\ref{MSPs}). 
    We note that this estimate holds strictly for pulsars formed in an isolated environment. 
    \item We find that approximately 90\% of NSBH binaries have M$_\mathrm{chirp} \le 2.5$\,M$_\odot$, where our models assume a maximum NS mass of 2.5M$_\odot$. For BHNS binaries  M$_\mathrm{chirp} \le 3$\,M$_\odot$ for 80\% of the systems (see Fig.~\ref{fig:chirp_mass_radio_total_NSBH_BHNS}).
    We show that 40\% of LIGO NS+BH binaries from isolated field evolution will have M$_\mathrm{chirp}\ge 3.0$ M$_\odot$, compared to 20\% of those detected with LISA, and 60\% of the SKA observable Galactic PSR+BHs will have M$_\mathrm{chirp}\le$ 2.5 M$_\odot$ (see Fig.~\ref{fig:chirp_mass_LIGO_LISA_SKA}).
    \item The mean mass-ratio for LIGO BHNSs is shown to be 0.19 compared to 0.20 for SKA BHNSs. 
    The LIGO NSBH population has a mean of 0.43 compared to 0.45 for SKA. Hence it can be said that in general, the mean mass-ratio for BHNSs and NSBHs remains around 0.20 and 0.44 respectively.
    \item 
    We derive the distribution of BH spins in merging NSBH binaries according to two models. 
    Model BH-Q \citep[based on the results of][]{Qin:2018nuz} assumes that BH progenitors in short orbital period binaries are tidally spun up, resulting in rapidly rotating BHs, while wide binaries leave behind slowly spinning BHs. 
    In model BH-Z all BHs are assumed to be non-spinning \citep[c.f.][]{Fuller:2019sxi}.
    The BHs in BHNS binaries are assumed to be non-spinning in all models.
    The mean effective spin $\chi_\mathrm{eff}$ --- that has contributions from both the NS (which may be recycled for NSBHs) and BH spins --- is shown to be higher in NSBHs for both models BH-Z and BH-Q (LIGO population: $5.90\times 10^{-3}$ and $6.51\times 10^{-1}$ respectively)  
    than BHNSs (LIGO population: $8.02\times 10^{-6}$).
    We find that 90\% of the LIGO BHNSs have log$_{10}\chi_{\mathrm{eff}}\leq -5$ and 80\% of the SKA BHNSs have log$_{10}\chi_{\mathrm{eff}}\leq -4$. 
    The NSBH BH-Q population shows log$_{10}\chi_{\mathrm{eff}}\leq -0.025$ for only 10\% of the population. 
    For the BH-Z model, 85\% of the LIGO NSBHs and 60\% of the SKA NSBHs have log$_{10}\chi_{\mathrm{eff}}\leq -2$ (see Fig.~\ref{fig:chi_cdf}). 
    \item The post-merger remnant BH spins $\chi_\mathrm{rem}$ of the LIGO observable systems are calculated using a polynomial fit to the results of \citet{Zappa:2019ntl}, neglecting any dependence of the NSs tidal deformability on its mass and spin.
    For both models BH-Z and BH-Q, the distribution of $\chi_\mathrm{rem}$ for the NSBH population is distinguishable from the BHNS population.
    For BHNSs $0.3 \lesssim \chi_\mathrm{rem}\lesssim 0.7$, peaking at about 0.35. 
    For NSBHs the distribution peaks at 0.6 for model BH-Z. 
    Around 60\% of the BHNS merger remnants are expected to have $\chi_\mathrm{rem} \leq 0.4$, while for model BH-Z 80\% of the NSBHs have $\chi_\mathrm{rem} \leq 0.7$ and for model BH-Q 90\% of the NSBH mergers have $\chi_\mathrm{rem} \leq 0.9$ (see Fig.~\ref{fig:chi_rem}).
\end{enumerate}


\section*{Acknowledgements}

We thank Rahul Sengar, Manjari Bagchi, Christopher Berry, Ryan Shannon, Renee Spiewak, Danny Price, Adam Deller, Ilya Mandel and the COMPAS collaboration for constructive discussions.
We thank the referee for insightful suggestions which helped improve this paper.
The authors are supported by the Australian Research Council Centre of Excellence for Gravitational Wave Discovery (OzGrav), through project number CE170100004. 
This work made  use  of  the  OzSTAR  high  performance  computer at  Swinburne  University  of  Technology. 
OzSTAR  is funded by Swinburne University of Technology and the National Collaborative Research Infrastructure Strategy (NCRIS).


\section*{Data availability}

This paper made use of an early version of the rapid binary population synthesis code COMPAS; the latest version is publicly available at \url{www.github.com/TeamCOMPAS/COMPAS}. 
Our data is being made publicly available at  \url{https://zenodo.org/communities/compas/}. The simulations can also be requested to the main author.



\bibliographystyle{mnras}
\bibliography{bib} 








\bsp	
\label{lastpage}
\end{document}

%% file: telescope_specs.tex
\begin{table}
    \caption{Telescope specifications assumed in this work, based on (a) Parkes Multibeam Survey \citep{Manchester:2001}, (b) TRansients and PUlsars with MeerKAT (TRAPUM), MeerKAT Large Survey Projects \citep{Stappers:2018Le}, (c) Square Kilometer Array Pulsar Search \citep{Grainge:2017,Levin:2017mkq}. We assume a central observing frequency of 1400\,MHz for all surveys. Columns $T_\mathrm{rec}$, $t_\mathrm{samp}$ and $t_\mathrm{int}$ signify receiver temperature, sampling time and integration time respectively.}
    \hspace{-1.0cm}
    \begin{tabular}{lrrrrrr}
    \hline
    Telescope & Bandwidth & Gain & $T_\mathrm{rec}$ & $t_\mathrm{samp}$ & $t_\mathrm{int}$ & Coverage \\
    survey & (MHz) &  & (K) & ($\mu$s) & (s) & - \\
    \hline

Parkes$^{(a)}$  & 288 & 0.65 & 24 & 256 & 2100 & all-sky\tablefootnote{We do not apply a cut-off to the sky coverage of our Parkes-like survey, since we are using it as a representative proxy for the multiple surveys that have discovered DNSs across the globe.
} \\
MeerKAT$^{(b)}$ & 400 & 1.80 & 18 & 64 & 637 & Galactic-plane\tablefootnote{cut-off point:$|b|<5.2$ and $-110^{\circ}<l<10^{\circ}$}   \\
MeerKAT$_\mathrm{F}$ & 800 & 2.80 & 18 & 64 & 2100 & all-sky-cutoff\tablefootnote{cut-off point: $\delta<30^{\circ} $}  \\
MeerKAT$_\mathrm{T}$ & 800 & 2.80 & 18 & 64 & 300 & all-sky-cutoff$^{3}$  \\
MeerKAT$_\mathrm{G}$ & 800 & 2.80 & 18 & 64 & 2100 & Galactic-plane\tablefootnote{cut-off point:  $-5^{\circ}<\delta<5^{\circ}$}  \\
MeerKAT$_\mathrm{GT}$ & 800 & 2.80 & 18 & 64 & 300 & Galactic-plane$^{4}$  \\
SKA$^{(c)}$     & 300 & 8.40 & 30 & 64 & 2100 & all-sky-cutoff$^{3}$ \\
    \hline
    \end{tabular}
    \label{table:telescope_specs}
\end{table}

%% file: modelsTable.tex

\begin{table*}
    \resizebox{0.85\textwidth}{!}{\begin{minipage}{\textwidth}
    \hspace{-1.9cm}
    \begin{tabular}{lrrrrrrrrrr}
    \hline
    Model & Mass Range & BH kick prescription & CE model & CE accretion &$\tau_d$ & $\Delta M_d$ & $B_\mathrm{birth}$ distribution & $B_\mathrm{spin}$ distribution & Metallicity & Supernovae Prescription\\
    & ($\mathrm{M_\odot}$) & - & - & - &(Myrs) & ($\mathrm{M_\odot}$) & - & - & (Z) & - \\
    \hline
Fiducial   & 4--100 & Fallback & Optimistic& Macleod+ &1000 & 0.15 & Uniform & Uniform & 0.0142  & Delayed\\
BHK-Z   & 4--100 & Zero & Optimistic & Macleod+ &1000 & 0.15 & Uniform & Uniform & 0.0142 & Delayed\\
BHK-F   & 4--100 & Full & Optimistic &  Macleod+& 1000 & 0.15 & Uniform & Uniform & 0.0142 & Delayed \\ 
CE-P   & 4--100 & Fallback & Pessimistic & Macleod+ &1000 & 0.15 & Uniform & Uniform & 0.0142 & Delayed \\
CE-Z & 4-100 & Fallback & Optimistic & Zero& 1000 & 0.15 & Uniform & Normal &0.0142  & Delayed \\
ZM-001 & 4--100 & Fallback & Optimistic & Macleod+ & 1000 & 0.15 & Uniform & Uniform & 0.001  & Delayed\\
ZM-02 & 4--100 & Fallback & Optimistic & Macleod+ & 1000 & 0.15 & Uniform & Uniform & 0.02  & Delayed\\
FDT-500  & 4--100 & Fallback & Optimistic & Macleod+ & 500 & 0.15 & Uniform & Uniform & 0.0142  & Delayed\\  
FDM-20 & 4--100 & Fallback & Optimistic & Macleod+ & 1000 & 0.2 & Uniform & Uniform & 0.0142 & Delayed \\  
BMF-FL & 4--100 & Fallback & Optimistic & Macleod+ & 1000 & 0.15 & Flat in Log & Uniform & 0.0142  & Delayed\\
RM-R   & 4--100 & Fallback & Optimistic& Macleod+ &1000 & 0.15 & Uniform & Uniform & 0.0142 & Rapid \\

    \hline
    \end{tabular}
    \end{minipage}}
    \caption{Suite of models used for the analysis. Each model following Fiducial has one parameter that has been varied from it.} 
    \label{tab:NSBH}
\end{table*}

%% file: ZAMS.tex
\begin{table*}
\begin{tabular}{lrrrrrrrr}
\hline 
Variable    & M$_\mathrm{NS}$ (ZAMS) & M$_\mathrm{BH}$ (ZAMS) & M$_\mathrm{NS}$ & M$_\mathrm{BH}$ & a (ZAMS) & a(DCO) & t$_\mathrm{birth}$ & $\Delta$t$_\mathrm{form}$ \\
            & $\mathrm{M_\odot}$             & $\mathrm{M_\odot}$             & $\mathrm{M_\odot}$      & $\mathrm{M_\odot}$      & AU      & AU & Gyrs   & Myrs              \\
            \hline
BHNS\_total   &      20.82            &    59.24              &    1.45       &   6.78        &    4.77     &  2.81  &   7.97     &      13.41             \\
BHNS\_radio &     37.93             &       69.35           &    1.50       &     7.88      &     3.34    &  0.94  &   12.78     &         9.38          \\
NSBH\_total   &     23.26             &        18.01          &     1.48      &    3.85       &    0.67     & 0.22   &   7.16     &      11.86             \\
NSBH\_radio &       23.01           &     17.00             &    1.49       &  3.42  &0.39       &   0.07      &  10.87  &    12.40    \\                
\hline
\end{tabular}
\caption{Mean values of binary parameters for NS+BH binaries in our Fiducial model. While "(ZAMS)" refers to the parameter mean value at zero-age  at sequence, "(DCO)" refers to the same at the formation of the double compact object. t$_\mathrm{birth}$ denotes the birth time of the pulsar (in a 13 Gyr old Milky Way) while $\Delta$t$_\mathrm{form}$ signifies the time taken by the binary to evolve from zero-age main sequence star to double compact object. The distributions of some these quantities are shown in Section~\ref{sec:results_radio}.}
\label{tab:ZAMS_radioVSnet}
\end{table*}

%% file: NetNumbers.tex
\begin{table*}
\label{table:models}
    \resizebox{0.95\textwidth}{!}{\begin{minipage}{\textwidth}
    \hspace{-1.3cm}
    \begin{tabular}{ |l|rrrrr|rrrrr|rrrrcc| }
    \hline
    Model &  \multicolumn{5}{c}{NSBH}  &  \multicolumn{5}{c}{BHNS} &  total & total & DNS & DNS& $\mathcal{R}$ & $\mathcal{R}_\mathrm{SKA}$ \\
    \hline
     & net & radio & Parkes & MeerKAT & SKA & net & radio & Parkes & MeerKAT & SKA & Parkes & SKA & Parkes  & SKA & &\\
   
    \hline
Fiducial$^*$   & 764 & 145 & 3 & 5 & 9 & 27599 & 518 & 5 & 10 & 21 & 8 & 30 & 27 & 78 & 0.29  & 0.38 \\
BHK-Z$^\dagger$    & 2231 & 299 & 18 & 23 & 50 & 123090  & 1747 & 22 & 33 & 72 & 40 & 122 & 27  & 71 & 1.48  & 1.72 \\

BHK-F$^*$    & 297 & 62 & 2 & 3 & 7 & 5074 & 60 & 1 & 1 & 2 & 3 & 9 & 25  & 72 & 0.12 & 0.13 \\
CE-P$^*$    & 36 & 1 & 0 & 0 & 0 & 23541 & 185 & 2 & 4 & 6 & 2 & 6 & 21  & 62 &0.10 & 0.10 \\
CE-Z$^\dagger$    & 734 & 7 & 0 & 0 & 0 & 27325 & 483 & 5 & 9 & 18 & 5 & 18 & 3  & 11 &1.67 & 1.64 \\
ZM-001$^*$    & 7754 & 103 & 1 & 1 & 3 & 45867 & 533 & 8 & 10 & 21 & 9 & 24 &  14 & 45 &0.64 & 0.53 \\
ZM-02$^*$    & 467 & 66 & 2 & 3 & 6 & 12895 & 120 & 2 & 2 & 4 & 4 & 10 & 40 & 121 &0.10 & 0.08 \\
FDT-500$^*$    & 823 & 94 & 3 & 3 & 6 & 26631 & 461 & 5 & 8 & 18  & 8 & 24 &  20 & 62 & 0.40 & 0.39 \\
FDM-20$^*$    & 902 & 93  & 3 & 4 & 9 & 27807 & 500 & 5 & 10 & 20 & 7 & 29 & 19  & 64 &0.37 &0.45 \\
BMF-FL$^\dagger$    & 7900 & 2939 & 90 & 125 & 264 & 275782 & 32421 & 551 & 877 & 1910 & 641 & 2174 & 78 & 239 &8.21 &9.10 \\
RM-R$^*$ & 480 & 26 & 0 & 0 & 0 & 77141 & 975 & 11 & 19 & 40 & 11 & 40 & 33 & 85 & 0.33 & 0.47\\

    \hline
    \end{tabular}
    \end{minipage}
    }
    \caption{Quantitative distributions of NS+BHs for each model scaled to a population representative of the Milky Way. The column `net' signifies the total count of binaries: both radio and non-radio. The survey observed data-sets are hence a subset to the radio population, after accounting for the radio observational biases. The column `total' signifies the combined survey-observed NSBH and BHNS systems, post-radio-selection effects.
    The ratio of the number of observations by Parkes, estimating the current detection rate of NS+BH per DNS is given by $\mathcal{R}$ = (Parkes: NS+BH)/(Parkes: DNS). To estimate the future detection rate we quote $\mathcal{R}_\mathrm{SKA}$ = (SKA: NS+BH)/(SKA: DNS).
    Superscripts $*$ and $\dagger$
    on model names denote feasible or unfeasible models respectively on the basis of order of magnitude disparity in detection rates. The Poisson uncertainty for $\mathcal{R}$ and $\mathcal{R}_\mathrm{SKA}$ remain $\lessapprox$0.2 for all models.}
    \label{tab:netNumbers}
\end{table*}

%% file: radioLifetime.tex
\begin{table}
    \begin{tabular}{lrrrrr}
    \hline
    Model  & \multicolumn{3}{c}{$r_\mathrm{birth}$ (Myrs$^{-1}$)} &  \multicolumn{2}{c}{$t_\mathrm{radio}$ (Myrs)}  \\
    &&&&&\\
     & NSBH & BHNS & NS+BH & NSBH & BHNS \\
    \hline
     Fiducial & 0.10 & 8.31 & 8.41 &1333 & 65\\
     BHK-Z & 0.13 & 27.74 & 27.87 &847 & 67 \\
     BHK-F & 0.05 & 1.05 & 1.10 &1197 & 64\\
     CE-P & 0.003 & 2.67 & 2.67 &137 & 74\\
     CE-Z & 0.08 & 8.23 & 8.31 &77 & 63 \\
     ZM-001 & 1.49 & 8.56 & 10.05 &87 & 72\\
     ZM-02 & 0.05 & 2.08 & 2.13 &1351 & 71\\
     FDT-500 & 0.09 & 8.04 & 8.13 & 873 & 59\\
     FDM-20 & 0.11 & 8.27& 8.38 &919 & 65\\
     BMF-FL & 0.10 & 8.15 & 8.25 &3252 & 435 \\
     RM-R  & 0.07 & 16.94 & 17.01 &476 &  63 \\
    \hline
    \end{tabular}
    \caption{Average birth rates $r_\mathrm{birth}$ and radio lifetimes $t_\mathrm{radio}$ of NSBH and BHNS binaries in each of our models. 
    In general, pulsars in BHNSs are non-recycled and have shorter lifetimes than the partially recycled pulsars in NSBHs.
    The birth rates $r_\mathrm{birth}$ are a function of the binary evolution parameters rather than the pulsar ones. 
    Hence the Fiducial model, FDT-500, FDM-20 and BMF-FL all have similar ranges of birth rates.
    }
    \label{table:radio_lifetime}
\end{table}

%% file: MeerKAT.tex
\begin{table}
    \caption{Estimates of the number of observed NSBHs and BHNSs from our Fiducial model under several different assumptions about our mock survey with MeerKAT. We report the number of detections relative to our default MeerKAT survey as $\mathcal{F}$, which we take to be model independent.}
    \hspace{0.5cm}
    \begin{tabular}{lcccc}
    \hline
    Telescope Survey & NSBH & BHNS & total & $\mathcal{F}$ \\
    
    \hline

MeerKAT & 5 & 10 & 15 & 1.0  \\
MeerKAT$_\mathrm{F}$ & 9 & 21 & 30 & 2.0   \\
MeerKAT$_\mathrm{T}$ & 9 & 15 & 24 & 1.6 \\
MeerKAT$_\mathrm{G}$ & 8 & 12  & 20 & 1.4 \\
MeerKAT$_\mathrm{GT}$ & 7 & 11 & 18 & 1.2  \\

    \hline
    \end{tabular}

    \label{tab:MeerKAT_rates}
\end{table}

%% file: meanValues_Observables.tex
\begin{table*}
\label{table:models}
    \resizebox{0.8\textwidth}{!}{\begin{minipage}{\textwidth}
    \hspace{-2cm}
    \begin{tabular}{|l|rrrrrrrr|rrrrrrrr}
    \hline
    Model &  \multicolumn{8}{c|} {NSBH}  &  \multicolumn{8}{c|}{BHNS}  \\
    \hline
     & $P$ & $\Dot{P}$ & $B$ & $P_{\mathrm{orb}}$ & $e$ & $m_{\mathrm{psr}}$ & $m_{\mathrm{cmp}}$ & $|Z|$ & $P$ & $\Dot{P}$ & $B$ & $P_{\mathrm{orb}}$ & $e$ & $m_{\mathrm{psr}}$ & $m_\mathrm{{cmp}}$ & $|Z|$  \\
     & s & s/s & Gauss & days & - & M$_\odot$ & M$_\odot$ & Kpc & s & s/s & Gauss & days & - & M$_\odot$ & M$_\odot$ & Kpc \\
   
    \hline
Fiducial   & 0.104 & 5.992e-17 & 5.424e10 & 10.702 & 0.511 & 1.491 & 3.417 &  0.385 & 3.135 & 5.420e-15 & 2.074e12 & 239.922 & 0.231  & 1.496 & 7.884 &  0.160\\
BHK-Z   & 0.079  & 4.358e-17 & 3.535e10 & 1473.933 & 0.882 & 1.516 & 3.272 & 0.134 & 3.112 & 5.596e-15  &2.029e12 & 178.434  & 0.250 & 1.474 & 5.747 & 0.227\\
BHK-F   &  0.081 & 1.012e-17 & 2.424e10 & 20.174 & 0.645 & 1.504 & 3.615 & 2.086 & 3.272 & 3.458e-15 & 2.002e12 & 904.641 & 0.305 & 1.427 & 6.025 &  0.194 \\
CE-P   & 0.925 & 4.956e-16 & 5.128e11 & 160.451 & 0.473 & 2.014 & 4.897 & 0.595 &2.946  &  3.559e-15 & 1.915e12 & 16653.155 & 0.389 & 1.493 & 6.474 & 0.215\\
CE-Z   & 2.465 & 1.471e-15 & 1.448e12 & 82.078 &  0.552 & 1.544  & 4.060 & 0.561 & 3.272 & 4.863e-15 & 2.149e12 & 1110.678 & 0.232 & 1.500 & 7.995 & 0.177 \\
ZM-001   & 1.673 & 1.384e-15 & 1.087e12 & 5.470 & 0.353 & 1.807 &  14.486 & 0.164 & 2.996 & 6.970e-15 & 2.175e12 & 9515.507 & 0.394 & 1.749 & 10.663 &  0.238 \\
ZM-02   & 0.120 & 3.361e-17 &  5.482e10 & 8.856 & 0.479 & 1.438 & 3.041 & 0.575 & 3.316 & 6.182e-15 & 2.070e12 & 490.496 & 0.367 & 1.365 & 5.027 & 0.175\\
FDT-500   & 0.061 & 2.450e-17 & 2.861e10 & 856.467 & 0.544 & 1.514 & 3.314 & 0.600 & 3.124  & 3.480e-15 & 2.090e12 & 214.027 & 0.228 & 1.491 & 7.951 & 0.164\\
FDM-20   & 0.191 & 4.420e-17  & 6.018e10 & 3.409 & 0.544 & 1.527 & 3.371 & 0.445 &  3.211 & 8.718e-15 & 2.150e12 & 445.976 & 0.237 & 1.502 & 7.936 & 0.169\\
BMF-FL   & 0.038 & 5.864e-18 & 9.134e09 & 2452.445 & 0.582 & 1.520 & 3.345 & 0.553 & 0.519 & 2.095e-16 &  1.849e11 & 847.694 & 0.255 & 1.460 & 7.592 & 0.179\\

RM-R & 0.334 & 1.335-16 & 1.475e-11 & 832.403 & 0.108 & 1.249 & 7.598 & 0.112 & 3.242 & 5.416e-15 & 1.953e-12 & 759.672 & 0.231 & 1.234 & 7.978 & 0.226\\
    \hline
    \end{tabular}
    \end{minipage}}
    \caption{Mean values of radio observables for PSR+BH binaries. We show the values for both NSBH and BHNS sub-populations for each of our models.}
    \label{tab:radioObservablesMean:Radio}
\end{table*}

%% file: meanValuesSKA.tex
 \begin{table*}
 \label{table:models}
 \resizebox{0.8\textwidth}{!}{
 
 \begin{minipage}[l]{\textwidth}
 \hspace{-2cm}
 \begin{tabular}{|l|rrrrrrrr|rrrrrrrr}
 \hline
 Model &  \multicolumn{8}{c|} {NSBH}  &  \multicolumn{8}{c|}{BHNS}  \\
 \hline
 & $P$ & $\Dot{P}$ & $B$ & $P_{\mathrm{orb}}$ & $e$ & $m_{\mathrm{psr}}$ & $m_{\mathrm{cmp}}$ & $|Z|$ & $P$ & $\Dot{P}$ & $B$ & $P_{\mathrm{orb}}$ & $e$ & $m_{\mathrm{psr}}$ & $m_\mathrm{{cmp}}$ & $|Z|$  \\
 & s & s/s & Gauss & days & - & M$_\odot$ & M$_\odot$ & Kpc & s & s/s & Gauss & days & - & M$_\odot$ & M$_\odot$ & Kpc \\
   
 \hline
Fiducial   & 0.047 & 1.952e-17 & 1.626e10 & 25.836 & 0.777 & 1.504 & 3.284 & 0.807 & 1.058 & 8.555e-15 & 1.217e12 & 80.252 & 0.262 & 1.504 & 7.688 & 0.186\\

BHK-Z & 0.025 & 9.699e-18 & 6.923e09 & 88.373 & 0.890 & 1.513 & 3.244 & 0.222 & 0.983 & 4.246e-15 & 1.102e12 & 136.315 & 0.274 & 1.505 & 5.699 & 0.209\\
BHK-F   & 0.028 & 9.560e-19 & 4.648e09 & 46.515 &  0.888 & 1.508 & 3.797 & 1.813 & 1.113 & 2.672e-15 & 1.160e12 & 604.350 & 0.322 & 1.427 & 5.963 & 0.245\\

CE-P   & 0.239 & 1.298e-16 & 1.039e11 & 5.017 &  0.304 & 2.067 & 5.276 & 0.324 & 0.866 & 3.616e-15 & 9.452e11 & 673.042 & 0.419 & 1.495 & 6.3933 & 0.232\\
CE-Z   & 0.823 & 9.082e-16 & 6.722e07 & 78.275 & 0.558 & 1.564  & 4.121  &  0.294 & 1.129 &  5.806e-15  & 1.275e12 & 1550.569 & 0.246 & 1.516 & 7.968 &  0.198\\

ZM-001   & 0.659 & 1.255e-15 & 6.689e11 & 4.884  & 0.431 & 1.605 & 14.653 & 0.213 & 0.954 & 1.311e-14 & 1.229e12 & 14093.453 & 0.421 & 1.741 & 10.678 & 0.262\\
ZM-02   & 0.045 & 5.165e-18 & 8.182e09 & 18.032 & 0.758 & 1.455 & 2.989 & 0.581 & 0.972 & 2.815e-15 & 1.024e12 & 1368.329 & 0.400 & 1.415 & 5.340 & 0.181\\

FDT-500   & 0.033 & 2.500e-17 & 1.485e10 & 1346.088 & 0.796 & 1.522 & 3.328 & 1.108 & 1.055 & 4.200e-15 & 1.204e12 & 142.553 & 0.235 & 1.491 & 7.863 & 0.172 \\
FDM-20   & 0.126 & 1.079e-17 & 2.217e10 & 3.887 & 0.637 & 1.518 & 3.411 & 0.467 & 1.101 & 2.793e-14  & 1.209e12 & 736.782 &  0.255 & 1.516 & 7.919 & 0.214\\
BMF-FL   & 0.022 & 3.423e-18 & 4.918e09 & 6311.351  & 0.813 & 1.558 & 3.256 &  0.855 & 0.323 & 1.985e-16 & 1.243e11 & 1276.610 &  0.359 & 1.485 & 7.131 & 0.233\\

RM-R & 0.448 & 2.243e-16 & 2.237e-11 & 1201.290 & 0.293 & 1.309 & 7.490 & 0.171 & 1.079 & 8.380e-15 & 1.115e-12 & 393.102 & 0.253 & 1.242 & 7.950 & 0.246\\
    \hline
    \end{tabular}
    
    \end{minipage}}
    \caption{Mean values of radio observables for PSR+BH binaries observed by the SKA. This population accounts for radio selection effects as described in Section~\ref{subsec:radio_selection_effects}. We show the values for both NSBHs and BHNSs for each of our models.}
    \label{tab:radioObservablesMean:SKA}
  \end{table*}